\documentclass[%
 reprint,
superscriptaddress,
twocolumn,
 amsmath,amssymb,
 aps,
prx,
]{revtex4-2}

\usepackage{booktabs}
\usepackage{graphicx}
\usepackage{dcolumn}
\usepackage{bm}
\usepackage{xcolor}
\usepackage{float}
\usepackage{microtype}
\usepackage{bbm}
\usepackage{amsmath}
\usepackage{amsfonts}
\usepackage{braket}
\usepackage{txfonts}
\usepackage[normalem]{ulem}

\usepackage{hyperref}
\hypersetup{colorlinks=true,breaklinks,linkcolor=blue,urlcolor=blue,citecolor=blue}
\usepackage{orcidlink}

\newcommand\AddAuthorComment[3]{
    {\color{#1} ({\bf #2}%
        \if\relax\detokenize{#3}\relax%
        \else%
            {\normalfont: #3}%
        \fi%
    )}%
}
\newcommand\AuthorReplace[5]{
    \AddAuthorComment{#1}{#2}{#3}%
    \if\relax\detokenize{#4#5}\relax%
    \else%
        { \color{#1}\sout{#4}\uwave{#5}}
    \fi%
}

\definecolor{darkgreen}{rgb}{0,0.5,0}

\newcommand\Dmax{\ensuremath{D_\mathrm{max}}}

\begin{document}

\preprint{APS/123-QED}

\title{
Entanglement across scales: Quantics tensor trains as a natural framework for renormalization}

\author{Stefan Rohshap\,\orcidlink{0009-0007-2953-8831}}
\email{stefan.rohshap@tuwien.ac.at}
\affiliation{%
    Institute of Solid State Physics, TU Wien, 1040 Vienna, Austria}%

\author{Jheng-Wei Li}
\affiliation{%
    Université Grenoble Alpes, CEA, Grenoble INP, IRIG, Pheliqs, F-38000 Grenoble, France}%

\author{Alena Lorenz\,\orcidlink{0000-0001-6831-3685}}
\affiliation{%
    Institute of Solid State Physics, TU Wien, 1040 Vienna, Austria}%
\affiliation{%
    Institut f\"ur Theoretische Physik und Astrophysik and W\"urzburg-Dresden Cluster of Excellence ct.qmat, Universit\"at W\"urzburg, 97074 W\"urzburg, Germany}%

\author{Serap Hasil\,}
\affiliation{%
    Institute of Solid State Physics, TU Wien, 1040 Vienna, Austria}%

\author{Karsten Held\,\orcidlink{0000-0001-5984-8549}}
\affiliation{%
    Institute of Solid State Physics, TU Wien, 1040 Vienna, Austria}

\author{Anna Kauch\,\orcidlink{0000-0002-7669-0090}}
\affiliation{%
    Institute of Solid State Physics, TU Wien, 1040 Vienna, Austria}%

\author{Markus Wallerberger\,\orcidlink{0000-0002-9992-1541}}
\affiliation{%
    Institute of Solid State Physics, TU Wien, 1040 Vienna, Austria}%

\date{\today}

\begin{abstract}
Understanding entanglement remains one of the most intriguing problems in physics. While particle and site entanglement have been studied extensively, the investigation of length or energy scale entanglement — quantifying the information exchange between different length scales — has received far less attention. 
Here, we identify the quantics tensor train (QTT) technique, a matrix product state-inspired approach for overcoming computational bottlenecks in resource-intensive numerical calculations, as a renormalization group method by analytically expressing an exact cyclic reduction-based real-space renormalization scheme in QTT language, which serves as a natural formalism for the method. In doing so, we precisely match the QTT bond dimension — a measure of length scale entanglement — to the number of rescaled couplings generated in each coarse-graining renormalization step. While QTTs have so far been applied almost exclusively to numerical problems in physics, our analytical calculations demonstrate that they are also powerful tools for mitigating computational costs in semi-analytical treatments. We present our results for the one-dimensional tight-binding model with $\mathbf{n}$-th-nearest-neighbor hopping, where the $\mathbf{2n}$ rescaled couplings generated in the renormalization procedure precisely match the QTT bond dimension of the one-particle Green's function.
\end{abstract}

\maketitle

\section{\label{sec:introduction}Introduction}


Romanesco broccoli and the neuron-based brain architecture~\cite{Smith2021} are two striking examples of naturally occurring fractals, exhibiting patterns that repeat across multiple scales. The scale invariance inherent in such structures directly connects to the concept of the renormalization group (RG) in physics, which leverages coarse-graining and rescaling to study complex systems. Since the early spin-blocking ideas~\cite{Kadanoff1966}, the formulation of real-space RG methods~\cite{Wilson1975} has laid the foundation for powerful techniques in quantum field theory, including the numerical renormalization group~\cite{Bulla2008, Weichselbaum2009, Kugler2021, Ritz2025}, functional renormalization group~\cite{Wetterich1993, Kugler2018, Hille2020, Bippus2024}, and density matrix renormalization group (DMRG)~\cite{White1992, White1993, Schollwoeck2005} approaches. Although the matrix product state (MPS)~\cite{Baxter1968} representation was originally developed independently, its integration into DMRG allowed the method to become the state-of-the-art for obtaining low-energy physics in one-dimensional lattice systems. The success of DMRG for treating one-dimensional short-ranged correlated systems relies on a bound on the strength of quantum entanglement dictated by the correlation length via the entropic area law~\cite{Cho2018}, which ensures efficient representation of ground states as MPSs with low bond dimensions.

However, near criticality, entanglement can grow large. This problem was addressed by the development of the multiscale entanglement renormalization ansatz (MERA)~\cite{Vidal2007,Vidal2008, Crosswhite2008}. MERA implements a quantum circuit with logarithmic depth that effectively coarse-grains the system, renormalizing entanglement organized in layers corresponding to different length scales. Beyond its role in quantum criticality~\cite{Pfeifer2009} and holography~\cite{Swingle2012}, entanglement renormalization has also been linked to wavelet transforms~\cite{Evenbly2016, Haegeman2018}. Notably, the development of wavelets~\cite{Mallat1989, Daubechies1990} was influenced by RG concepts from the outset, and the scope of this technique was only recently extended to efficiently compress correlation functions in a numerical study~\cite{Moghadas2024}. 

In a parallel development, the compressibility of correlators was studied in the same model in Ref.~\onlinecite{Rohshap2024}, following a different approach with the quantics tensor train (QTT)~\cite{Oseledets2009, Oseledets2011, Khoromskij2011, Dolgov2012, Khoromskij2018} technique. 
Besides studying the compressibility of correlators with QTTs, also involved two-particle calculations, namely the parquet equations, were performed within the QTT framework. Given that QTTs were identified as algebraic wavelet transforms with adaptively determined filters~\cite{Oseledets2011b}, it is unsurprising that both approaches allow the tackling of similar problems. This connection suggests a deeper link between QTTs and MERA, as both exploit the separation of exponentially different length or energy scales.

\begin{figure}
    \centering
    \includegraphics[width=0.95\linewidth]{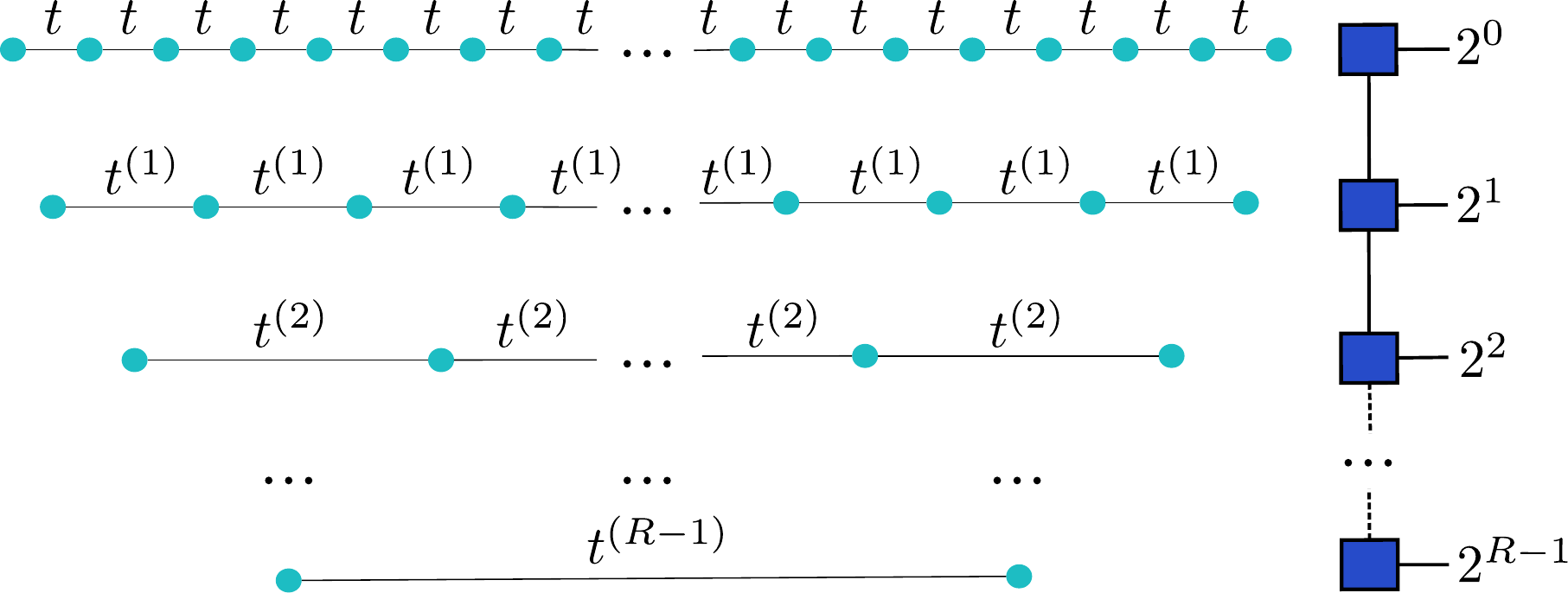}
    \caption{QTT decomposition with separation of exponentially different length scales as a natural framework of real-space RG methods.
    }
    \label{fig:qtt-renormalization}
\end{figure}

QTTs have rapidly emerged as practical tools across various fields. The method has proven useful not only in the study of turbulence~\cite{Gourianov2022-vn,peddinti2023complete, kornev2023,holscher2024, Gourianov2024, GomezLozada2025}, plasma physics~\cite{Ye2022} and quantum chemistry~\cite{Jolly2023}, but especially in quantum many-body physics~\cite{Shinaoka2023,Erpenbeck2023,Ritter2024, Ishida2024, Murray2024, Takahashi2025, Eckstein2024, OteroFumega2024, Frankenbach2025,Niedermeier2025, Inayoshi2025, Mizuno2025}.  In this context, the QTT approach enables the treatment of highly resolved grids by approximating input data with only a logarithmic number of parameters, offering the potential to outperform state-of-the-art sparse modeling approaches~\cite{Eckhardt2020, Astretsov2019, Wentzell2020,Krien2020a, Krien2020b} or compact representations in a suitable basis~\cite{Wallerberger2021,Shinaoka2020,Shinaoka2018, Kiese2024}. 

The success of the QTT method depends on the amount of length scale entanglement in the system, which characterizes the flow of information between different scales and, thus, the separability of those length scales. Despite its importance, this notion of entanglement, distinct from site or particle entanglement~\cite{Li2024, Bellomia2024, Bippus2025a, Bippus2025b}, has received little attention. Hitherto only in a recent work has time scale entanglement been shown to be a system-inherent property that peaks at phase transitions, enabling QTT-based diagnostics of critical points~\cite{Rohshap2025}. 

In this paper, we bridge the gap between length scale entanglement, QTTs, and RG by analytically demonstrating that the QTT framework naturally represents a coarse-graining real-space renormalization procedure~\cite{Bini1999} for the $n$-th-nearest-neighbor tight-binding model (see Fig.~\ref{fig:qtt-renormalization} for comparison). Leveraging mathematical methods for analyzing the QTT bond dimension of Toeplitz matrices~\cite{Zamarashkin2009, Oseledets2011c, Kazeev2012, Kazeev2013, Vysotsky2023}, we recast this RG scheme into QTT language in an exact manner, showing that the QTT bond dimension of the single-particle Green's function precisely matches the number of rescaled couplings generated in the RG procedure. While QTTs have proven powerful in compressing data, their approximation accuracy relative to the achieved compression rate is generally not known a priori in numerical treatments. Here, we derive analytical error bounds for the approximate QTT construction of the one-particle Green’s function, specifically for infinitely decaying couplings, as a function of the chosen maximum bond dimension. Finally, we complement our analytical results with detailed numerical studies, which also extend the analysis to higher-dimensional and interacting systems.

This paper is organized as follows: In Sec.~\ref{sec:renormalization}, we introduce the tight-binding model and calculate the one-particle Green's function through a cyclic reduction-based RG scheme. Next, QTTs and the relevant formalisms are reviewed (Sec.~\ref{sec:qtt}). Subsequently, in Sec.~\ref{sec:qtt-renormalization}, the renormalization procedure is recast into the QTT language, which we find to be a natural formalism of the RG method. Next, analytical error bounds for the QTT construction in the case of infinitely decaying couplings are derived in Sec.~\ref{sec:approximate-construction}. Last, Sec.~\ref{sec:numerical-experiments} presents numerical verifications and extensions to higher-dimensional and interacting cases. We conclude in Sec.~\ref{sec:conclusions}. Detailed calculations are provided in the appendix and the Supplemental Material~\cite{supp}.

\section{\label{sec:renormalization}Green's function renormalization}
\subsection{Model}
Let us consider a one-dimensional chain of atoms consisting of $2^R$ lattice sites arranged in a ring. In the case of tightly bound non-interacting electrons, this system can be described by the tight-binding model with periodic boundary conditions with the Hamiltonian $\hat H$ in second quantization of the form
\begin{align}
    \hat H=-\sum_{ij} t_{ij} \hat c_i^{\dagger} \hat c_j,
\end{align}
with the fermionic annihilation (creation) operators $\hat c^{(\dagger)}_i$, annihilating (creating) an electron at site $i$ and the parameter $t_{ij}$ describing the hopping amplitude between sites $i$ and $j$. Although spin indices will be neglected in this calculation, the approach is also applicable to this case by interpreting site $i$ instead as a spin orbital or, more generally, as flavor. The one-particle Green's function $G_{ij}(\omega)$ describing the propagation of an added electron between sites $j$ and $i$ can then be obtained from the following equation
\begin{align} \label{eq:general-GF}
    \sum_j \left( \omega \delta_{ij} + t_{ij} \right) G_{jk}(\omega) = \delta_{ik}.
\end{align}
Due to periodic boundary conditions, the system is translationally invariant, which allows for the simplification $G_i := G_{i0}(\omega) = G_{(i+j)j}(\omega)$. Generally, $G_{i}$ can then be obtained by inverting $\left( \omega \delta_{ij} + t_{ij} \right)$. For large systems, this can become computationally demanding, especially if we are interested in analytical results. In the following, we present an alternative route, a renormalization-based approach for obtaining the Green's function.  In this RG procedure, the system will gradually be coarse-grained generating renormalized couplings at each step of this cyclic reduction approach~\cite{Bini1999}, until only two sites remain. This will allow us to easily obtain the local Green's function $G_0$ and, additionally, some of the nonlocal ones $G_i$.

\subsection{Nearest-neighbor hopping\label{sec:nn-exact-greens-function-renormalization} }
In order to motivate the idea of the cyclic reduction-based renormalization procedure, we first focus on the simplest case of nearest-neighbor hopping only, where $t_{ij} := t (\delta_{j,i-1}+\delta_{j,i+1})$. Then, Eq.~\eqref{eq:general-GF} simplifies to
\begin{align} \label{eq:nn-general-GF}
    t G_{i-1} + \omega G_{i}+ t G_{i+1} = \delta_i,
\end{align}
which connects the Green's function of neighboring sites, where we define $\delta_i \equiv \delta_{i0}$ here and in the following. In order to coarse-grain the system, we aim at connecting the Green's functions of next neighboring sites instead. Therefore, we can shift Eq.~\eqref{eq:nn-general-GF} and arrive at the following three equations
\begin{subequations}
\begin{align} 
    &t G_{i-2} \ \ + &\omega  G_{i-1} \ \ &+ t G_{i} \ \  \ &\ &\  \ &=& \ \delta_{i-1}, \label{eq:nn-minus1-GF} \\
    &\ &t G_{i-1} \ \  &+ \omega G_{i}   &+ t G_{i+1} \ \ \ &  \ &=& \ \delta_i, \label{eq:nn-minus0-GF} \\
    &\ &\ &\ \ \ \ \ t G_{i}   &+ \omega G_{i+1} \ \   &+ t G_{i+2}  &=& \ \delta_{i+1}, \label{eq:nn-plus1-GF} 
\end{align}
\end{subequations}
Assuming $\omega \neq 0$ from this point onward, we add $-\frac{t}{\omega}$ times Eqs. (\ref{eq:nn-minus1-GF}, \ref{eq:nn-plus1-GF}) to Eq.~\eqref{eq:nn-minus0-GF}, which leads to the cancellation of the $G_{i-1}$ and $G_{i+1}$ terms resulting in the following equation 
\begin{align}
    -\frac{t^2}{\omega} G_{i-2} + \left (\omega-\frac{2t^2}{\omega} \right) G_i - \frac{t^2}{\omega} G_{i+2} = \delta_i -\frac{t}{\omega} \left (\delta_{i-1}+\delta_{i+1}\right).
\end{align}
Now, this first cyclic reduction step can be completed by separating out the even parts, leading to the renormalized equation
\begin{align} \label{eq:nn-general-GF-first-renormalization}
    \underbrace{-\frac{t^2}{\omega}}_{t^{(1)}} G_{2(i-1)} + \underbrace{\left(\omega-\frac{2t^2}{\omega}\right)}_{\omega^{(1)}} G_{2i} \underbrace{- \frac{t^2}{\omega}}_{t^{(1)}} G_{2(i+1)} = \delta_i ,
\end{align}
where the renormalized couplings $t^{(1)}, \omega^{(1)}$ are generated. Since Eq.~\eqref{eq:nn-general-GF-first-renormalization} is now of the same form as Eq.~\eqref{eq:nn-general-GF}, we can iterate this procedure until the $r$-th renormalization step leads to
\begin{subequations}
    \begin{align} 
    \delta_i &= t^{(r)} G_{2^r(i-1)} + \omega^{(r)} G_{2^ri} + t^{(r)} G_{2^r(i+1)}, \label{eq:nn-r-GF}\\
    \omega^{(r)} &= \omega^{(r-1)}-\frac{2|t^{(r-1)}|^2}{\omega^{(r-1)}}, \\
    t^{(r)} &= -\frac{|t^{(r-1)}|^2}{\omega^{(r-1)}},
\end{align}
\end{subequations}
with the rescaled couplings $\omega^{(r)}$ and $t^{(r)}$ expressed recursively through the couplings of the previous coarse-graining step. We note that if any $\omega^{(r)}$ is zero, the system becomes singular at this point. 
Repetition of the cyclic reduction procedure until $r=R-1$, a coarse-grained system consisting of only two rescaled sites is obtained. Inserting $i=0,1$ we obtain the following set of equations 
\begin{subequations}
    \begin{align}
    1 &= \omega^{(R-1)} G_0 + 2t^{(R-1)} G_{2^{(R-1)}} , \label{eq:equation-R1-1} \\
    0 &= 2t^{(R-1)} G_0 + \omega^{(R-1)} G_{2^{(R-1)}} , \label{eq:equation-R1-2}
\end{align}
\end{subequations}
where we used $G_{2^{(R-1)}}=G_{2^{R}+2^{(R-1)}}=G_{-2^{(R-1)}}$ due to periodic boundary conditions. These two coupled equations can easily be solved, leading to the results
\begin{subequations}
    \begin{align} 
    G_0 &= \frac{\omega^{(R-1)}}{\left(\omega^{(R-1)}\right)^2-4\left(t^{(R-1)}\right)^2}, \label{eq:result-G0} \\
    G_{2^{(R-1)}} &= -\frac{2t^{(R-1)}}{\left(\omega^{(R-1)}\right)^2-4\left(t^{(R-1)}\right)^2}. \label{eq:result-GR1}
\end{align}
\end{subequations}
Therefore, just by coarse-graining the system and rescaling the couplings accordingly, a general solution for the local Green's function $G_0$ and the nonlocal $G_{2^{(R-1)}}$ can be derived. 
After obtaining these results, we can fine-grain the system and insert the four sites ($i=0,1,2,3$) in Eq.~\eqref{eq:nn-r-GF} for $r=R-2$ leading to the following set of equations
\begin{subequations}
    \begin{align}
    1 &= t^{(R-2)}  G_{2^{(R-1)}+2^{(R-2)}} + \omega^{(R-2)}  G_0 + t^{(R-2)}  G_{2^{(R-2)}}  \\
    0 &= t^{(R-2)}  G_0 + \omega^{(R-2)}  G_{2^{(R-2)}} + t^{(R-2)}  G_{2^{(R-1)}}  \label{eq:nn-R2-eq-2}\\
    0 &= t^{(R-2)}  G_{2^{(R-2)}} + \omega^{(R-2)}  G_{2^{(R-1)}} + t^{(R-2)}  G_{2^{(R-1)}+2^{(R-2)}}  \\
    0 &= t^{(R-2)}  G_{2^{(R-1)}} + \omega^{(R-2)}  G_{2^{(R-1)}+2^{(R-2)}} + t^{(R-2)}  G_0 . \label{eq:nn-R2-eq-4}
\end{align}
\end{subequations}
These four equations now determine only two unknown nonlocal Green's functions $G_{2^{(R-2)}}, G_{2^{(R-1)}+2^{(R-2)}}$. However, inserting the previous results in Eqs. (\ref{eq:result-G0}, \ref{eq:result-GR1}) and making use of $t^{(R-1)}= -\frac{|t^{(R-2)}|^2}{\omega^{(R-2)}}$ and $\omega^{(R-1)} = \omega^{(R-2)}-\frac{2|t^{(R-2)}|^2}{\omega^{(R-2)}}$, we find that the equations can be solved leading to
\begin{align}
   G_{2^{(R-2)}} = G_{2^{(R-1)}+2^{(R-2)}}=\frac{t^{(R-2)}}{4(t^{(R-2)})^2-(\omega^{(R-2)})^2},
\end{align}
where $G_{2^{(R-2)}} = G_{2^{(R-1)}+2^{(R-2)}}$ due to periodic boundary conditions and mirror symmetry. This procedure can be continued, and in the $r$-th step $2^{R-r}$ equations are obtained determining $2^{R-r-1}$ unknowns. However, half of the equations are redundant, if the results from the previous renormalization steps are inserted. By considering only a subset of the equations in the $r$-th renormalization step, we arrive at a general solution (see App.~\ref{app:cyclic-reduction-proof} for details) for the Green's function $G_{2^{(r)}}$, with $0 \leq r \leq R-1$: 
\begin{align}
    G_{2^{( r)}} = G_{-2^{( r)}} = G_{2^{R-1} + 2^{R-2} + ... +2^{( r)}} &= \frac{1-G_0 \omega^{( r)}}{2 t^{( r)}}.
\end{align}
Therefore, the corresponding equations do not have to be included in the calculation of the Green's function of the subsequent step on a finer-grained lattice, hence, reducing the complexity of the problem.  

This idea can also be leveraged to the other subsets of equations determining other Green's functions. This, for example, means that $G_{2^{R-2}+2^{R-3}+2^{R-4}}=G_{2^{R-1}+2^{R-4}}$ structurally has the same result as $G_{2^{R-2}+2^{R-3}}=G_{2^{R-1}+2^{R-3}}$ just differing by the rescaled couplings
\begin{subequations}
    \begin{align}
    G_{2^{R-2}+2^{R-3}}&= -\frac{\omega^{(R-3)}}{2 t^{(R-3)}} G_{2^{(R-1)}}, \\
    G_{2^{R-2}+2^{R-3}+2^{R-4}}&=-\frac{\omega^{(R-4)}}{2 t^{(R-4)}} G_{2^{(R-1)}}.
\end{align}
\end{subequations}
Hence, the Green's functions can be classified into groups that have the same structural form. This means that if one function of the class is determined, the rest is known as well. Therefore, in using a cyclic reduction-based renormalization approach, it is possible to reduce the number of equations necessary to obtain the Green's functions of the nearest-neighbor tight-binding model significantly. 

\begin{figure}
    \centering
    \includegraphics[width=1.0\linewidth]{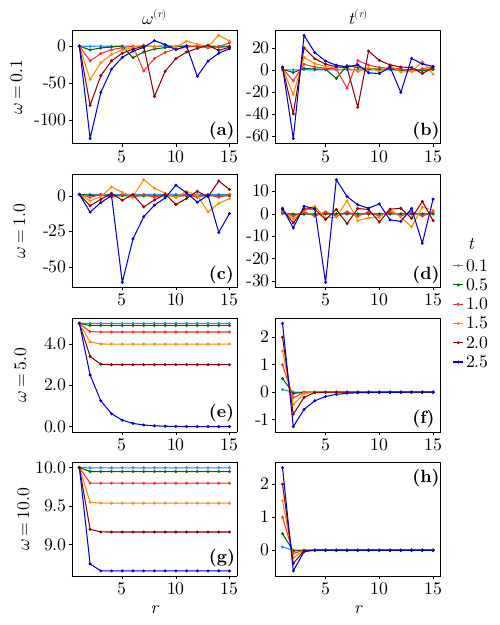}
    \caption{``Running'' of the renormalized couplings ($\omega^{(r)}$ (a),(c),(e),(g); $t^{(r)}$ (b),(d),(f),(h)) in every step of the renormalization procedure in the case of $R=15$ for different values of $\omega,t$.}
    \label{fig:nn-running-couplings}
\end{figure}

Fig.~\ref{fig:nn-running-couplings} shows how the renormalized couplings evolve in every step of the renormalization procedure. It can be seen that as long as $\omega < 2 t$, the rescaled couplings oscillate and do not converge towards a specific value. At $\omega = 2t$, both renormalized couplings $\omega^{(r)}, t^{(r)}$ slowly converge towards zero. In the case of $\omega > 2t$, $t^{(r)}$ quickly approaches zero, while $\omega^{(r)}$ converges to a finite value. An explanation for this behavior is provided in Sec.~\ref{sec:exponential-couplings}.

\subsection{\texorpdfstring{$n$}{n}-th-nearest-neighbor hopping}
The cyclic reduction-based RG idea introduced in the previous section can now be generalized to the more complicated case of $n$-th-nearest-neighbor hopping with $2^R$ sites. For this case, the following defining equation
\begin{align} \label{eq:n-general-general-GF}
     &t_n G_{i-n}+...+ t_2 G_{i-2} + t_1 G_{i-1} + \omega G_{i}+ t_1 G_{i+1} \nonumber \\
     &\quad +t_2 G_{i+2} + ...+ t_n G_{i+n} \nonumber \\
    &\quad = \delta_i + a_1 (\delta_{i-1} + \delta_{i+1}) \nonumber \\
    &\quad + ... +  a_{n-1} (\delta_{i-n+1} + \delta_{i+n-1}),
\end{align}
is found, with the hoppings $t_1,...,t_n$ and, where the additional couplings $a_1,a_2,..., a_{n-1}=0$ that are necessary to find a consistent renormalization scheme, are introduced. In the $r$-th renormalization step, we then arrive at 
\begin{align}
    &t_n^{(r)} G_{2^r(i-n)}+...+ t_2^{(r)} G_{2^r(i-2)} + t_1^{(r)} G_{2^r(i-1)} + \omega G_{(2n)^r i} \nonumber \\
     &\quad+ t_1^{(r)} G_{2^r(i+1)}  +t_2^{(r)} G_{2^r(i+2)} + ...+ t_n^{(r)} G_{2^r(i+n)} \nonumber \\
    &\quad = \delta_i + a_1^{(r)} (\delta_{i-1} + \delta_{i+1}) \nonumber \\
    &\quad + ... +  a_{n-1}^{(r)} (\delta_{i-n+1} + \delta_{i+n-1}).
\end{align}
For $n$-th-nearest-neighbor hopping, $2n$ renormalized couplings ($\omega^{(r)},t_1^{(r)},...,t_n^{(r)}, a_1^{(r)}, ..., a_{n-1}^{(r)}$) are generated for coarse-graining $2$ sites in the cyclic reduction step by only keeping the even $(\mod{2})$ parts. Similarly to the previous section, due to the structure of the renormalization procedure, the Green's functions can be classified into groups that are structurally equal, which will reduce the number of equations necessary to be solved to obtain all Green's functions. In App.~\ref{app:nnn-renormalization}, the renormalization procedure will be explicitly derived in the case of next-nearest-neighbor hopping, in order to get a deeper understanding of the presented idea. More details on the RG scheme for the general $n$-th-nearest-neighbor hopping case with an explanation of the origin of the additional renormalized couplings $a_1^{(r)}, ..., a_{n-1}^{(r)}$  can be found in App.~\ref{app:n-general-renormalization}. Additionally, let us note that the choice of a quantics base is not unique, and also a different base (number of coarse-grained sites) can be chosen in the renormalization procedure (see App.~\ref{app:nnn-base-4} for more details).

\section{\label{sec:qtt}Quantics tensor train representation}
\subsection{Quantics tensor trains}
Accurate resolution and sparse memory usage are two conflicting desiderata in numerical computations, especially in multi-dimensional systems. A promising method for overcoming the ``curse of dimensionality'' in such systems is the quantics tensor train (QTT) approach, which allows the approximation of input data with a logarithmic number of parameters only~\cite{Oseledets2009, Oseledets2011, Khoromskij2011, Dolgov2012, Khoromskij2018, Shinaoka2023}.
A general $d$-dimensional tensor $T$ depending on $d$ variables $i^{(1)}, ..., i^{(d)}$ in this $d$-dimensional space with $N=2^R$ grid points in every dimension can be reshaped to quantics representation into a $2^d\times 2^d\times ...\times 2^d$ ($R$-way) tensor $\mathbf{T} := T_{(i^{(1)}_1 i^{(2)}_1 ... i^{(d)}_1), (i^{(1)}_2...i^{(d)}_2),...,(i^{(1)}_R...i^{(d)}_R)}$  with $R$ external legs, ``mode indices'' or quantics indices $(i^{(1)}_1 i^{(2)}_1 ... i^{(d)}_1), (i^{(1)}_2...i^{(d)}_2),...,(i^{(1)}_R...i^{(d)}_R)$. These quantics indices correspond to bits in a binary representation of the corresponding variable. This is defined in the following way
\begin{equation}
    i^{(j)} = (i^{(j)}_1 i^{(j)}_2 \dots i^{(j)}_R)_2 = \sum_{\ell=1}^{R} 2^{R-\ell} i^{(j)}_{\ell},
    \quad
    i^{(j)}_\ell \in \{0, 1\}
    \,, \label{eq:quanticsrep}
\end{equation}
where each quantics index now corresponds to a distinct length scale of dimension $j$ of the system. The coarsest length scale of dimension $j$ of the system is represented by $i^{(j)}_1$, while the finest length scale is characterized by $i^{(j)}_R$. After reshaping the tensor $\mathbf{T}$ into this so-called quantics representation, it can be decomposed into a  tensor train (TT), also known as matrix product state (MPS),
\begin{align}
    \mathbf{T} = \sum_{\alpha_1 = 1}^{D_1} \sum_{\alpha_2 = 1}^{D_2} ... \sum_{\alpha_{R-1} = 1}^{D_{R-1}} t^{(1)}_{(i^{(1)}_1...i^{(d)}_1), 1 \alpha_1} \cdot t^{(2)}_{(i^{(1)}_2...i^{(d)}_2), \alpha_1 \alpha_2} \cdot ... \cdot t^{(R)}_{(i^{(1)}_R...i^{(d)}_R), \alpha_{R-1} 1}, \label{eq:general-qtt}
\end{align}
with the internal, bond or ``rank'' indices $\alpha_1,...,\alpha_{R-1}$ (see Fig.~\ref{fig:qtt-mpo}(a)). The bond dimensions $D_1, ..., D_{R-1}$ of this QTT indicate how much information different energy or length scales exchange, thus characterizing the amount of length scale entanglement in the system. This notion of scale entanglement differing from the ``conventional'' ideas of site or particle entanglement has recently been further examined in Ref.~\onlinecite{Rohshap2025}. Here, it was shown that the amount of length or time scale entanglement is a system-inherent property of physical systems, which peaks at phase transitions and crossovers allowing QTTs to diagnose such critical points. Although the QTT decomposition in Eq.~\eqref{eq:general-qtt} can, in general, be performed exactly, in numerical computations, truncated TT factorization based on singular value decomposition (SVD) or tensor cross interpolation (TCI) \cite{NunezFernandez2022, Ritter2024, NunezFernandez2024} is applied. The truncation can then either be performed with respect to a fixed maximum bond dimension $D_{\mathrm{max}} := \textrm{max}_\ell(D_\ell)$, with the bond $\ell$, or to a specified error tolerance $\epsilon$. For many applications in many-body physics, a low maximum bond dimension, which governs the compression rate of this approach, has been shown to be sufficient to find a good representation of the input data or function \cite{Rohshap2024,Erpenbeck2023,Ritter2024, Ishida2024, Murray2024, Takahashi2025, Eckstein2024, OteroFumega2024, Frankenbach2025, Shinaoka2023}. However, if we do not use truncated TT factorizations, many functions already allow for an analytic representation by a QTT with a low bond dimension. For example, an exponential function can be exactly represented by a QTT of bond dimension one
\begin{align}
    T = e^{i} = e^{\sum_{\ell=1}^{R} 2^{R-\ell} i_{\ell}} = \prod_{\ell=1}^{R} e^{2^{R-\ell} i_{\ell}}, \quad i =0,1,...,2^R-1,
\end{align}
with the mode index $i_\ell=0,1$. Similarly, a diagonal matrix can be represented by a QTT of bond dimension one, while the sine or cosine function (sum of two exponentials) can be decomposed into a QTT of bond dimension two. 

\subsection{Matrix product operators}
\begin{figure}
    \centering
    \includegraphics[width=0.95\linewidth]{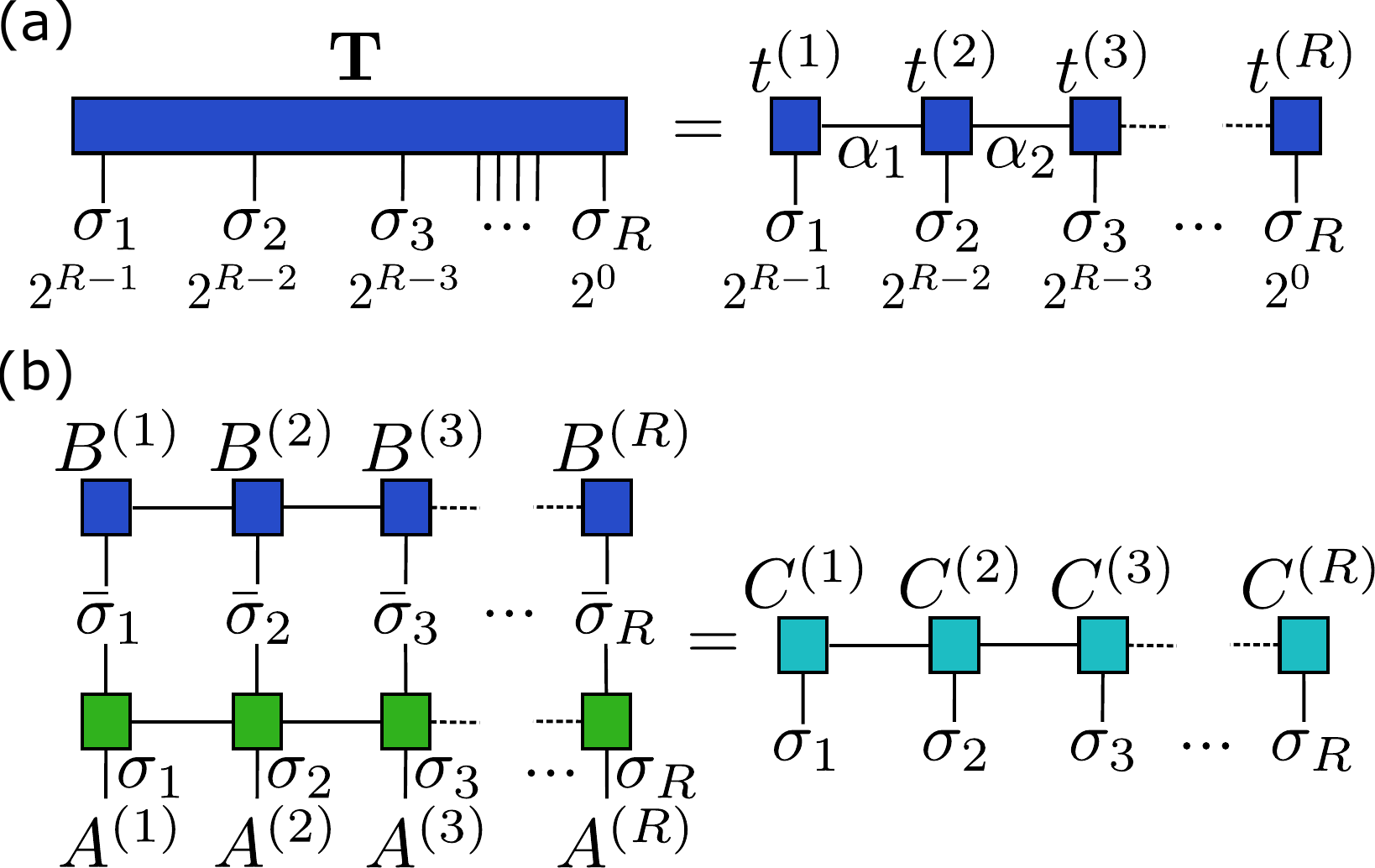}
    \caption{(a)~Decomposition of tensor $\mathbf{T}$ in QTT form with the indices $\sigma_j := (i_j^{(1)},i_j^{(2)},...,i_j^{(d)})$ and the connection to the corresponding length scales $2^{R-j}$. (b)~Contraction of MPO $\mathbf{A}$ with QTT $\mathbf{B}$ leading to QTT $\mathbf{C}$.
    }
    \label{fig:qtt-mpo}
\end{figure}

Matrix product operators (MPOs) are used to perform operations on QTTs (MPSs), such as matrix-vector products. This is shown in Fig.~\ref{fig:qtt-mpo}(b), where the MPO $\mathbf{A}$ representing a matrix is contracted with the QTT $\mathbf{B}$ representing a vector resulting in the QTT $\mathbf{C}$. The MPO $\mathbf{A}$ is defined in the following
\begin{align}
    \mathbf{A} = \sum_{\alpha_1 = 1}^{D_1} ... \sum_{\alpha_{R-1} = 1}^{D_{R-1}} A^{(1)}_{\sigma_1 \bar \sigma_1, 1 \alpha_1} \cdot A^{(2)}_{\sigma_2 \bar \sigma_2, \alpha_1 \alpha_2} \cdot ... \cdot A^{(R)}_{\sigma_R \bar \sigma_R, \alpha_{R-1} 1},
\end{align}
with two external legs or mode indices $\sigma_j, \bar \sigma_j$ connected to each tensor $A^{(j)}$. The MPO-MPS contraction over the shared mode indices $\bar \sigma_j$ is defined in the following way
\begin{align}
    \mathbf{C} = &\sum_{\bar \sigma_1 ... \bar \sigma_R} \sum_{\alpha_1 ... \alpha_{R-1}} \sum_{\bar \alpha_1 ... \bar \alpha_{R-1}} A^{(1)}_{\sigma_1 \bar \sigma_1, 1 \alpha_1} \cdot A^{(2)}_{\sigma_2 \bar \sigma_2, \alpha_1 \alpha_2} \cdot ... \cdot A^{(R)}_{\sigma_R \bar \sigma_R, \alpha_{R-1} 1} \nonumber \\
    &\quad \cdot B^{(1)}_{\bar \sigma_1, 1 \bar \alpha_1} \cdot B^{(2)}_{\bar \sigma_2, \bar \alpha_1 \bar \alpha_2} \cdot ... \cdot B^{(R)}_{\bar \sigma_R, \bar \alpha_{R-1} 1}.
\end{align}
In practical numerical computations, this contraction is often truncated to a specified maximum bond dimension \Dmax{} or tolerance $\epsilon$.

\subsection{\label{sec:qtt-formalism}QTT formalism}
\begin{figure}
    \centering
    \includegraphics[width=0.88\linewidth]{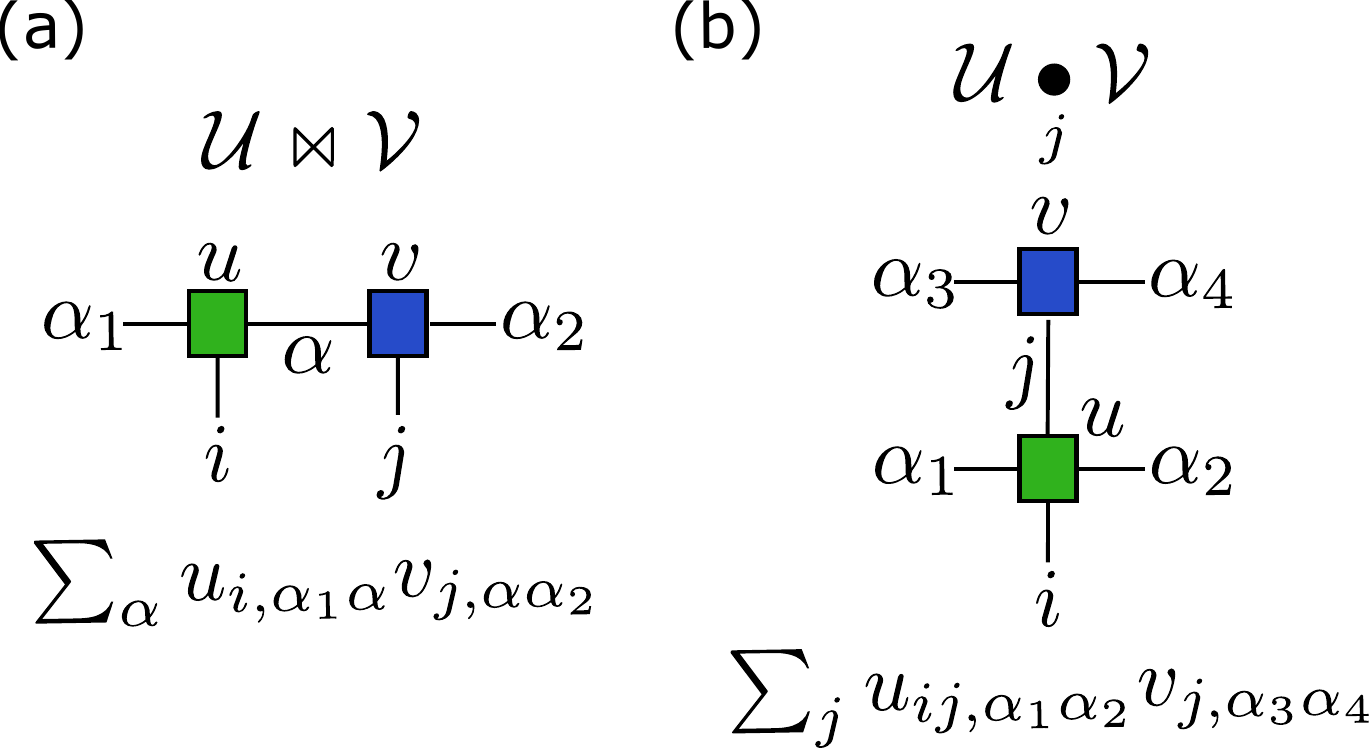}
    \caption{``Rosetta stone'' of QTT formalisms. (a)~Rank product of two tensor cores $\mathcal{U}, \mathcal{V}$ defined with the help of the strong Kronecker product ($\lrtimes$), as a tensor network diagram and in index notation. (b)~Contraction of two tensor cores $\mathcal{U}, \mathcal{V}$ representing a MPO-MPS contraction defined with the help of the mode product ($\underset{*}{\bullet}$), as a tensor network diagram and in index notation.
    }
    \label{fig:qtt-formalism}
\end{figure}

\begin{figure}
    \centering
    \includegraphics[width=0.96\linewidth]{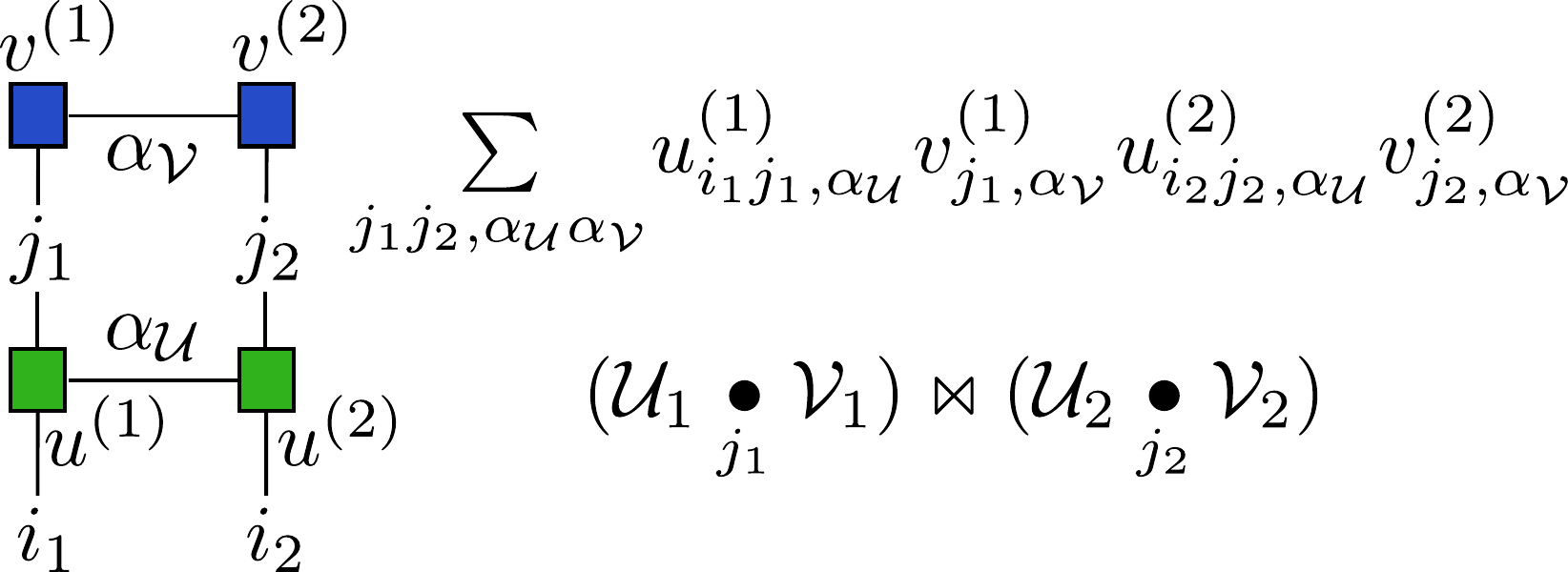}
    \caption{``Rosetta stone'' of MPO-MPS contractions in different QTT formalisms. Contraction of MPO $\mathbf{U} = \mathcal{U}_1 \lrtimes \mathcal{U}_2$ and the MPS $ \mathbf{V} = \mathcal{V}_1 \lrtimes \mathcal{V}_2$ in QTT form over the shared mode indices $j=(j_1,j_2)_2$.
    }
    \label{fig:qtt-formalism-2}
\end{figure}

In the remainder of the paper, we will use a more convenient QTT formalism better suited for our purposes, which will be introduced in the following section. Starting from Eq.~\eqref{eq:general-qtt}, we can define so-called $(D_{\ell-1}\times D_\ell)$ ``core matrices'' \cite{Kazeev2012, Kazeev2013}
\begin{align}
    \mathcal{T}_\ell := \left[\begin{matrix}
		T^{(\ell)}_{1 1} & T^{(\ell)}_{1 2} & ... &          T^{(\ell)}_{1 D_\ell} \\
            T^{(\ell)}_{2 1} & T^{(\ell)}_{2 2} & ... & T^{(\ell)}_{2 D_\ell} \\
            ...& ... & ...& ... \\
            T^{(\ell)}_{D_{\ell-1} 1} & T^{(\ell)}_{D_{\ell-1} 2} & ... & T^{(\ell)}_{D_{\ell-1} D_\ell} 
		\end{matrix}\right], 
\end{align}
with the bond dimension $D_\ell$ of the $\ell$-th bond, $D_0 = D_R = 1$ and where every element of this core matrix is a $2\times 2 \times ... \times 2$ ($d$-way) tensor itself
\begin{align}
    T^{(\ell)}_{\alpha_{\ell-1} \alpha_\ell} := t^{(\ell)}_{i^{(1)}_r...i^{(d)}_r, \alpha_{\ell-1} \alpha_\ell}
\end{align}
with $\alpha_{\ell-1},\alpha_\ell$ taking a value between $1$ and $D_{\ell-1}$, $D_\ell$ respectively. With the help of these core matrices, we can then rewrite Eq.~\eqref{eq:general-qtt} in the following way
\begin{align}
    \mathbf{T} = \mathcal{T}_1 \lrtimes \mathcal{T}_2 \lrtimes ... \lrtimes \mathcal{T}_R,
\end{align}
where we use the ``strong Kronecker product'' ($\lrtimes$)\cite{Launey1994} to represent the rank product of the cores. The rank product of two cores $\mathcal{U}, \mathcal{V}$
\begin{align}
    \mathcal{U} = \left[ \begin{matrix}
        U_{11} & U_{12} \\
        U_{21} & U_{22}
    \end{matrix} \right] , \qquad
    \mathcal{V} = \left[ \begin{matrix}
        V_{11} & V_{12} \\
        V_{21} & V_{22}
    \end{matrix} \right],
\end{align}
is defined in the following way
\begin{align}
    \mathcal{U} \lrtimes \mathcal{V} &= \left[ \begin{matrix}
        U_{11} & U_{12} \\
        U_{21} & U_{22}
    \end{matrix} \right] \lrtimes
    \left[ \begin{matrix}
        V_{11} & V_{12} \\
        V_{21} & V_{22}
    \end{matrix} \right] \\
    &= \left[ \begin{matrix}
        U_{11} \otimes V_{11} + U_{12} \otimes V_{21} & U_{11} \otimes V_{12} + U_{12} \otimes V_{22} \\
        U_{21} \otimes V_{11} + U_{22} \otimes V_{21} & U_{21} \otimes V_{12} + U_{22} \otimes V_{22}
    \end{matrix} \right], \nonumber
\end{align}
with $U_{\alpha_1, \alpha_2}:=u_{i,\alpha_1 \alpha_2}, V_{\alpha_3 \alpha_4}:=v_{j,\alpha_3 \alpha_4}$, and with $i,j$ mode or physical indices and $\alpha_1, \alpha_2, \alpha_3, \alpha_4$ rank or bond indices. Fig.~\ref{fig:qtt-formalism}(a), shows the rank product in index notation and as a tensor network diagram in comparison.
Similarly, if, for example, we want to treat MPO-MPS contractions, we can define a mode product ``$\underset{*}{\bullet}$'', where $*$ denotes the mode indices summed over, in the following way
\begin{align}
    \mathcal{U} \underset{j}{\bullet} \mathcal{V} &= \left[ \begin{matrix}
        U_{11} & U_{12} \\
        U_{21} & U_{22}
    \end{matrix} \right] \underset{j}{\bullet}
    \left[ \begin{matrix}
        V_{11} & V_{12} \\
        V_{21} & V_{22}
    \end{matrix} \right] \\
    &= \left[ \begin{matrix}
        U_{11}  V_{11} & U_{11}  V_{12} & U_{12}  V_{11} & U_{12}  V_{12}  \\
        U_{11}  V_{21} & U_{11}  V_{22} & U_{12}  V_{21} & U_{12}  V_{22}  \\
        U_{21}  V_{11} & U_{21}  V_{12} & U_{22}  V_{11} & U_{22}  V_{12}  \\
        U_{21}  V_{21} & U_{21}  V_{22} & U_{22}  V_{21} & U_{22}  V_{22}  
    \end{matrix} \right], \nonumber
\end{align}
where mode $j$ is summed over. E.g. in the case of an MPO-MPS contraction ($U_{\alpha_1, \alpha_2}:=u_{ij,\alpha_1 \alpha_2}, V_{\alpha_3 \alpha_4}:=v_{j,\alpha_3 \alpha_4}$, with $i,j$ mode indices and $\alpha_1, \alpha_2, \alpha_3, \alpha_4$ rank/bond indices) $U_{\alpha_1 \alpha_2} V_{\alpha_1 \alpha_2}$ denotes a matrix-vector product 
\begin{align}
    U_{\alpha_1 \alpha_2} V_{\alpha_3 \alpha_4} = \sum_j u_{ij,\alpha_1 \alpha_2} v_{j,\alpha_3 \alpha_4},
\end{align}
where the corresponding tensor network diagram as well as the formalism translated to index notation can be found in Fig.~\ref{fig:qtt-formalism}(b).
Moreover, the mode product of two rank products can be recast corewise for any combination of mode indices leading to the following relation
\begin{align}
    (\mathcal{U}_1 \lrtimes \mathcal{U}_2) \underset{j}{\bullet} (\mathcal{V}_1 \lrtimes \mathcal{V}_2) = (\mathcal{U}_1 \underset{j_1}{\bullet} \mathcal{V}_1) \lrtimes (\mathcal{U}_2 \underset{j_2}{\bullet} \mathcal{V}_2), \label{eq:mode-rank-product-relation}
\end{align}
for the matrix-vector product $\mathbf{U} \underset{j}{\bullet} \mathbf{V}$ with the MPO in QTT form $\mathbf{U} = \mathcal{U}_1 \lrtimes \mathcal{U}_2$ and the MPS in QTT form $ \mathbf{V} = \mathcal{V}_1 \lrtimes \mathcal{V}_2$ and $j=(j_1,j_2)_2$ (denoting quantics representation). This relation can be easily understood when investigating the corresponding ``Rosetta stone'' in Fig.~\ref{fig:qtt-formalism-2} showing the correspondence to index formalism and tensor network diagrams.

\section{\label{sec:qtt-renormalization}Renormalization based QTT construction}
In the previous sections, we showed that the Green's function of the $n$-th-nearest-neighbor tight-binding model can be calculated using a cyclic reduction-based renormalization procedure, where $2n$ rescaled couplings are generated (see Sec.~\ref{sec:renormalization}), and introduced the QTT approach (see Sec.~\ref{sec:qtt}). In this section, we investigate the connection between the bond dimension of the QTT representation and the proposed renormalization scheme and derive a direct link between those two concepts. We will apply the concept to the nearest-neighbor (Sec.~\ref{sec:nn-qtt-renormalization}), next-nearest-neighbor (Sec.~\ref{sec:nnn-qtt-renormalization}) and $n$-th-nearest-neighbor (Sec.~\ref{sec:n-general-qtt-renormalization}) hopping cases.

\subsection{Setting the stage for the Green's function calculations}
First, let us rewrite Eq.~\eqref{eq:general-GF}
\begin{align} 
    \sum_j \left( \omega \delta_{ij} + t_{ij} \right) G_{jk}(\omega) = \delta_{ik}
\end{align}
with the help of $G_i := G_{i0}(\omega)$ with periodic boundary conditions implying translational invariance and $G_{i-j} := G_{(i-j)0}(\omega)=G_{i j}(\omega)$ to 
\begin{align} \label{eq:general-GF-QTT}
    \sum_j \left( \omega \delta_{ij} + t_{ij} \right) G_{j}(\omega) = \mathbf{M} \underset{j}{\bullet} \mathbf{G} = \mathbf{\Delta} := \delta_{i},
\end{align}
with $\mathbf{M} := \left( \omega \delta_{ij} + t_{ij} \right)$. We can now decompose $\mathbf{M}, \mathbf{G}, \mathbf{\Delta}$ into QTTs
\begin{subequations}
    \begin{align}
    \mathbf{M} &= \sum_{\alpha_1 = 1}^{D_1} \sum_{\alpha_2 = 1}^{D_2} ... \sum_{\alpha_{R-1} = 1}^{D_{R-1}} m^{(1)}_{i_1 j_1, 1 \alpha_1} \cdot m^{(2)}_{i_2 j_2, \alpha_1 \alpha_2} \cdot ... \cdot m^{(R)}_{i_R j_R, \alpha_{R-1} 1} \nonumber \\
    &= \mathcal{M}_1 \lrtimes \mathcal{M}_2 \lrtimes ... \lrtimes \mathcal{M}_R, \\
    \mathbf{G} &= \sum_{\alpha_1 = 1}^{D_1} \sum_{\alpha_2 = 1}^{D_2} ... \sum_{\alpha_{R-1} = 1}^{D_{R-1}} g^{(1)}_{j_1, 1 \alpha_1} \cdot g^{(2)}_{j_2, \alpha_1 \alpha_2} \cdot ... \cdot g^{(R)}_{j_R, \alpha_{R-1} 1} \nonumber \\
    &= \mathcal{G}_1 \lrtimes \mathcal{G}_2 \lrtimes ... \lrtimes \mathcal{G}_R, \\
    \mathbf{\Delta} &= \delta_{i_1} \cdot \delta_{i_2} \cdot ... \cdot \delta_{i_R} = \underbrace{\Delta \lrtimes \Delta \lrtimes ... \lrtimes \Delta}_{R \textrm{ times}} \equiv \Delta^{\lrtimes (R)}, 
    \end{align}
    \label{eq:nn-qtt-representation-M-G-delta}
\end{subequations}

\noindent where the superscript ($\lrtimes (R)$) defines the $R$ times strong Kronecker product of the corresponding tensor core with itself. Moreover, we explicitly express $\delta_i$ as a QTT with a bond dimension of 1 with $\delta_{i_k} = \left(\begin{matrix} 1 \\ 0 \end{matrix}\right)$ and define the corresponding tensor core $\Delta := \left[ \delta_{i_k} \right]$. Hence, Eq.~\eqref{eq:general-GF-QTT} can be expressed in terms of QTTs in the following way
\begin{align}
    &(\mathcal{M}_1 \lrtimes \mathcal{M}_2 \lrtimes ... \lrtimes \mathcal{M}_R) \underset{j}{\bullet} (\mathcal{G}_1 \lrtimes \mathcal{G}_2 \lrtimes ... \lrtimes \mathcal{G}_R) = \Delta^{\lrtimes (R)},
\end{align}
which can be rewritten with the help of Eq.\eqref{eq:mode-rank-product-relation} to
\begin{align} \label{eq:nn-M-G-contraction}
    &(\mathcal{M}_1 \underset{j_1}{\bullet} \mathcal{G}_1) \lrtimes (\mathcal{M}_2 \underset{j_2}{\bullet} \mathcal{G}_2) \lrtimes ... \lrtimes (\mathcal{M}_R\underset{j_R}{\bullet} \mathcal{G}_R)  = \Delta^{\lrtimes (R)} , 
\end{align}
in the quantics representation $j=(j_1,j_2,...,j_R)_2$.

\subsection{\label{sec:nn-qtt-renormalization}Nearest-neighbor hopping}
From the renormalization procedure in Section~\ref{sec:renormalization}, we know that $\mathbf{G}$ can be obtained with two renormalized couplings connecting each of the consecutive coarse-graining steps in the case of nearest-neighbor hopping. 
Here, we will now connect the renormalization procedure to the construction of the QTT and its bond dimension. In order to do this, we start by finding an explicit QTT construction of $\mathbf{M}$ which plays the role of an MPO.

\subsubsection{Explicit QTT construction of \texorpdfstring{$\mathbf{M}$}{M}}
Since $\mathbf{M}$ can be represented by a circulant matrix, it can be decomposed into
\begin{align}
    \mathbf{M}^{(R)} = \omega \mathbf{\mathbb{I}}^{(R)} + t (\mathbf{P}_1^{(R)} + \mathbf{P}_{2^{R}-1}^{(R)}) \label{eq:nn-qtt-mpo-M-definition}
\end{align}
with $\mathbf{P}_{2^{R}-1}^{(R)} = (\mathbf{P}_1^{(R)})^T = (\mathbf{P}_1^{(R)})^{2^{R}-1}$ and the translational operator
\begin{align}
    \mathbf{P}_1^{(R)} = \left(\begin{matrix}
        0 & 0& 0 & ... & 0 & 0 & 1 \\
        1 & 0& 0 & ... & 0 & 0 & 0 \\
        0 & 1& 0 & ... & 0 & 0 & 0 \\
        ...& ...& ...& ...& ...& ...& ...\\
        0 & 0& 0 & ... & 1 & 0 & 0 \\
        0 & 0& 0 & ... & 0 & 1 & 0 
    \end{matrix} \right),
\end{align}
where the superscript ${(R)}$ indicates that these are $2^R \times 2^R$ matrices. $\mathbf{P}_{1}^{(R)}$ and $\mathbf{P}_{2^{R}-1}^{(R)}$ can be explicitly written as a QTT with bond dimension 2, while $\mathbf{\mathbb{I}}^{(R)}$ can be represented by a QTT with bond dimension 1, which is shown in the following
\begin{subequations}
    \begin{align}
    \mathbf{P}_{1}^{(R)} &= \mathcal{W}_1 \lrtimes \mathcal{W}_2^{\lrtimes (R-2)} \lrtimes \mathcal{W}_3, \\
    \mathbf{P}_{2^{R}-1}^{(R)} &= \mathcal{W}_4 \lrtimes \mathcal{W}_5^{\lrtimes (R-2)} \lrtimes \mathcal{W}_3, \\
    \mathbf{\mathbb{I}}^{(R)} &= \mathcal{W}_6^{\lrtimes (R)},
    \end{align}
\end{subequations}
with the cores
\begin{subequations}
    \begin{align}
    \mathcal{W}_1 &= \left[ \begin{matrix}
        I & H
    \end{matrix} \right] \qquad &\mathcal{W}_4 &= \left[ \begin{matrix}
        H & I
    \end{matrix} \right] \\
    \mathcal{W}_2 &= \left[ \begin{matrix}
        I & J' \\
          & J
    \end{matrix} \right] \qquad &\mathcal{W}_5 &= \left[ \begin{matrix}
        J' &  \\
        J & I
    \end{matrix} \right] \\
    \mathcal{W}_3 &= \left[ \begin{matrix}
         J' \\
         J
    \end{matrix} \right] \qquad &\mathcal{W}_6 &= \left[ I \right]
\end{align}
\end{subequations}
and the following matrices.
\begin{subequations}
    \begin{align}
    I &= \left( \begin{matrix}
        1 & 0 \\
        0 & 1 
    \end{matrix} \right) \qquad 
    &J &= \left( \begin{matrix}
        0 & 1 \\
        0 & 0 
    \end{matrix} \right) \qquad 
    &J' &= \left( \begin{matrix}
        0 & 0 \\
        1 & 0 
    \end{matrix} \right) \\
    H &= \left( \begin{matrix}
        0 & 1 \\
        1 & 0 
    \end{matrix} \right) \qquad 
    &K &= \left( \begin{matrix}
        -1 & 1 \\
        1 & -1 
    \end{matrix} \right) 
\end{align}
\end{subequations}
The explicit proof can be found in Ref. \cite{Kazeev2013}. Now we want to explicitly construct the QTT for the circulant matrix $\mathbf{M}$ (Eq.~\eqref{eq:nn-qtt-mpo-M-definition}). We find that this tensor can be analytically decomposed into a QTT with bond dimension 3 in the following way
\begin{align}
    \mathbf{M}^{(R)} = \left[ \begin{matrix}
        I & K & K
    \end{matrix} \right] \lrtimes
    \left[ \begin{matrix}
        I & K & K \\
          & J &   \\
          &   & J'
    \end{matrix}\right]^{\lrtimes (R-2)} \lrtimes 
    \left[ \begin{matrix}
        \omega I + 2t H \\
        t J \\
        t J'
    \end{matrix}\right]. \label{eq:nn-qtt-mpo-M-explicit-formula}
\end{align}
The explicit proof of this QTT representation is shown in App.~\ref{app:nn-explicit-formula-M}.

\subsubsection{Renormalization in QTT framework}
\begin{figure}
    \centering
    \includegraphics[width=0.95\linewidth]{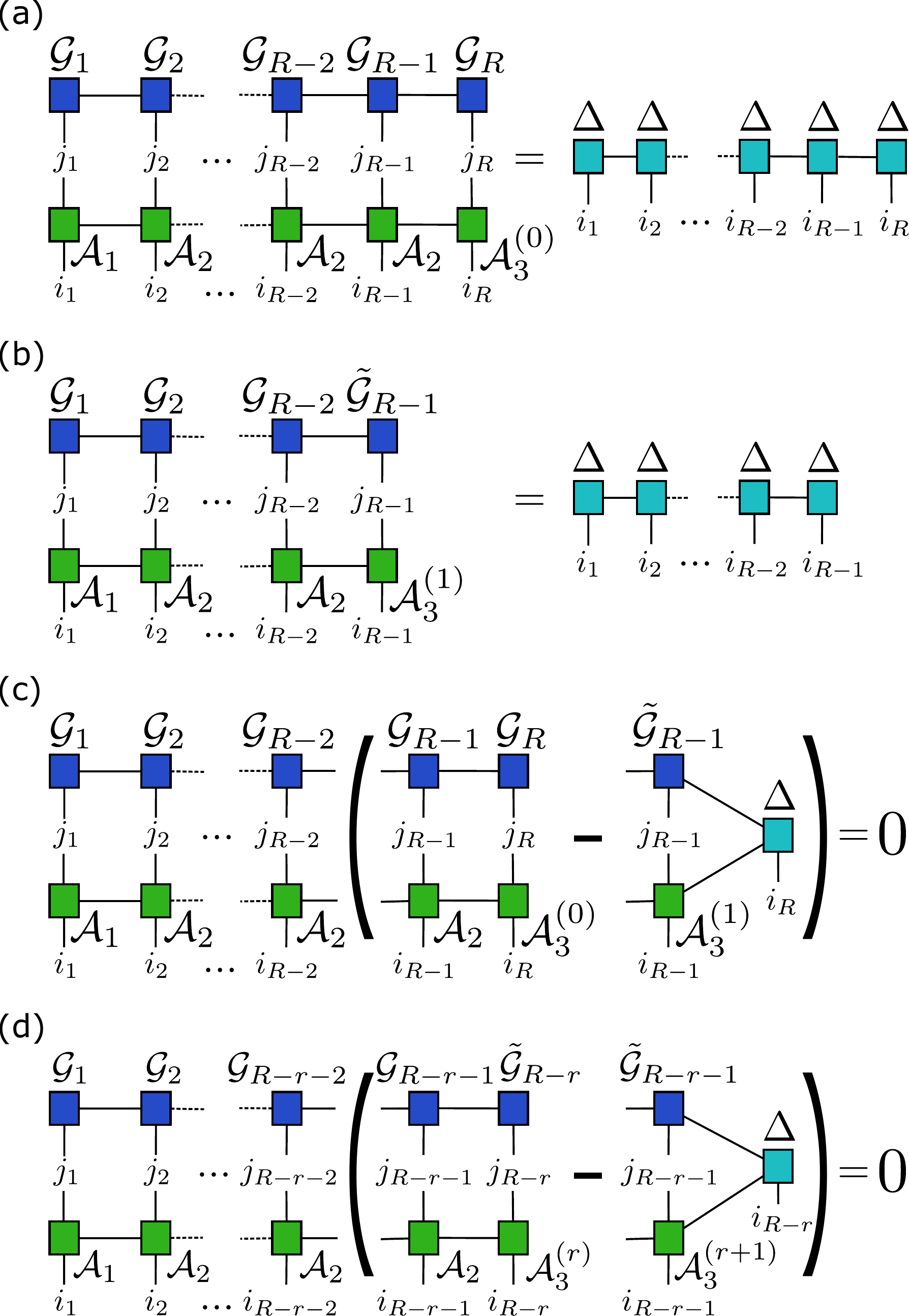}
    \caption{Renormalization in QTT framework: (a) Original equation (Eq.~\eqref{eq:qtt-greens-function-zeroth-renormalization}) in QTT formalism with the Green's function $\mathbf{G} = \mathcal{G}_1 \lrtimes \mathcal{G}_2 \lrtimes ... \lrtimes \mathcal{G}_{R-1} \lrtimes \mathcal{G}_R$ and the MPO $\mathbf{M}_{0} = \left[ \omega \delta_{ij} + t (\delta_{j,i-1}+\delta_{j,i+1}) \right] = \mathcal{A}_1 \lrtimes \mathcal{A}_2^{\lrtimes (R-2)} \lrtimes \mathcal{A}_3^{(0)}$. (b) First renormalization step (Eq.~\eqref{eq:qtt-greens-function-first-renormalization}) in QTT formalism with the renormalized Green's function $\mathbf{G_1} = \mathcal{G}_1 \lrtimes \mathcal{G}_2 \lrtimes ... \lrtimes \tilde{\mathcal{G}}_{R-1}$ and the MPO $\mathbf{M}_{1} = \left[ \omega \delta_{ij} + t (\delta_{j,i-1}+\delta_{j,i+1}) \right] = \mathcal{A}_1 \lrtimes \mathcal{A}_2^{\lrtimes (R-3)} \lrtimes \mathcal{A}_3^{(1)}$. (c) Eq.~\eqref{eq:nn-qtt-renormalization-1} connecting the zeroth and first renormalization steps. (d) Eq.~\eqref{eq:nn-qtt-renormalization-r} connecting the $r$-th and $r+1$-th renormalization steps leading to structurally the same equations and solutions in every RG step due to the fixed relation of $\mathcal{A}_3^{(r)}$ and $\mathcal{A}_3^{(r+1)}$.
    }
    \label{fig:qtt-greens-function-renormalization}
\end{figure}

Starting from Eqs. (\ref{eq:nn-qtt-representation-M-G-delta}-\ref{eq:nn-M-G-contraction}), conducting the first coarse-graining step in the renormalization procedure, we find a new MPO $\mathbf{M}$ that is defined in the same way except for a change to the dimensions $2^{R-1} \times 2^{R-1}$ and the renormalized couplings $t \rightarrow t^{(1)}, \omega \rightarrow \omega^{(1)}$. With the results of the previous section, we can define the cores
\begin{align}
    \mathcal{A}_1 = \left[ \begin{matrix}
        I & K & K
    \end{matrix} \right],  
    \mathcal{A}_2 = \left[ \begin{matrix}
        I & K & K \\
          & J &   \\
          &   & J'
    \end{matrix}\right],   
    \mathcal{A}_3^{(r)} = \left[ \begin{matrix}
        \omega^{(r)} I + 2t^{(r)} H \\
        t^{(r)} J \\
        t^{(r)} J'
    \end{matrix}\right],
\end{align}
with $\mathcal{M}_1 = \mathcal{A}_1; \mathcal{M}_2,...,\mathcal{M}_{R-1} = \mathcal{A}_2; \mathcal{M}_R = \mathcal{A}_3^{(0)}$ leading to
\begin{align}
    \mathbf{M}_{r} = \mathcal{A}_1 \lrtimes \mathcal{A}_2^{\lrtimes (R-2-r)} \lrtimes \mathcal{A}_3^{(r)},
\end{align}
where $\mathbf{M}_{r}$ is now associated with the MPO of the $r$-th renormalization step, where $t^{(r)}, \omega^{(r)}$ are the corresponding renormalized couplings. Therefore, the QTT representation of $\mathbf{M}_{r}$ has the same form, differing just by the renormalized couplings and the dimension in each coarse-graining step. E.g. in the case of $r=R-2$, we obtain
\begin{align}
    \mathbf{M}_{R-2} &= \mathcal{A}_1 \lrtimes \mathcal{A}_3^{(R-2)} = 
    \left[ \begin{matrix}
        I & K & K
    \end{matrix} \right] \lrtimes 
    \left[ \begin{matrix}
        \omega^{(R-2)} I + 2t^{(R-2)} H \\
        t^{(R-2)} J \\
        t^{(R-2)} J'
    \end{matrix}\right] \nonumber \\
    &= \omega^{(R-2)} I \otimes I + 2t^{(R-2)} I \otimes H + t^{(R-2)} K \otimes J + t^{(R-2)} K \otimes J' \nonumber \\
    &= \left( \begin{matrix}
        \omega^{(R-2)} & t^{(R-2)} & 0 & t^{(R-2)} \\
        t^{(R-2)} & \omega^{(R-2)} & t^{(R-2)} & 0 \\
        0 & t^{(R-2)} & \omega^{(R-2)} & t^{(R-2)} \\
        t^{(R-2)} & 0 & t^{(R-2)} & \omega^{(R-2)}
    \end{matrix}\right).
\end{align}
Similarly, the Green's function can be decomposed in every step of the renormalization procedure in the following way
\begin{align}
    \mathbf{G}_r = \mathcal{G}_1 \lrtimes \mathcal{G}_2 \lrtimes ... \lrtimes  \mathcal{\tilde G}_{R-r},
\end{align}
where $\mathcal{\tilde G}_{R-r}$ denotes the last tensor that needs to be further resolved, in order to obtain the full Green's function (and $\mathcal{\tilde G}_{R}= \mathcal{G}_{R}$ in the last step). Hence, we can start from the original equation (see Fig.~\ref{fig:qtt-greens-function-renormalization}(a) for tensor network representation)
\begin{align}
    &(\mathcal{A}_1 \underset{j_1}{\bullet} \mathcal{G}_1) \lrtimes (\mathcal{A}_2 \underset{j_2}{\bullet} \mathcal{G}_2) \lrtimes  ... \lrtimes (\mathcal{A}_2 \underset{j_{R-1}}{\bullet} \mathcal{G}_{R-1}) \lrtimes (\mathcal{A}_3^{(0)} \underset{j_R}{\bullet} \mathcal{G}_R)  \nonumber \\
    &= \underbrace{\Delta \lrtimes \Delta \lrtimes ... \lrtimes \Delta}_{R \textrm{ times}} = \Delta^{\lrtimes (R)} \label{eq:qtt-greens-function-zeroth-renormalization}
\end{align}
and then apply the first renormalization step leading to
\begin{align} 
    &(\mathcal{A}_1 \underset{j_1}{\bullet} \mathcal{G}_1) \lrtimes (\mathcal{A}_2 \underset{j_2}{\bullet} \mathcal{G}_2) \lrtimes  ... \lrtimes (\mathcal{A}_2 \underset{j_{R-2}}{\bullet} \mathcal{G}_{R-2}) \lrtimes (\mathcal{A}_3^{(1)} \underset{j_{R-1}}{\bullet} \mathcal{\tilde G}_{R-1})  \nonumber \\
    &= \underbrace{\Delta \lrtimes \Delta \lrtimes ... \lrtimes \Delta}_{R-1 \textrm{ times}} = \Delta^{\lrtimes (R-1)}, \label{eq:qtt-greens-function-first-renormalization}
\end{align}
where Fig.~\ref{fig:qtt-greens-function-renormalization}(b) shows a tensor network diagram of the equation.
We can now insert the result of the first coarse-graining step into the equation of the zeroth renormalization step, and after simplification we obtain

\begin{align}
    &0= (\mathcal{A}_1 \underset{j_1}{\bullet} \mathcal{G}_1) \lrtimes (\mathcal{A}_2 \underset{j_2}{\bullet} \mathcal{G}_2) \lrtimes  ... \lrtimes (\mathcal{A}_2 \underset{j_{R-2}}{\bullet} \mathcal{G}_{R-2}) \lrtimes \label{eq:nn-qtt-renormalization-1} \\
    & \left(  (\mathcal{A}_2 \underset{j_{R-1}}{\bullet} \mathcal{G}_{R-1}) \lrtimes (\mathcal{A}_3^{(0)} \underset{j_R}{\bullet} \mathcal{G}_R) - (\mathcal{A}_3^{(1)} \underset{j_{R-1}}{\bullet} \mathcal{\tilde G}_{R-1}) \lrtimes \Delta \right) , \nonumber
\end{align}
connecting the zeroth and first renormalization steps (see Fig.~\ref{fig:qtt-greens-function-renormalization}(c)). Similarly, we can connect the $r$-th to the $r+1$-th renormalization step in the following way 
\begin{small}
    \begin{align}
    &0= (\mathcal{A}_1 \underset{j_1}{\bullet} \mathcal{G}_1) \lrtimes (\mathcal{A}_2 \underset{j_2}{\bullet} \mathcal{G}_2) \lrtimes  ... \lrtimes (\mathcal{A}_2 \underset{j_{R-r-2}}{\bullet} \mathcal{G}_{R-r-2}) \lrtimes \label{eq:nn-qtt-renormalization-r}\\
     &\Bigg(  (\mathcal{A}_2 \underset{j_{R-r-1}}{\bullet} \mathcal{G}_{R-r-1}) \lrtimes (\mathcal{A}_3^{(r)} \underset{j_{R-r}}{\bullet} \mathcal{\tilde G}_{R-r}) - (\mathcal{A}_3^{(r+1)} \underset{j_{R-r-1}}{\bullet} \mathcal{\tilde G}_{R-r-1}) \lrtimes \Delta \Bigg) . \nonumber
\end{align}
\end{small}
These equations are structurally the same in each RG step (see Fig.~\ref{fig:qtt-greens-function-renormalization}(d)), due to the fixed relation of $\mathcal{A}_3^{(r)}$ and $\mathcal{A}_3^{(r+1)}$ set by the relation of the rescaled couplings $t^{(r+1)}= -\frac{|t^{(r)}|^2}{\omega^{(r)}}$, $\omega^{(r+1)} = \omega^{(r)}-\frac{2|t^{(r)}|^2}{\omega^{(r)}}$. Hence, if the bond dimension $D_{R-r-1}$ (determined by this equation) matches that of the previous step $D_{R-r-2}$ then the bond dimension saturates for the consecutive steps and the system is completely solved, since the degrees of freedom and the equations remain the same. Therefore, the equations in subsequent fine-graining steps lead to the same solution in each step. Let us emphasize again that at $r=R-4$, we arrive at the equations
\begin{align}
    &(\mathcal{A}_1 \underset{j_1}{\bullet} \mathcal{G}_1) \lrtimes (\mathcal{A}_2 \underset{j_2}{\bullet} \mathcal{G}_2) \lrtimes \\
    & \left(  (\mathcal{A}_2 \underset{j_{3}}{\bullet} \mathcal{G}_{3}) \lrtimes (\mathcal{A}_3^{(R-4)} \underset{j_{4}}{\bullet} \mathcal{\tilde G}_{4}) - (\mathcal{A}_3^{(R-3)} \underset{j_{3}}{\bullet} \mathcal{\tilde G}_{3}) \lrtimes \Delta \right) =0, \nonumber
\end{align}
which are now structurally the same as in the next finer-grained step ($r=R-5$)
\begin{align}
    &(\mathcal{A}_1 \underset{j_1}{\bullet} \mathcal{G}_1) \lrtimes (\mathcal{A}_2 \underset{j_2}{\bullet} \mathcal{G}_2) \lrtimes (\mathcal{A}_2 \underset{j_3}{\bullet} \mathcal{G}_3) \lrtimes \\
    & \left(  (\mathcal{A}_2 \underset{j_{4}}{\bullet} \mathcal{G}_{4}) \lrtimes (\mathcal{A}_3^{(R-5)} \underset{j_{5}}{\bullet} \mathcal{\tilde G}_{5}) - (\mathcal{A}_3^{(R-4)} \underset{j_{4}}{\bullet} \mathcal{\tilde G}_{4}) \lrtimes \Delta \right) =0. \nonumber
\end{align}
Hence, starting from $r=R-4$, the QTT bond dimension saturates at two, precisely matching the number of renormalized couplings. Therefore, the analytical solutions for the tensor components remain structurally identical, differing only in the superindices corresponding to each renormalization step. Once these solutions are derived, the components of all tensors in the QTT, and consequently all Green's functions, are known analytically for arbitrary system sizes. In this way, the QTT construction can be understood as a natural RG framework, as the tensors not only show similar patterns across different length scales but are, in fact, identical up to the rescaled couplings.
In App.~\ref{app:nn-renormalization-solutions}, the explicit analytical solutions for the individual tensors of the QTTs are shown, derived by solving Eq.~\eqref{eq:nn-qtt-renormalization-r}.

\subsection{\label{sec:nnn-qtt-renormalization}Next-nearest-neighbor hopping}
Due to the generation of the additional renormalized coupling $a^{(r)}$ (defined in Eq.~\eqref{eq:nnn-renormalized-coupling-a}) in the next-nearest-neighbor hopping case, the combination of renormalization scheme with the QTT approach slightly differs from the nearest-neighbor case discussed in the previous section. While Eq.~\eqref{eq:nn-M-G-contraction} still holds for the entire non-renormalized system, after applying the first coarse-graining step the equation needs to be modified in the following way
\begin{align} \label{eq:nnn-M-G-contraction-1st-renormalization}
    &(\mathcal{M}_1 \underset{j_1}{\bullet} \mathcal{G}_1) \lrtimes (\mathcal{M}_2 \underset{j_2}{\bullet} \mathcal{G}_2) \lrtimes ... \lrtimes (\mathcal{M}_{R-1}\underset{j_{R-1}}{\bullet} \mathcal{\tilde G}_{R-1})  =\mathbf{D}_1. 
\end{align}
where $\mathbf{D}_1=\left(1, a^{(1)},0,...,0,a^{(1)} \right)^T$ differs from $\Delta^{\lrtimes (R-1)}$ due to the generated renormalized coupling $a^{(1)}$. In the following, we will derive an explicit QTT representation of $\mathbf{M}$ and $\mathbf{D}$, followed by combining the renormalization and QTT concepts.

\subsubsection{Explicit QTT construction of \texorpdfstring{$\mathbf{M}$}{M}}
Since $\mathbf{M}$ is a circulant matrix, we can decompose it into
\begin{align}
    \mathbf{M}^{(R)} = \omega \mathbf{\mathbb{I}}^{(R)} + t (\mathbf{P}_1^{(R)} + \mathbf{P}_{2^{R}-1}^{(R)}) + t' (\mathbf{P}_2^{(R)} + \mathbf{P}_{2^{R}-2}^{(R)}) \label{eq:nnn-qtt-mpo-M-definition}
\end{align}
with $\mathbf{P}_{2^{R}-1}^{(R)} = (\mathbf{P}_1^{(R)})^T = (\mathbf{P}_1^{(R)})^{2^{R}-1}, \mathbf{P}_{2^{R}-2}^{(R)} = (\mathbf{P}_2^{(R)})^T = (\mathbf{P}_1^{(R)})^{2^{R}-2}$ and the translational operator
\begin{align}
    \mathbf{P}_2^{(R)} = \left(\begin{matrix}
        0 & 0& 0 & ... & 0 & 1 & 0 \\
        0 & 0& 0 & ... & 0 & 0 & 1 \\
        1 & 0& 0 & ... & 0 & 0 & 0 \\
        0 & 1& 0 & ... & 0 & 0 & 0 \\
        ...& ...& ...& ...& ...& ...& ...\\
        0 & 0& 0 & ... & 0 & 0 & 0 \\
        0 & 0& 0 & ... & 1 & 0 & 0 
    \end{matrix} \right),
\end{align}
where the superscript ${(R)}$ indicates that these are $2^R \times 2^R$ matrices. $\mathbf{P}_{1}^{(R)}$ and $\mathbf{P}_{2^{R}-1}^{(R)}$ can be explicitly written as a QTT with bond dimension 2, while $\mathbf{\mathbb{I}}^{(R)}$ can be represented by a QTT with bond dimension 1 (see Ref. \cite{Kazeev2013}). Now we want to explicitly construct the QTT for the circulant matrix $\mathbf{M}$ (Eq.~\eqref{eq:nnn-qtt-mpo-M-definition}) and its coarse-grained versions. Therefore, we find that the tensor of the $r$-th renormalization step ($\mathbf{M}_{r}$) can be decomposed into a QTT with bond dimension 5 in the following way
\begin{align} \label{eq:nnn-qtt-mpo-M-explicit-formula-1}
    \mathbf{M}_{r} = \mathcal{A}_1 \lrtimes \mathcal{A}_2^{\lrtimes (R-2-r)} \lrtimes \mathcal{A}_3^{(r)},
\end{align}
with 
\begin{align}
    &\mathcal{A}_1 = \left[ \begin{matrix}
        I & K & K & K & K 
    \end{matrix} \right], \quad  
    \mathcal{A}_2 = \left[ \begin{matrix}
        I & K & K & K & K \\
          & J &   &   &   \\
          &   & J'&   &   \\
          &   &   & J &   \\
          &   &   &   & J'
    \end{matrix}\right], \nonumber \\ 
    &\mathcal{A}_3^{(r)} = \left[ \begin{matrix}
        (\omega^{(r)} + 2t'^{(r)}) I + 2t^{(r)} H \\
        t^{(r)} J \\
        t^{(r)} J' \\
        t'^{(r)} I \\
        t'^{(r)} I
    \end{matrix}\right], \label{eq:nnn-qtt-mpo-M-explicit-formula-2}
\end{align}
and the renormalized couplings $\omega^{(r)}, t^{(r)}, t'^{(r)}$. The explicit proof of this construction can be found in App.~\ref{app:nnn-proof-M}.

\subsubsection{Explicit QTT construction of \texorpdfstring{$\mathbf{D}$}{D}}
$\mathbf{D}_r$ can be analytically represented by a QTT (MPS) of bond dimension 2 in the following way
\begin{align} \label{eq:nnn-qtt-mps-D-explicit-formula-1}
    \mathbf{D}_{r} = \mathcal{D}_1 \lrtimes \mathcal{D}_2^{\lrtimes (R-2-r)} \lrtimes \mathcal{D}_3^{(r)},
\end{align}
with the tensor cores
\begin{align}
    \mathcal{D}_1 &= \left[ \begin{matrix}
        C & F  
    \end{matrix} \right], \quad
    \mathcal{D}_2 &= \left[ \begin{matrix}
        C & F  \\
          & B 
    \end{matrix}\right], \quad 
    \mathcal{D}^{(r)}_3 &= \left[ \begin{matrix}
        C + 2a^{(r)}B \\
        a^{(r)} B 
    \end{matrix}\right], \label{eq:nnn-qtt-mps-D-explicit-formula-2}
\end{align}
with the renormalized coupling $a^{(r)}$ and the vectors
\begin{align}
    B &= \left( \begin{matrix}
        0  \\
        1  
    \end{matrix} \right), \qquad
    &C &= \left( \begin{matrix}
        1  \\
        0  
    \end{matrix} \right), \qquad
    &F &= \left( \begin{matrix}
        -1 \\
        1 
    \end{matrix} \right) .
\end{align}
The explicit proof is provided in App.~\ref{app:nnn-proof-Delta}.

\subsubsection{Renormalization in QTT framework}
Starting from Eq.~\eqref{eq:nnn-M-G-contraction-1st-renormalization}, a similar relation connecting the $r$-th and the $r+1$-th renormalization step as in the nearest-neighbor hopping case can be derived. From the equations of the $r$-th
\begin{align}
    &(\mathcal{A}_1 \underset{j_1}{\bullet} \mathcal{G}_1) \lrtimes (\mathcal{A}_2 \underset{j_2}{\bullet} \mathcal{G}_2) \lrtimes  ... \lrtimes (\mathcal{A}_2 \underset{j_{R-r-1}}{\bullet} \mathcal{G}_{R-r-1}) \lrtimes (\mathcal{A}_3^{(r)} \underset{j_r}{\bullet}  \mathcal{\tilde G}_{R-r})  \nonumber \\
    &= \mathcal{D}_1 \lrtimes \mathcal{D}_2^{\lrtimes (R-2-r)} \lrtimes \mathcal{D}_3^{(r)}
\end{align}
and $r+1$-th renormalization step
\begin{align}
    &(\mathcal{A}_1 \underset{j_1}{\bullet} \mathcal{G}_1) \lrtimes (\mathcal{A}_2 \underset{j_2}{\bullet} \mathcal{G}_2) \lrtimes  ... \lrtimes (\mathcal{A}_2 \underset{j_{R-r-2}}{\bullet} \mathcal{G}_{R-r-2}) \nonumber \\
    &\qquad \lrtimes (\mathcal{A}_3^{(r+1)} \underset{j_{R-r-1}}{\bullet}  \mathcal{\tilde G}_{R-r-1})  \nonumber \\
    &= \mathcal{D}_1 \lrtimes \mathcal{D}_2^{\lrtimes (R-3-r)} \lrtimes \mathcal{D}_3^{(r+1)}
\end{align}
we obtain
\begin{align}
    &(\mathcal{A}_1 \underset{j_1}{\bullet} \mathcal{G}_1) \lrtimes (\mathcal{A}_2 \underset{j_2}{\bullet} \mathcal{G}_2) \lrtimes  ... \lrtimes (\mathcal{A}_2 \underset{j_{R-r-2}}{\bullet} \mathcal{G}_{R-r-2}) \lrtimes \nonumber \\
    & \left(  (\mathcal{A}_2 \underset{j_{R-r-1}}{\bullet} \mathcal{G}_{R-r-1}) \lrtimes (\mathcal{A}_3^{(r)} \underset{j_{R-r}}{\bullet} \mathcal{\tilde G}_{R-r}) - (\mathcal{A}_3^{(r+1)} \underset{j_{R-r-1}}{\bullet} \mathcal{\tilde G}_{R-r-1}) \lrtimes \Delta \right)  \nonumber \\
    &= \mathcal{D}_1 \lrtimes \mathcal{D}_2^{\lrtimes (R-3-r)} \lrtimes  \left( \mathcal{D}_3^{(r+1)} \lrtimes \Delta - \mathcal{D}_2  \lrtimes \mathcal{D}_3^{(r)} \right).   \label{eq:nnn-qtt-renormalization-r}
\end{align}
connecting the $r$-th and $r+1$-th renormalization step. These equations are structurally the same in every renormalization step, due to the fixed relation of $\mathcal{A}_3^{(r)}, \mathcal{D}_3^{(r)}$ and $\mathcal{A}_3^{(r+1)}, \mathcal{D}_3^{(r+1)}$ defined by the renormalized couplings $\omega^{(r)}, t^{(r)}, t'^{(r)}, a^{(r)}$ (see Eqs.~\eqref{eq:nnn-renormalized-couplings}). Hence, if the bond dimension $D_{R-r-1}$ (determined by this equation) is the same as in the previous step $D_{R-r-2}$ then the bond dimension saturates, additionally giving the same solution in every coarse-graining step just differing by the superscripts of the regarding renormalization step. In the analytical calculations, the bond dimension saturates at four matching the number of couplings connecting the different length scales in the cyclic reduction-based renormalization procedure.

\subsubsection{\label{sec:n-general-qtt-renormalization}\texorpdfstring{$n$}{n}-th-nearest-neighbor hopping}
Naturally, this QTT-based renormalization procedure can be extended to arbitrary $n$-th-nearest-neighbor hoppings. 
The QTT-based renormalization procedure can then be performed as in the previous sections, and once the bond dimension of the resulting QTT representing the Green's function saturates at $\mathbf{2n}$ (see Section~\ref{sec:n-th-nearest-neighbor-bond-dimension}), also the solutions of the tensors representing finer length scales have the same solution, just differing by the renormalization step, meaning that the entire system is solved for arbitrary grid sizes.

\subsection{Discussion}
In this section, we have demonstrated that the QTT approach provides a natural and effective framework for describing the renormalization procedure introduced in Section~\ref{sec:renormalization}. While QTTs have so far primarily been regarded as powerful tools for numerical computations, our results reveal that the QTT formalism can also greatly simplify analytical calculations. This opens the door to a potentially entirely new line of research, where QTTs might enable analytical solutions to problems previously deemed intractable. Furthermore, we found that the QTT bond dimension precisely matches the number of renormalized couplings generated in the procedure — a correspondence that is not surprising, as both quantities reflect the flow of information between different length scales, i.e., the degree of length scale entanglement. Additionally, it was shown that once the bond dimension saturates, the solution of the individual QTT tensors also saturate, differing only by the rescaled couplings. With these insights, we can interpret the QTT framework as an RG method in its own right, since QTTs serve as the natural language of the proposed procedure.


\section{\label{sec:approximate-construction}Approximate analytical construction}
In the previous section, it was argued that the maximum bond dimension of the analytical QTT of the one-particle Green's function is equal to $2n$ in the case of $n$-th-nearest-neighbor hopping. Therefore, in the case of $n \rightarrow N$, where $2N+1$ is the number of lattice sites and $N \rightarrow \infty$ the maximum bond dimension of the analytical QTT will diverge. To overcome this problem, an approximate analytical (or numerical) QTT with a maximum bond dimension bound by $2m$ can be constructed, where the hoppings $t_n$ are set to zero for $n>m$. In this section, a priori error bounds of such an approximate QTT construction in the case of the tight-binding model with decaying, but nonzero, couplings on the whole lattice will be studied.

\subsection{Model and error bounds}
\subsubsection{General error bounds}
We consider the model with $2N+1$ sites with decaying couplings $t_n$ along the chain that satisfy $\sum_{n=-\infty}^{\infty} |t_n| < \infty$ and aim to construct an approximate QTT for the one-particle Green's function within some error $\epsilon$. We define the Toeplitz matrix $A:=(\omega \mathbb{I} -H)$ from which the Green's function can be derived $G=A^{-1}$. Instead of analytically decomposing $A^{-1}$ into a QTT with diverging \Dmax{}, we approximate matrix $A$ by $B$, which contains only nonzero hoppings $t_n$ for $n \leq m$. All hoppings for $n>m$ are set to zero, from which we can construct an exact QTT of $B^{-1}$ with maximum bond dimension $2m$ (see Sec.~\ref{sec:qtt-renormalization}). From $A = B+R$, where $R$ is the rest matrix containing the neglected hoppings ($t_n, n>m$) and
\begin{align}
    \frac{1}{x+\delta} = \frac{1}{x} - \frac{1}{x^2}\delta +....
\end{align}
 follows the following a priori error bound
\begin{align}
    \epsilon &= ||A^{-1}-B^{-1}|| \leq ||R|| \cdot ||B^{-1}||^2 \cdot \exp \left(||R|| \cdot ||  B^{-1}|| \right) \label{eq:general-error-bound},
\end{align}
for the approximation of $A^{-1}$ by $B^{-1}$, and, thus, for the approximation of $A^{-1}$ by a QTT with \Dmax $=2m$, with respect to an arbitrary error norm $||.||$. The derivation of this error bound is detailed in App.~\ref{app:general-error-bounds}.

\subsubsection{Spectral norm}
In order to derive more explicit error bounds, we will consider the spectral norm $||.||_2 = \max |\lambda_i|$, which corresponds to the largest eigenvalue of the respective matrices since $A,R,B$ (and their inverses) are Hermitian. Because $||B||_2 = \max_{\lambda_i \in \mathrm{sp}(B)} |\lambda_i|$ it follows that $||B^{-1}||_2 = (\min_{\lambda_i \in \mathrm{sp}(B)} |\lambda_i|)^{-1}$, where $\mathrm{sp}(B)$ denotes the spectrum of $B$. Hence, in order to derive error bounds, we need to identify bounds on the spectra of $R$ and $B$. Therefore, we define a Fourier series
\begin{align}
    f(\phi) = \sum_{n = -\infty}^{\infty} t_n e^{i n \phi} = \sum_{n = -N}^{N} t_n e^{i n \phi},
\end{align}
where the sum is truncated to $2N+1$ contributions for a finite system with $2N+1$ sites. Following Lemma 4.1 in Ref.~\onlinecite{Gray2005}, we can derive upper (lower) bounds on the eigenvalues of the corresponding matrices by obtaining the maximum $M_f$ (minimum $m_f$) values of $f(\phi)$.
\begin{align}
    m_f \leq \lambda_n \leq M_f
\end{align}
With these bounds we can determine the a priori error bound
\begin{align}
    \epsilon &\leq ||R||_2 \cdot ||B^{-1}||_2^2 \cdot \exp \left(||R||_2 \cdot ||  B^{-1}||_2 \right) \nonumber \\
    &\leq M_{f_R} (m_{f_B})^{-2} e^{M_{f_R} (m_{f_B})^{-1}}, \label{eq:error-bounds}
\end{align}
where $M_{f_R}$ and $m_{f_B}$ denote the essential supremum and infimum of the corresponding Fourier series of matrices $R$ and $B$.

\subsection{\label{sec:exponential-couplings}Exponentially decaying couplings}
First, we consider exponentially decaying couplings $t_n = t_{-n} = C e^{-\alpha n}, C,\alpha \in \mathbb{R}, n \in \mathbb{N}$ and $t_0=\omega$. To a priori calculate a bound on the error for the construction of a QTT of the Green's function with bond dimension $2m$ and, thus, neglecting all hoppings for $|n|>m \ (t_n =0)$, the Fourier series of the corresponding matrices have to be studied. From the Fourier series of matrix $B$ ($f_B(\phi)$) the lower bound on the eigenvalues can be obtained
\begin{align}
    m_{f_B} = f_B(\phi=\pi) = \omega + \frac{2C [-1 + e^{-\alpha m} \cos(m \pi)]}{1+e^{\alpha}} \label{eq:error-bounds-exp-minimum-eigenvalues}
\end{align}
as the lower bound $m_{f_B}$ on the spectrum of $B$. Similarly, we can derive the upper bound $M_{f_R}$ on the spectrum of $R$ from the respective Fourier series $f_R(\phi)$ leading to
\begin{align}
    M_{f_R}= f(\phi=0) = \frac{2C \left(e^{-\alpha m} - e^{-\alpha N}\right)}{e^{\alpha}-1}. \label{eq:error-bounds-exp-maximum-eigenvalues}
\end{align}
The explicit derivations can be found in App.~\ref{app:error-bounds-exponential}. We obtain the explicit a priori error bound by inserting the results from Eqs. (\ref{eq:error-bounds-exp-minimum-eigenvalues},\ref{eq:error-bounds-exp-maximum-eigenvalues}) into the error bounds derived in Eq.~\eqref{eq:error-bounds}.

Moreover, by deriving a lower bound on the spectrum of $B$, we also determine a ``problematic'' region. For small values of $\omega$, if $m_{f_b} = 0 \Leftrightarrow \omega = -\frac{2C [-1 + e^{-\alpha m} \cos(m \pi)]}{1+e^{\alpha}}$, eigenvalues of matrix $B$ can become zero meaning that matrix $B$ becomes singular and is no longer invertible. Numerical experiments suggest that $B$ becomes singular exactly at $m_{f_b}=0$ leading the first eigenvalue to become zero. Moreover, for even smaller values of $\omega$ ($m_{f_b}<0$), additional eigenvalues cross zero, becoming negative, meaning that in this case the matrix becomes singular (not invertible) at many points. Therefore, in this regime $m$ needs to be chosen larger for an accurate representation of the Green's function with the help of QTTs. However, if we are only interested in Green's functions with values of $\omega$ so that $m_{f_b}>0$, we do not encounter any singularities, thus, implying that $B$ is invertible and that we can conveniently work with the derived error bounds in Eq.~\eqref{eq:error-bounds}. Let us emphasize that we not only have derived explicit a priori error bounds for the inversion of general Toeplitz and circulant matrices with exponentially decaying hoppings, but have, additionally, found the first crossing of an eigenvalue through zero making the matrix singular and not invertible, after which more and more eigenvalues become negative. From this point we can also gain a better understanding of the behavior of the renormalized couplings in Fig.~\ref{fig:nn-running-couplings}, where we see strongly oscillating couplings in the case of $\omega < 2t$. Inserting $m=1$ and $\alpha=0$ (nearest-neighbor hopping case with $t=C$) into Eq.~\eqref{eq:error-bounds-exp-minimum-eigenvalues}, we obtain $m_{f_B}=\omega-2t$, meaning that the system becomes singular at $\omega=2t$. Therefore, in the case of $\omega>2t$ the renormalized couplings show convergence, while for $\omega < 2t$ the system is ill-behaved.

\subsection{Power-law decay}
In the case of a power-law decay of the hoppings $t_n = t_{-n} = C n^{-\alpha}, C \in \mathbb{R}, n \in \mathbb{N}, \alpha > 1$ and $t_0=\omega$, we can calculate the Fourier series for the matrices $B$ and $R$ in the same way as before and arrive at 
\begin{align}
    m_{f_B} &= f_B(\phi=\pi) = \omega + 2C \sum_{n=1}^{m} (-1)^n n^{-\alpha}\\
    M_{f_R}&= f(\phi=0) = 2C \sum_{n=m+1}^{N} n^{-\alpha} ,
\end{align}
which can then be inserted into Eq.~\eqref{eq:error-bounds} to obtain the explicit a priori error bounds. Further details on the derivation can be found in App.~\ref{app:error-bounds-power-law}. The performance of the derived bounds will be investigated in Sec.~\ref{sec:numerical-error-bounds}.

\section{\label{sec:numerical-experiments}Numerical analysis}
\subsection{Verification of 1d analytics}
First, we verify the maximum bond dimension of \Dmax$=2n$ for the $n$-th-nearest-neighbor hopping tight-binding one-particle Green's function QTT with a different analytical derivation and, second, the results are numerically verified. 
\subsubsection{\label{sec:qtt-ranks}QTT ranks}
The bond dimension of the QTT form of the one-particle Green's function of the tight-binding model is derived using an approach that is general to long-ranged and complex hoppings. Therefore, we make the following ansatz with a general tensor train of the form
\begin{align}
    &G_{2^{(R-1)}i_{1}+2^{(R-2)}i_{2}+...+2^{0}i_{R}} =  \sum_{\alpha_1 ... \alpha_{R-1}} g_{i_{1},1\alpha_1}^{(1)} g_{i_{2},\alpha_1 \alpha_2}^{(2)} ... g_{i_{R},\alpha_{R-1}1}^{(R)},
\end{align}
with $\alpha_{\ell} \in (1,...,D_{\ell})$, where $D_{\ell}$ is the bond dimension of the $\ell$-th bond. The only requirement is the cyclic property of the translational operator, $P^N \propto \mathbb{I}$ (the identity) 

For simplicity, let us consider $N$-lattice sites (with $N=2^R$), where the translational operator reads 
\begin{align}
 P = \sum_{i=1}^{N}\ket{i}\bra{i+1},
\end{align}
with $\ket{N+1}\equiv \ket{1}$ under the periodic boundary condition.
Then, from translational invariance of the Hamiltonian $H$ follows translational invariance of its Green's function $G_{\omega,\eta} = \dfrac{1}{\omega-i\eta - H }$.
Moreover, it is possible to express the single-particle Green's function as a function of the translational operator, i.e., 
\begin{align}
G_{\omega, \eta} = \sum_{n=0}^{N-1} g_{\omega, \eta}(n) P^{-n}.
\label{Eq:gn0}
\end{align}
Based on this observation, we can see that the QTT expressibility of the Green's function boils down to estimating the rank of $g(n)$.

\subsubsection{Rank 2: the nearest-neighbor tight-binding Hamiltonian}
\label{Sec:A}

For our first example, we consider the nearest-neighbor tight-binding Hamiltonian, $H = P + P^{\dagger} = P + P^{-1}$. Then, we have 
\begin{align}
G_{\omega, \eta = 0} = \frac{1}{\omega - P - P^{-1}} = \frac{P}{(P-z_+)(P-z_-)} = \sum_{n=0}^{\infty}b_n P^{-n},
\label{Eq:gn1}
\end{align}
with $z_\pm = \frac{1}{2}(\omega \pm \sqrt{\omega^2 - 4})$, and $b_n = \dfrac{z_{+}^n - z_{-}^n}{z_+-z_-}$ by invoking the residual theorem.
Next, taking into account the cyclic property $P^{kN+n} \equiv P^n$, we can further reduce the infinite sum to a finite one, that allows us to express $g(n)$ in Eq.~\eqref{Eq:gn0} in terms of $b_n$.
Doing that gives
\begin{align}
  g(n) = \sum_{k=0}^{\infty} b_{n+kN} = \frac{z_+^n / (z_+ - z_-)}{1-z_+^N} - \frac{z_-^n / (z_+ - z_-) }{1-z_-^N}.
  \label{Eq:gn2}
\end{align}

Since a single exponential function has a QTT representation of rank $1$, Eq.~\eqref{Eq:gn2} having two exponentials is of rank $2$ \cite{Kazeev2012}. 
Importantly, the QTT rank of $g(n)$ is independent of system size ($N$) and frequency ($\omega$), except when the Green's function is singular, e.g., at the band edges when $\omega = \pm 2$ and $z_+ = z_-$.

One comment follows.
In Eq.~\eqref{Eq:gn2}, the finite-size scaling factor enters the denominator due to the geometric sum $\sum_{k=0}^{\infty}z^{kN} = 1/ (1-z^N)$. 
This evidently works when $|z| < 1$.
However, for $|z|>1$ this begs for explanations.
For this reason, in the following we inspect the expansion of $1/ (P-z)$ when $|z|>1$:  
\begin{align}
\dfrac{1}{P-z} &= \dfrac{-1/z}{1-P/z} = \frac{-1}{z} 
\left[ 
1 + \frac{P}{z} + \frac{P^2}{z^2} + \cdots
\right] \nonumber \\
&\overset{P^{kN+n} \equiv P^n}{=} \frac{-1}{z}\frac{1}{1-z^{-N}} \left(1 + \frac{P}{z} + \cdots \frac{P^{N-1}}{z^{N-1}} \right) \nonumber \\
&\overset{P^{n} \equiv P^{n-N}}{=} \frac{1}{1-z^N} \sum_{n=0}^{N-1} z^{n} / P^{n+1},
\label{Eq:gn3}
\end{align}
where we first do a Taylor expansion over $1/z$, and apply the geometric sum on the second line where the common ratio $r=|z^{-N}|$ smaller than one guarantees the convergence. 
We then recast the polynomial of $P$ in terms of its inverse $P^{-1}$ by applying the cyclic rule.
Indeed, the final expression in Eq.~\eqref{Eq:gn3} proves that the finite-size scaling factor $(1-z^N)$ remains identical for both $|z|<1$ and $|z|>1$.

\subsubsection{Rank \texorpdfstring{$2n$}{2n}: the \texorpdfstring{$n$}{n}-th-nearest-neighbor tight-binding Hamiltonian \label{sec:n-th-nearest-neighbor-bond-dimension}} 

The finite-ranged tight-binding Hamiltonian can be written as 
$H = \sum_{r=1}^{n} h_r (P^r + P^{-r})$, where $h_r$ describes the hopping amplitudes as a function of distance $r$.
Then, for the Green's function, we have 
\begin{align}
G_{\omega} = \frac{1}{\omega - \sum_{r=1}^{n} h_r (P^r + P^{-r})} = \frac{P^n}{f(P)}, 
\label{Eq:gn4}
\end{align}
where $f(P)$ is a polynomial function of degree $2n$.

In the simplest scenario, the function, $f(z) \propto \prod_{i=1}^{2n} (z-z_i) $, will have $2n$ simple roots with no degeneracy.
Then repeating the procedure in Sec.~\ref{Sec:A}, we obtain
\begin{align}
  g(r) = \sum_{i=1}^{2n} F_i z_i^r, 
  \label{Eq:gn44}
\end{align}
where $F_i = \dfrac{z_i^{n-1}} {f_i(z_i) (1-z_i^N)}$ is the form factor, which characterizes the contribution of each exponent, and $f_i(z) = \dfrac{f(z)}{z-z_i}$.
Following Eq.~\eqref{Eq:gn4}, we conclude that the maximal QTT rank is $2n$ in the case of $n$-th-nearest neighbor hopping verifying the results of the previous sections. Let us emphasize that the bond dimension of $2n$ precisely matches the number of renormalized couplings generated in the renormalization procedure in Sec.~\ref{sec:renormalization} since both characterize the flow of information between different length scales of the system. 

\subsubsection{Numerical verification}
\begin{figure}
    \centering
    \includegraphics[width=1.0\linewidth]{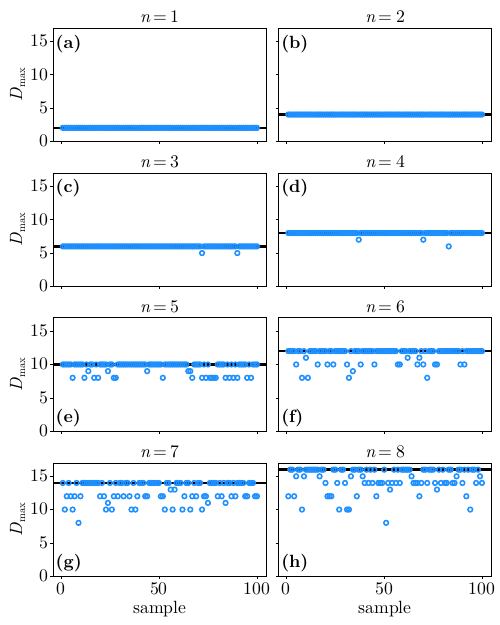}
    \caption{Maximum bond dimension \Dmax{} of the QTT of the one-particle Green's function for $n=1,2,...,8$-th-nearest-neighbor hopping with periodic boundary conditions and with randomly generated couplings $\omega, t_n \in [0,1]$ with $t_n>t_{n+1}$ and an SVD cutoff of $\epsilon=10^{-12}$. Blue circles show the \Dmax{} for 100 samples in each case and the black solid line (covered by the blue circles for small $n$) is the exact analytical result \Dmax$=2n$.}
    \label{fig:bond-dimension-periodic}
\end{figure}

\begin{figure}
    \centering
    \includegraphics[width=1.0\linewidth]{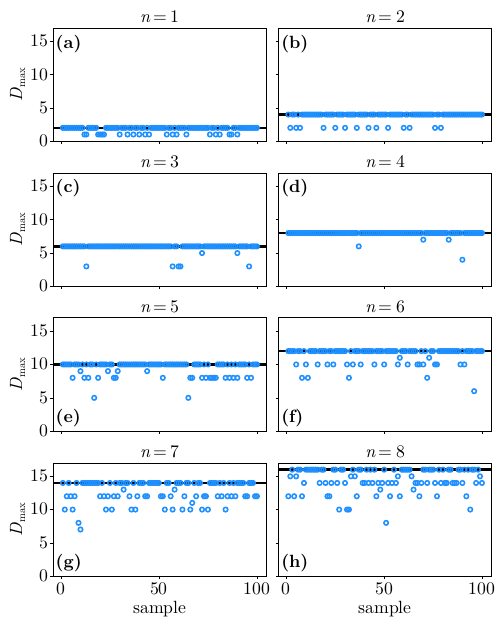}
    \caption{Maximum bond dimension \Dmax{} of the QTT of the one-particle Green's function (first site) for $n=1,2,...,8$-th-nearest-neighbor hopping with open boundary conditions and with randomly generated couplings $\omega, t_n \in [0,1]$ with $t_n>t_{n+1}$ and an SVD cutoff of $\epsilon=10^{-12}$. Blue circles show the \Dmax{} for 100 samples in each case and the black solid line (covered by the blue circles for small $n$) is the exact analytical result \Dmax$=2n$.}
    \label{fig:bond-dimension-open}
\end{figure}

In the following, numerical experiments are conducted as checks for the derived results of the previous sections. For the case of the $n-$th-nearest-neighbor hopping tight-binding model, we randomly generate 100 samples of couplings for $n=1,2,...,8$ each (with $\omega, t_n \in [0,1]$ and $t_n>t_{n+1}$), construct the corresponding Hamiltonian, generate the numerical Green's function by inversion $G=(\omega \mathbbm{1} - H)^{-1}$ and decompose one column of the resulting matrix (translational symmetry) into a QTT with SVD applying a cutoff of $\epsilon=10^{-12}$. In Fig.~\ref{fig:bond-dimension-periodic}, the results are reported. As can be seen, the maximum bond dimension of the various samples (blue circles) is bound from above by $2n$ indicated by the black solid line. Since in the numerical QTT construction, some information is truncated, some samples lead to maximum bond dimensions below $2n$, but never above, verifying the analytical results from the previous section. 
Similarly, in Fig.~\ref{fig:bond-dimension-open}, we show numerical results for open boundary conditions instead of periodic ones. Since the system loses its translational symmetry, we construct a QTT of the Green's function with respect to the first site of the chain. It can be seen that the same bound as in the case of periodic boundary conditions (\Dmax$=2n$) is maintained. However, an analytical derivation of this bound for the open boundary condition case will be left for future work.

\subsection{Error bounds \label{sec:numerical-error-bounds}}
\subsubsection{Error bounds in Eq.~(\ref{eq:error-bounds})}
\begin{figure}
    \centering
    \includegraphics[width=1.0\linewidth]{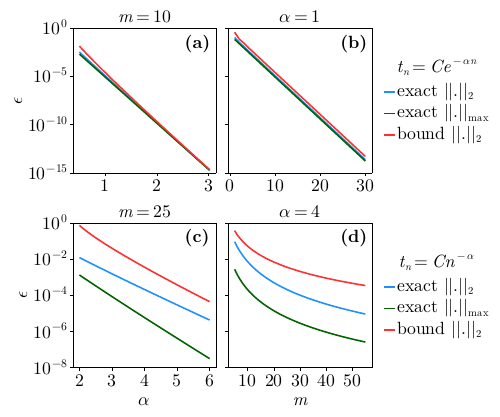}
    \caption{ Approximation error of $G=A^{-1}:=(\omega \mathbb{I} -H)^{-1}$ by a QTT with maximum bond dimension \Dmax$=2m$. (a) and (b) show the case of exponentially decaying couplings $t_n=Ce^{-\alpha n}$, while (c) and (d) present results for power-law decaying hoppings $t_n=Cn^{-\alpha}$ for the tight-binding model with $C=1, \omega=2$ and 1001 sites, where $m$ denotes the number of nonzero hoppings in the approximating matrix. (a) and (c) show the errors for fixed $m$ as a function of $\alpha$ and (b) and (d) for fixed $\alpha$ as a function of $m$. The calculated a priori error bound (red) and the exact approximation error (blue), both with respect to the spectral norm, as well as the exact maximum normalized error (green) are shown.}
    \label{fig:error-bound}
\end{figure}

Revisiting the derived error bounds in Sec.~\ref{sec:approximate-construction}, we numerically check their validity and performance in Fig.~\ref{fig:error-bound}. In Figs.~\ref{fig:error-bound}(a) and (b) the case of exponentially decaying couplings $t_n=Ce^{-\alpha n}$, while in (c) and (d) the case of power-law decaying hoppings $t_n=Cn^{-\alpha}$ is presented for the tight-binding model with $C=1, \omega=2$ and 1001 sites, where $m$ denotes the number of nonzero hoppings in the approximating matrix $B$. We show the calculated error bound computed with the spectral norm (red) in accordance with Sec.~\ref{sec:approximate-construction} in comparison to the exact error $\epsilon=||A^{-1}-B^{-1}||$ for both the spectral norm $||.||_2$ (blue) and the maximum normalized error $||A^{-1}-B^{-1}||_{\max}:= \max(A^{-1}-B^{-1})/\max(A^{-1})$ (green). It can be seen that in the case of exponentially decaying couplings the error bound is very tight and allows a good a priori estimate of the true error of the approximation of the one-particle Green's function by a QTT with maximum bond dimension \Dmax$=2m$. Moreover, the error decreases exponentially resulting in a good QTT approximation already with small $m$. In the case of power-law decaying hoppings, the a priori bound on the error is looser, but still gives a correct upper estimate on the error of the corresponding Green's function QTT with \Dmax$=2m$. 
The qualitative difference between the upper (exponential decay) and lower (power-law decay) panels in Fig.~\ref{fig:error-bound} can be understood in the following way. First of all, the error in panel (a) is decaying much faster than in panel (c), because in the exponential case, hoppings with $t_n = C e^{-\alpha n}, n > m=10$ are neglected which are much smaller than in the power law case, where $t_n = C n^{-\alpha}, n>m=25$ are omitted. Additionally, this leads to a tighter bound in the exponential case, which is further elaborated on following the derivation in App.~\ref{app:general-error-bounds}. Due to the variation of the ``cutoff'' parameter $m$ in panel (d), the power-law behavior of the decaying couplings becomes visible in the form of the error deviating from an exponential decrease, thus, differing from the other panels. Additionally, we extend panels (a) and (c) to larger values of $\alpha$ in Fig.~\ref{fig:error-bound-reply} in App.~\ref{app:error-bounds-extended-figure}.

These results are important from different perspectives. Numerically computing the QTT representation by SVD or tensor cross interpolation (TCI) \cite{NunezFernandez2022, Ritter2024, NunezFernandez2024} is typically a ``black box'' with respect to the approximation error. This means that in practical calculations we can, in general, not a priori be certain how well the QTT approximation will represent the original function or data. This is especially important when applying TCI, where the function or data is only sampled instead of evaluated everywhere as with SVD. Hence, important features might be missed by the algorithm resulting in a smaller maximum bond dimension. Therefore, a priori error estimates connected to the maximum bond dimension would be very helpful for applying this method. Hence, we hope that the results of this section inspire new research into this direction to overcome some of these problems.

\subsubsection{Compressibility of Eq.~(\ref{Eq:gn4})}
Is the QTT representation of Eq.~\eqref{Eq:gn4} compressible?
The answer depends largely on the form factor.
As a concrete example, consider a system with $1/r^2$ decaying long-range hopping.
Under the periodic boundary condition, the hopping amplitude reads $h_r = \dfrac{(\pi/N)^2}{{\rm{sin}}^2 (\pi r/N)}$.
Since the hopping amplitude decays, it is natural to introduce a maximal cutoff distance $n=r_{\rm{max}}$ to the Hamiltonian, which in turn allows us to limit the QTT rank, which is $2r_{\rm{max}}$ according to Eq.~\eqref{Eq:gn4}

\begin{figure}[!htb]
  \includegraphics[width=1.0\linewidth,trim = 0.0in 0.0in 0.0in 0.0in,clip=true]{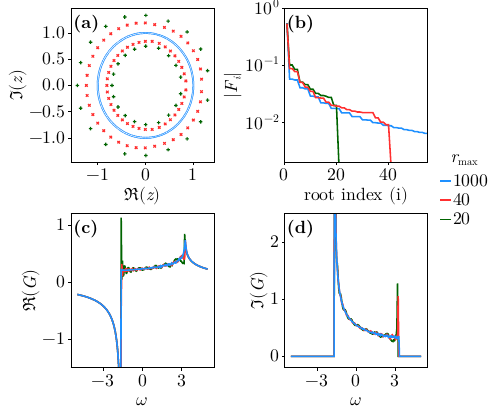}
  \caption{QTT approximation for single-particle Green's function of $1/r^2$ decaying hopping with $N=2^{30}$.
   (a) The root distribution of $f(T)$ in Eq.~\eqref{Eq:gn3} under different maximal distance cutoffs ($r_{\rm{max}}$) at $\omega = 0$.
   (b) Spectra of the leading form factors of Eq.~\eqref{Eq:gn4} at $\omega = 0$.
   (c,d) The real and imaginary parts of the local Green's function at $i=0$ for $\omega \in [-5,5]$ with $\eta = 10^{-8}$.
  }
  \label{fig:r2}
\end{figure}

In Fig.~\ref{fig:r2}, we show the results by approximately solving Eq.~\eqref{Eq:gn4} at a large system size $N=2^{30}$ using different choices of $r_{\rm{max}}$.
The numerical exact results shown in green are obtained at $r_{\rm{max}}=1000$, where $h_{r=1000} \sim 10^{-6}$ is negligible.
We then choose a rather small cutoff $r_{\rm{max}}=20$ and $40$ to inspect the consequence of the Hamiltonian truncation.

Looking at $\omega=0$ as a demonstration, we see [Fig.~\ref{fig:r2}(a)] that at the numerical exact limit, all the roots of $f(T)$ condense uniformly close to the unit circle, while the effect of a short $r_{\rm{max}}$ cutoff is to pull them away from the unit circle, separating them into two disparate groups $|z|\lessgtr 1$. 
This appears to have an interesting consequence.

Recalling our derivation of $g(n)$, it shows that the finite-size effect figures in as a scaling factor $(1-z_i^N)$ in the denominator for each root $z_i$. 
This implies roots sitting outside the unit circle will eventually become irrelevant in the large-$N$ limit.
Fig.~\ref{fig:r2}(b) succinctly corroborates this observation.
There, by sorting the $F_i$ [see Eq.~\eqref{Eq:gn4}] in descending order, a sharp drop in its magnitude is clearly seen when $i \sim r_{\rm{max}}$ (It is not difficult to verify those carrying finite weights are from roots within the unit circle in Fig.~\ref{fig:r2}(a)).

In short, for the long-ranged hopping Hamiltonian, its QTT rank of the single-particle Green's function, $g(n)$, is compressible. Given the maximal hopping distance $r_{\rm{max}}$, the QTT rank can be reduced to about $r_{\rm{max}}$ instead of $2r_{\rm{max}}$.
Note this finding is not specific to the form of hopping (we have used truncated $1/r^2$ merely for numerical demonstration), since $z_i$ and $1/z_i$ are always simultaneous solutions for $f(z)=0$. 

Having established the QTT rank with respect to the maximal hopping distance $r_{\rm{max}}$, we are now in the position to study its consequence on the Green's function.
In Fig.~\ref{fig:r2}(c,d), we plot the real and imaginary parts of the on-site element ($i=0$) over an energy window $-5 < \omega < 5$.
Note the rank of QTT is energy-independent, while to avoid the singularity, particularly at band edges, we introduce a small broadening with $\eta = 10^{-8}$.
Clearly, even with a short-range cutoff $r_{\rm{max}}=20$, the approximated Green's function is not so bad as to reproduce the rough shape of the numerical exact result, which we computed at $r_{\rm{max}}=1000$. 
Truncating the long-range hopping, however, causes small fluctuations that are especially noticeable when close to the singularity.

\subsection{\label{sec:interacting-model}Interacting model and higher dimensions}
In this section, we will explore the QTT compressibility using TCI beyond one dimension for the interacting case. So far, an analytical solution for both the non-interacting and interacting cases in two or more dimensions has not been found. However, numerically generating QTTs and subsequently analyzing the bond dimension, can clarify their behavior and dependence on variables like filling or temperature. 

Using Matsubara formalism, we start with the next-nearest-neighbor tight-binding model in momentum space and the dispersion relation
\begin{math}
    \epsilon(\mathbf {k}) = -2t\sum_i\cos(k_i)- 4 t'\prod_i \cos{k_i}
\end{math}
, chemical potential $\mu$, and the self-energy inspired by a self-energy deep in the Mott-phase
\begin{math}
    \Sigma(\mathbf{k},\omega_n) = U/2 + \frac{U^2}{4}\frac{1}{i\omega_n+\alpha\epsilon(\mathbf{k})}
\end{math} \cite{wagner2023mott}
with the fermionic Matsubara frequencies $\omega_n$. The parameters $U$, $t$, $t'$ and $\alpha$ are set to $4.0$, $2.0$, $-0.3t$ and $1.0$ respectively.
Although the analytical Mott-phase inspired self energy is independent of the chemical potential, tuning $\mu$ will start to shift spectral weight resulting in a transition to a metal. Due to the $\mu$-independent self energy, this transition is closer related to the
metal-insulator transition in the Falicov-Kimball model \cite{Falicov1969,Freericks2003} than to a Mott transition. Moreover, when moving away from half-filling, the Green's function $G(\mathbf {k}, \omega_n) = (G_0(\mathbf{k},\omega_n)^{-1}-\Sigma(\mathbf{k},\omega_n))^{-1}$ changes rapidly at the first Matsubara frequency, $\omega_n=\frac{\pi}{\beta}$, and sharp peaks in the momentum dependence emerge, which are naturally costly to compress. Thus, the maximum bond dimension is expected to increase for $\mu>U/2$. Since increasing temperature smears out peaks, while lowering temperature leads to sharper features, increasing the inverse temperature $\beta=1/k_BT$ is expected to result in a rise in the bond dimension in the numerical QTT construction up to a certain error tolerance $\epsilon$. 
\begin{figure}[!ht]
  \includegraphics[width=1.0\linewidth,trim = 0.0in 0.0in 0.0in 0.0in,clip=true]{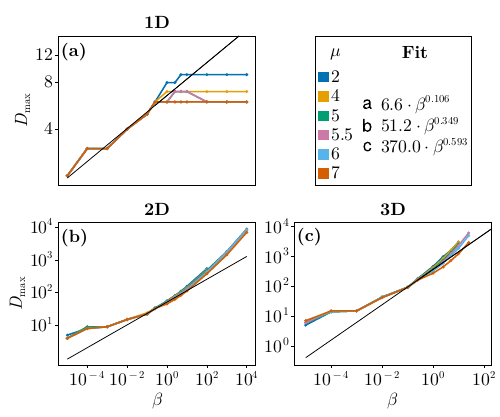}
  \caption{Maximum bond dimension of the QTT of the non-interacting one-particle Green's function $G_0(\mathbf {k},\omega_n=\frac{\pi}{\beta})$ as a function of inverse temperature $\beta$ for various values of the chemical potential $\mu$ in one (a), two (b) and three (c) dimensions computed with TCI at the tolerance $\epsilon=10^{-5}$ for $2^{24}$ points in momentum space in each dimension. The black line indicates a power law fit with the parameters in the upper right corner.}
  \label{fig:Dmax_tol1e-5_interdim_non_interacting}
\end{figure}

Let us first investigate the maximum bond dimension of the non-interacting problem. In Fig.~\ref{fig:Dmax_tol1e-5_interdim_non_interacting}, the maximum bond dimension $D_{\max}$ of the QTT of the one-particle Green's function in momentum space (we set $\omega_n=\frac{\pi}{\beta}$) is shown as a function of inverse temperature $\beta$ for different chemical potentials, in a one- (a), two- (b) and three-dimensional non-interacting ($U=0$) system (c). The QTT is computed using TCI with $2^{24}$ k-points in each direction and a tolerance of $\epsilon=10^{-5}$ and we fit the $\beta$-scaling of the resulting maximum bond dimensions to a power law \cite{Lorenz2024}. It can be seen that in the non-interacting case the bond dimension only weakly depends on the chemical potential, while the expected growth with lowering the temperature is observed in the two- and three-dimensional cases. However, in the one-dimensional case the bond dimension saturates at a low \Dmax{} value. This can be understood from the results of the previous sections, where in the next-nearest-neighbor hopping case a QTT with bond dimension four in real space can be analytically derived. To obtain the Green's function in momentum space, Fourier transformation has to be applied. In QTT language, a Fourier transform can be represented by a low bond dimension MPO. Therefore, applying this low bond dimension MPO to the real space Green's function QTT with bond dimension four, will result in a low upper bound on \Dmax{} of the momentum QTT, which is observed in Fig.~\ref{fig:Dmax_tol1e-5_interdim_non_interacting}(a). Hence, the saturation of the bond dimension reflects the QTT as a natural representation of a renormalization procedure in one dimension. Moreover, the temperature dependence of \Dmax{} stems from the numerical QTT construction, where truncation up to a tolerance $\epsilon$ is applied. In the exact QTT construction the bond dimension does not depend on temperature. However, in higher dimensions, in such a QTT-based renormalization procedure, additional couplings will be generated at each renormalization step, where the maximum bond dimension reflects the disregarding of some of these couplings below the tolerance. 

\begin{figure}[!ht]
  \includegraphics[width=1.0\linewidth,trim = 0.0in 0.0in 0.0in 0.0in,clip=true]{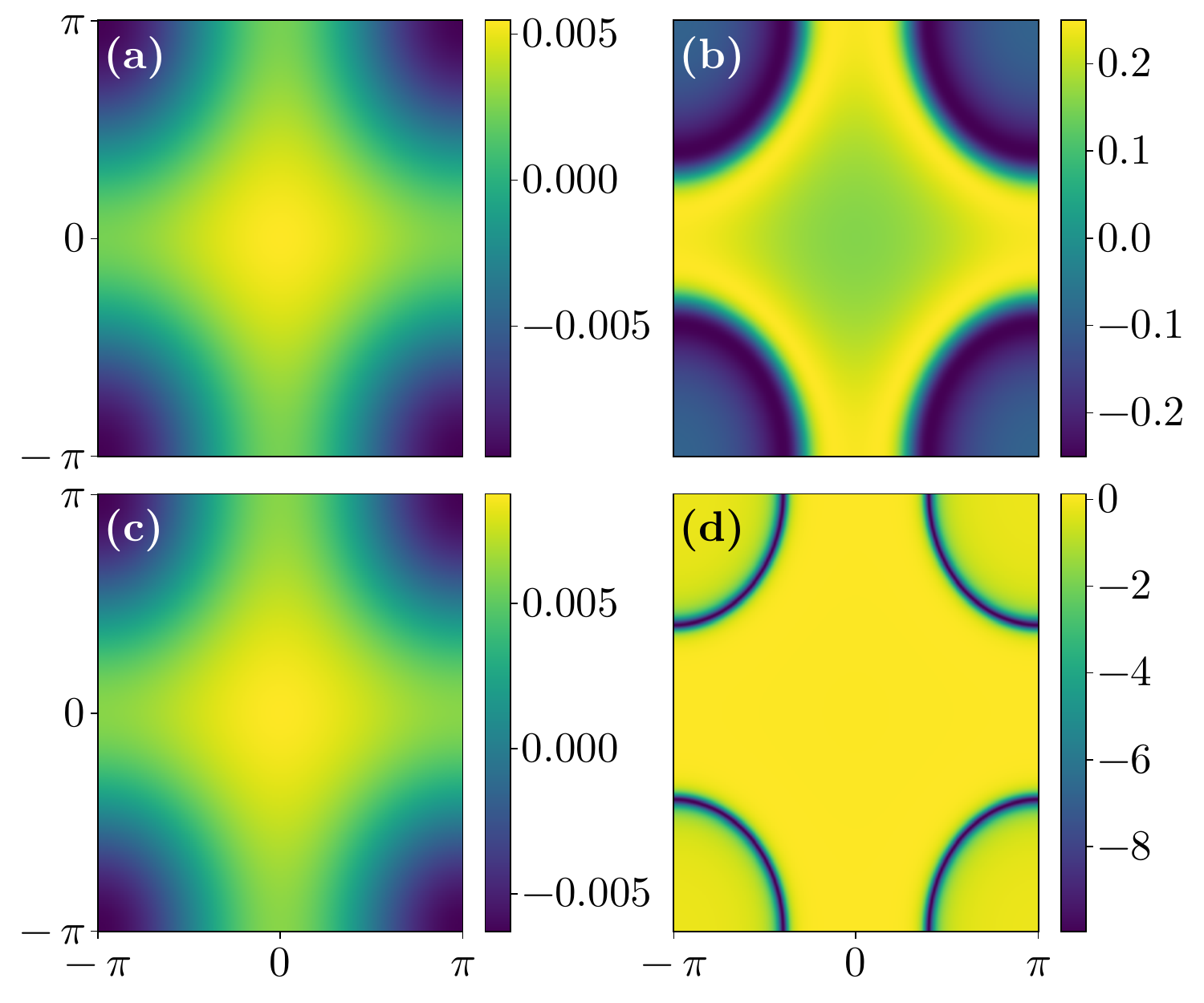}
  \caption{Momentum dependence of the real part of the 2D interacting Green's function at the first Matsubara frequency and $U=4.0$ (other parameters as specified in the text) for two different values of the chemical potential (a),(b) $\mu = U/2$ and (c),(d) $\mu = U/2+3.9$ and inverse temperature (a),(c) $\beta=0.1$ and (b),(d) $\beta=10^3$.}
  \label{fig:green_entang}
\end{figure}
We now turn to the interacting case. Fig.~\ref{fig:green_entang} shows the real part of $G(\mathbf {k}, \omega_n = \frac{\pi}{\beta})$ in two dimensions for two different fillings and two different temperatures. It demonstrates that at high temperatures the Green's function hardly differs for different chemical potentials, whereas when the inverse temperature ($\beta=10^3$) is increased, sharp structures appear away from half-filling. 
\begin{figure}[!ht]
  \includegraphics[width=1.0\linewidth,trim = 0.0in 0.0in 0.0in 0.0in,clip=true]{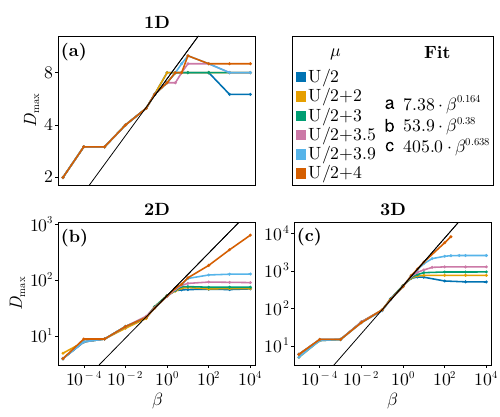}
  \caption{Maximum bond dimension of the QTT of the interacting one-particle Green's function $G(\mathbf {k},\omega_n=\frac{\pi}{\beta})$ as a function of inverse temperature $\beta$ for various fillings in one (a), two (b) and three (c) dimensions computed with TCI at the tolerance $\epsilon=10^{-5}$ for $2^{24}$ points in momentum space in each dimension. The black line indicates a power law fit with the parameters in the upper right corner.}
  \label{fig:Dmax_tol1e-5_interdim}
\end{figure}

Analogously to the non-interacting case, we plot in Fig.~\ref{fig:Dmax_tol1e-5_interdim}   \Dmax{} as a function of temperature for the same parameters as in Fig.~\ref{fig:Dmax_tol1e-5_interdim_non_interacting} but for $U=4.0$. At high temperatures ($\beta=[0.001,1]$) $D_{\max}$ is similar for all fillings. However, when lowering the temperature ($\beta=[10,10^4]$) $D_{\max}$ saturates for different fillings at different values (except for $\mu=U/2+4$). This suggests that, starting from a certain threshold value \(\beta_a\), the complexity of the Green's function remains constant, and that the specific value of the saturated maximum bond dimension $D_{\max}^a$ depends on the system filling. The saturation is caused by the interaction leading to a broadening of the poles, where larger interaction values (not shown) lead to saturation at lower values of \Dmax{}. The case of $\mu=U/2+4$ requires more attention. When tuning the chemical potential, spectral weight in the analytically continued and $k$-integrated spectral function gets shifted to lower values of $\omega$. At $\mu=U/2+4$ spectral weight is shifted to $\omega=0$ resulting in a transition to a metal (see Fig.~\ref{fig:spectral-function} in App.~\ref{app:spectral-function}). This becomes visible in the different bond dimension behavior in this case, where \Dmax{} does not saturate for the analyzed temperatures. Since QTTs can be interpreted as a one-dimensional renormalization method, it might be an interesting research question for the future, whether QTT-based projected entangled-pair states (PEPS)~\cite{Verstraete2004} approaches can play a similar role in higher dimensional cases, resulting in an exact analytical QTT construction and, thus, saturating bond dimensions also in the non-interacting case. 

\section{\label{sec:conclusions}Conclusion and outlook}
The interplay between MPSs, entanglement and renormalization has been extensively studied in the past. This work pioneers a new direction by identifying QTTs as the natural framework for one-dimensional real-space renormalization. By investigating the one-dimensional $n-$th-nearest-neighbor hopping tight-binding model we were able to recast a cyclic reduction-based real-space RG procedure into QTT language, establishing a direct correspondence between the QTT bond dimension of the one-particle Green's function and the number of renormalized couplings generated during the RG process. Both, the bond dimension and the number of renormalized couplings, represent the flow of information between different length or energy scales of the system, reflecting the degree of length scale entanglement in the system. Therefore, it is unsurprising that the number of rescaled couplings ($2n$) precisely matches the maximum bond dimension of the QTT, supporting the interpretation of QTTs themselves as renormalization procedures in one dimension.

While QTTs have predominantly been applied in numerical contexts within many-body physics, our analytical results provide a fundamentally new perspective. We showed that QTTs can drastically reduce the complexity of analytical calculations, opening opportunities to tackle problems previously deemed analytically intractable. Thus, this work paves the way for employing QTTs as analytical tools, potentially inspiring an entirely new research direction. Furthermore, the application of QTTs has so far relied on “trial-and-error” assessments of approximation accuracy relative to compression rates, without a priori analytical guidance. In this paper, we derived an analytical error bound for the QTT approximation with a specific maximum bond dimension for tight-binding models with decaying couplings, thereby making the QTT method more transparent and accessible for broader applications by providing conceptual guidance for the typically black-box QTT decomposition.

Our work brings significant advancement to the understanding of length scale entanglement in physical systems. In this way, the investigation of the connection of the outlined approach to entanglement renormalization schemes like MERA, will be an interesting route to follow in future work. Additionally, direct applications of this technique in momentum space or systems with open boundary conditions may yield further insights into the nature of energy scale entanglement. Finally, the investigation of higher-dimensional problems led to the conclusion that a QTT-based PEPS approach might be beneficial in such models to incorporate the additional dimensions efficiently, presenting exciting opportunities for both analytical and numerical research moving forward.

\begin{acknowledgments}
This work was funded in part  by the Austrian Science Fund (FWF) Projects No.~P~36332 (Grant DOI 10.55776/P36332), V~1018 (Grant DOI 10.55776/V1018) and PIN~4372024 (Grant DOI 10.55776/PIN4372024). J.-W. L. acknowledges funding support from the Plan France 2030 ANR-22-PETQ-0007 ``EPIQ''. For open access purposes, the authors have applied a CC BY public copyright license to any author-accepted manuscript version arising from this submission. The raw data for the figures reported are available at Ref.~\onlinecite{RohshapData2025}.
\end{acknowledgments}

\appendix
\begin{widetext}
\section{\label{app:cyclic-reduction-proof} Nearest-neighbor renormalization procedure}
The cyclic reduction-based renormalization procedure leads to $2^{R-r}$ equations determining $2^{R-r-1}$ unknowns in the $r$-th coarse-graining step. However, half of the equations are redundant, if the results from the previous renormalization steps are inserted. This can be understood in the following way.

In the $r$-th renormalization step, let us for simplicity just consider the sites $i=0,1,2^{R- r}-1$, where $i=1,2^{R- r}-1$ lead to $G_{2^{( r)}}=G_{2^{R-1} + 2^{R-2} + ... +2^{( r)}}$ (periodic boundary conditions). Then these equations reduce to the following two equations determining only one unknown ($G_{2^{( r)}}$)
\begin{subequations}
    \begin{align}
    1 &= 2t^{( r)}  G_{2^{( r)}} + \omega^{( r)}  G_0 \label{eq:nn-equation1-rth-step} \\
    0 &= t^{( r)}(G_0 + G_{2^{( r+1)}}) + \omega^{( r)} G_{2^{( r)}} \label{eq:nn-equation2-rth-step}.
\end{align}
\label{eq:nn-equation-rth-step} \end{subequations}
Solving these equations separately, gives the following solution.
\begin{align}
     G_{2^{( r)}} &= \frac{1-G_0 \omega^{( r)}}{2 t^{( r)}} = -\frac{(G_0 + G_{2^{( r+1)}})t^{( r)}}{\omega^{( r)}} \label{eq:result-rth-step}
\end{align}
This means that this system is only solvable, if these two results coincide. Solving this equation for $G_{2^{( r+1)}}$ gives
\begin{align}
    G_{2^{( r+1)}} = -\frac{\omega^{( r)}}{2\left(t^{( r)}\right)^2}\left(1-G_0 \left[\omega^{( r)}-\frac{2 \left(t^{( r)}\right)^2}{\omega^{( r)}} \right] \right). 
\end{align}
Now the ``couplings'' $\omega^{( r)}, t^{( r)}$ can be expressed in terms of $\omega^{( r+1)}, t^{( r+1)}$, which leads to
\begin{align} \label{eq:result-r+1th-step}
    G_{2^{( r+1)}} &= \frac{1-G_0 \omega^{( r+1)}}{2 t^{( r+1)}}.
\end{align}
This means that both Eqs.~\eqref{eq:nn-equation-rth-step} are fulfilled, if the above equation holds. Comparing this result to Eq.~\eqref{eq:result-rth-step}, it can be seen that the two equations are identical, only differing by the renormalization step. Hence, the condition in Eq.~\eqref{eq:result-r+1th-step} is automatically fulfilled if there exists a solution for the previous renormalization step $r +1$. Moreover, with the help of the renormalization procedure a general solution for the Green's function $G_{2^{(r)}}$, with $0 \leq r \leq R-1$ is found: 
\begin{align}
    G_{2^{( r)}} = G_{2^{R-1} + 2^{R-2} + ... +2^{( r)}} &= \frac{1-G_0 \omega^{( r)}}{2 t^{( r)}}.
\end{align}
This means that Eqs.~\eqref{eq:nn-equation-rth-step} do not need to be included in the calculation of the Green's function of the subsequent step on a finer-grained lattice, hence, reducing the complexity of the problem.

\section{\label{app:nnn-renormalization}Next-nearest-neighbor renormalization procedure}
In this section, the idea of the cyclic reduction-based RG scheme will be explicitly extended to the next-nearest-neighbor hopping case, where $t_{ij} := t (\delta_{j,i-1}+\delta_{j,i+1}) + t' (\delta_{j,i-2}+\delta_{j,i+2})$. Then, Eq.~\eqref{eq:general-GF} simplifies to
\begin{align} \label{eq:nnn-general-GF}
    &t' G_{i-2} + t G_{i-1} + \omega G_{i}+ t G_{i+1} + t' G_{i+2}  = \delta_i + a (\delta_{i-1} + \delta_{i+1}),
\end{align}
where the parameter $a=0$ is introduced in order to derive consistent equations in the renormalization procedure (see App.~\ref{app:nnn-additional-couplings} for an extended derivation). After completing this first cyclic reduction step by separating out the even parts, we arrive at
\begin{align}
    &t'^{(1)} G_{2(i-2)} + t^{(1)} G_{2(i-1)} + \omega^{(1)} G_{2i} +t^{(1)} G_{2(i+1)}  + t'^{(1)} G_{2(i+2)}  = \delta_i + a^{(1)} (\delta_{i-1} + \delta_{i+1}) ,
\end{align}
which is consistent with Eq.~\eqref{eq:nnn-general-GF} with the four renormalized couplings $t'^{(1)}, t^{(1)}, \omega^{(1)}, a^{(1)}$. Hence, the coarse-graining of the system can be continued and for the $r$-th step we arrive at
\begin{align}
    &t'^{(r)} G_{2^r(i-2)} + t^{(r)} G_{2^r(i-1)} + \omega^{(r)} G_{2^ri} +t^{(r)} G_{2^r(i+1)}  + t'^{(r)} G_{2^r(i+2)}  = \delta_i + a^{(r)} (\delta_{i-1} + \delta_{i+1}) ,
\end{align}
where the four renormalized couplings are defined recursively in the following way
\begin{subequations}
    \begin{align}
    t'^{(r)}&=-\frac{\left(t'^{(r-1)}\right)^2}{2 a^{(r-1)} t^{(r-1)}-\omega^{(r-1)}}, \\
    t^{(r)}&= -\frac{2 t'^{(r-1)} \omega^{(r-1)}-\left(t^{(r-1)}\right)^2}{2 a^{(r-1)} t^{(r-1)}-\omega^{(r-1)}}, \\
    \omega^{(r)}&=-\frac{2 \left(t'^{(r-1)}\right)^2-2 \left(t^{(r-1)}\right)^2+\left(\omega^{(r-1)}\right)^2}{2 a^{(r-1)} t^{(r-1)}-\omega^{(r-1)}}, \\
    a^{(r)} &= \frac{t'^{(r-1)}-a^{(r-1)} t^{(r-1)}}{\omega^{(r-1)}-2 a^{(r-1)} t^{(r-1)}}. \label{eq:nnn-renormalized-coupling-a}
\end{align}
\label{eq:nnn-renormalized-couplings}
\end{subequations}
Coarse-graining the system until only two renormalized sites are remaining ($r=R-1$), leads to the following two equations ($i=0,1$)
\begin{subequations}
    \begin{align}
    1 &= \left(\omega^{(R-1)}+2t'^{(R-1)}\right) G_0 + 2t^{(R-1)} G_{2^{(R-1)}} ,  \\
    2 a^{(R-1)} &= 2t^{(R-1)} G_0 + \left(\omega^{(R-1)}+2t'^{(R-1)}\right) G_{2^{(R-1)}} ,
\end{align}
\label{eq:nnn-renormalization-R-1}
\end{subequations}
where the factor $2$ in $2 a^{(R-1)}$ in the second equation is due to the $a^{(R-1)} (\delta_{i-1} + \delta_{i+1})$ term and periodic boundary conditions ($i=2 \equiv i=0 \rightarrow \delta_{2} \equiv \delta_0$). These equations can then be solved for $G_0, G_{2^{(R-1)}}$ and the renormalization procedure to obtain the Green's functions can then be continued in the same way as in the nearest-neighbor hopping case (see App.~\ref{app:nnn-solutions}). Similarly, some Green's functions are of the same structure and can be classified into groups, which reduces the computational effort to obtain the Green's functions by reducing the number of equation.

We want to note that the choice of a quantics base corresponding to a renormalization of two sites in every step is not unique and other bases can be used as well. This will be shown for base 4 in App.~\ref{app:nnn-base-4}, which corresponds to renormalizing four sites in each step of the renormalization procedure. Still, regardless of the base used always four renormalized couplings are generated in the case of the next-nearest-neighbor tight-binding model.

\subsection{\label{app:nnn-additional-couplings}Origin of coupling parameter \texorpdfstring{$a$}{a} in consistent renormalization}
In this section, the origin of the coupling parameter $a$ in Eq.~\eqref{eq:nnn-general-GF} will be derived. Let us again start from Eq.~\eqref{eq:general-GF} and insert $t_{ij} := t (\delta_{j,i-1}+\delta_{j,i+1}) + t' (\delta_{j,i-2}+\delta_{j,i+2})$. This will lead to
\begin{align} 
    t' G_{i-2} + t G_{i-1} + \omega G_{i}+ t G_{i+1} + t' G_{i+2} = \delta_i , \label{eq:nnn-general-GF-naive}
\end{align}
which connects the Green's functions of neighboring sites. Similarly to the nearest-neighbor hopping case, we want to derive structurally the same equation, but with Green's functions of next-neighboring sites. Therefore, we shift this equation in the following way:
\begin{subequations}
    \begin{align} 
    &t' G_{i-4}+&t G_{i-3} &+ &\omega  G_{i-2} &+ &t G_{i-1} &+&t' G_{i} \ & \ & \ & \ & \ & \ & \ & \ & \ &= \delta_{i-2}  \\
    &\ &t' G_{i-3}&+&t G_{i-2} &+ &\omega  G_{i-1} &+ &t G_{i} &+&t' G_{i+1} \ & \ & \ & \ & \ & \ & \ &= \delta_{i-1}  \\
    &\ &\ &\ &t' G_{i-2} &+ &t G_{i-1} &+ &\omega G_{i} &+ &t G_{i+1} &+&t' G_{i+2} \ & \ & \ & \ & \ &= \delta_i  \\
    &\ &\ &\ &\ &\ &t' G_{i-1} &+ &t G_{i} &+ &\omega G_{i+1} &+ &t G_{i+2} &+ &t' G_{i+3} \ & \ & \  &= \delta_{i+1}  \\
    &\ &\ &\ &\ &\ &\ &\ &t' G_{i} &+ &t G_{i+1} &+ &\omega G_{i+2} &+ &t G_{i+3} &+ &t' G_{i+4}  &= \delta_{i+2} 
\end{align}
\label{eq:nnn-shifted-equations-naive}
\end{subequations}
After eliminating $G_{i \pm 3}, G_{i \pm 1}$ and separating out the even parts, we arrive at
\begin{align}
    &t'^{(1)} G_{2(i-2)} + t^{(1)} G_{2(i-1)} + \omega^{(1)} G_{2i} +t^{(1)} G_{2(i+1)}  + t'^{(1)} G_{2(i+2)} = \delta_i + \alpha (\delta_{i-1} + \delta_{i+1}) ,
\end{align}
with the three renormalized couplings $t'^{(1)}, t^{(1)}, \omega^{(1)}$ and $\alpha$, a function of $t,t',\omega$. However, due to the $\delta_{i \pm 2}$ terms in Eqs.~\eqref{eq:nnn-shifted-equations-naive} the term $\alpha (\delta_{i-2} + \delta_{i+2})$ is generated, where the even parts need to be kept in the cyclic reduction leading to $\alpha (\delta_{2i-2} + \delta_{2i+2}) = \alpha (\delta_{i-1} + \delta_{i+1})$. Therefore, this equation is not consistent with Eq.~\eqref{eq:nnn-general-GF-naive} due to this additional term that is generated. In order to make the renormalization scheme consistent, Eq.~\eqref{eq:nnn-general-GF-naive} needs to be modified to Eq.~\eqref{eq:nnn-general-GF}
\begin{align} 
    &t' G_{i-2} + t G_{i-1} + \omega G_{i}+ t G_{i+1} + t' G_{i+2}  = \delta_i + a (\delta_{i-1} + \delta_{i+1}),
\end{align}
where $a=0$ at this point, but will become non-zero in the renormalization procedure due to the $\alpha (\delta_{i-1} + \delta_{i+1})$ term as shown before. This equation is then shifted in the following way:
\begin{subequations}
\begin{align} 
    &t' G_{i-4}+&t G_{i-3} &+ &\omega  G_{i-2} &+ &t G_{i-1} &+&t' G_{i} \ & \ & \ & \ & \ & \ & \ & \ & \ &= \delta_{i-2} &+ a & (\delta_{i-3} + \delta_{i-1}) \label{eq:nnn-minus2-GF}  \\
    &\ &t' G_{i-3}&+&t G_{i-2} &+ &\omega  G_{i-1} &+ &t G_{i} &+&t' G_{i+1} \ & \ & \ & \ & \ & \ & \ &= \delta_{i-1} &+ a & (\delta_{i-2} + \delta_{i}) \label{eq:nnn-minus1-GF}  \\
    &\ &\ &\ &t' G_{i-2} &+ &t G_{i-1} &+ &\omega G_{i} &+ &t G_{i+1} &+&t' G_{i+2} \ & \ & \ & \ & \ &= \delta_i &+ a & (\delta_{i-1} + \delta_{i+1})\label{eq:nnn-minus0-GF}  \\
    &\ &\ &\ &\ &\ &t' G_{i-1} &+ &t G_{i} &+ &\omega G_{i+1} &+ &t G_{i+2} &+ &t' G_{i+3} \ & \ & \  &= \delta_{i+1} &+ a & (\delta_{i} + \delta_{i+2})\label{eq:nnn-plus1-GF}  \\
    &\ &\ &\ &\ &\ &\ &\ &t' G_{i} &+ &t G_{i+1} &+ &\omega G_{i+2} &+ &t G_{i+3} &+ &t' G_{i+4}  &= \delta_{i+2} &+ a & (\delta_{i-1} + \delta_{i+3}) \label{eq:nnn-plus2-GF}  
\end{align}
\end{subequations}
After eliminating $G_{i \pm 3}, G_{i \pm 1}$ and completing this first cyclic reduction step by separating out the even parts, we arrive at
\begin{align}
    &t'^{(1)} G_{2(i-2)} + t^{(1)} G_{2(i-1)} + \omega^{(1)} G_{2i} +t^{(1)} G_{2(i+1)} + t'^{(1)} G_{2(i+2)} = \delta_i + a^{(1)} (\delta_{i-1} + \delta_{i+1}) ,
\end{align}
with the four renormalized couplings $t'^{(1)}, t^{(1)}, \omega^{(1)}, a^{(1)}$ which is now consistent with Eq.~\eqref{eq:nnn-general-GF}. Therefore, an additional coupling $a$ is generated in deriving a consistent cyclic reduction-based renormalization scheme.

\subsection{\label{app:nnn-solutions}Solutions}
Solving Eqs.~\eqref{eq:nnn-renormalization-R-1} leads to the results
\begin{subequations}
    \begin{align}
    G_0 &= \frac{-4 a^{(R-1)} t^{(R-1)}+2 t'^{(R-1)}+\omega^{(R-1)}}{4 \left(t'^{(R-1)}\right)^2+4 t'^{(R-1)} \omega^{(R-1)}-4 \left(t^{(R-1)}\right)^2+\left(\omega^{(R-1)}\right)^2}, \\
    G_{2^{(R-1)}} &= \frac{4 a^{(R-1)} t'^{(R-1)}+2 a^{(R-1)} \omega^{(R-1)}-2 t^{(R-1)}}{4 \left(t'^{(R-1)}\right)^2+4 t'^{(R-1)} \omega^{(R-1)}-4 \left(t^{(R-1)}\right)^2+\left(\omega^{(R-1)}\right)^2}.
    \end{align}
\end{subequations}
We can then fine-grain our system and at the next step $r=R-2$ we insert the renormalized couplings
\begin{subequations}
    \begin{align}
    t'^{(R-1)}&=-\frac{\left(t'^{(R-2)}\right)^2}{2 a^{(R-2)} t^{(R-2)}-\omega^{(R-2)}}, &\qquad 
    t^{(R-1)}&= -\frac{2 t'^{(R-2)} \omega^{(R-2)}-\left(t^{(R-2)}\right)^2}{2 a^{(R-2)} t^{(R-2)}-\omega^{(R-2)}} \\
    \omega^{(R-1)}&=-\frac{2 \left(t'^{(R-2)}\right)^2-2 \left(t^{(R-2)}\right)^2+\left(\omega^{(R-2)}\right)^2}{2 a^{(R-2)} t^{(R-2)}-\omega^{(R-2)}}, &\qquad 
    a^{(R-1)} &= \frac{t'^{(R-2)}-a^{(R-2)} t^{(R-2)}}{\omega^{(R-2)}-2 a^{(R-2)} t^{(R-2)}}
\end{align} 
\end{subequations}
into the previous results and solve the following four equations ($i=0,1,2,3$)
    \begin{subequations}
        \begin{align}
    1 &= 2t'^{(R-2)} G_{2^{(R-1)}} + t^{(R-2)} G_{2^{(R-1)}+2^{(R-2)}} + \omega^{(R-2)} G_{0} +t^{(R-2)} G_{2^{(R-2)}}  , \\
    a^{(R-2)} &= 2t'^{(R-2)} G_{2^{(R-1)}+2^{(R-2)}} + t^{(R-2)} G_{0} + \omega^{(R-2)} G_{2^{(R-2)}} +t^{(R-2)} G_{2^{(R-1)}}  , \\
    0 &= 2t'^{(R-2)} G_0 + t^{(R-2)} G_{2^{(R-2)}} + \omega^{(R-2)} G_{2^{(R-1)}}  +t^{(R-2)}   G_{2^{(R-1)}+2^{(R-2)}} , \\
    a^{(R-2)} &= 2t'^{(R-2)} G_{2^{(R-2)}} + t^{(R-2)} G_{2^{(R-1)}} + \omega^{(R-2)} G_{2^{(R-1)}+2^{(R-2)}}  +t^{(R-2)}   G_0,
\end{align}
    \end{subequations}
    which leads to the following results
    \begin{subequations}
        \begin{align}
    G_{2^{(R-2)}} &= \frac{2 a^{(R-2)} t'^{(R-2)}+a^{(R-2)} \omega^{(R-2)}-t^{(R-2)}}{4 \left(t'^{(R-2)}\right)^2+4 t'^{(R-2)} \omega^{(R-2)}-4 \left(t^{(R-2)}\right)^2+\left(\omega^{(R-2)}\right)^2},\\
    G_{2^{(R-1)}+2^{(R-2)}} &= G_{2^{(R-2)}}.
\end{align}
    \end{subequations}
Similar to the nearest-neighbor case, the Green's functions can be grouped in classes the structurally have the same form due to the renormalization procedure. E.g.
\begin{subequations}
    \begin{align}
    G_{2^{(R-2)}} &= \frac{1-2t'^{(R-2)} G_{2^{(R-1)}}- \omega^{(R-2)} G_0}{2 t^{(R-2)}}, &\qquad
    G_{2^{(R-3)}} &= \frac{1-2t'^{(R-3)} G_{2^{(R-2)}}- \omega^{(R-3)} G_0}{2 t^{(R-3)}},\\
    G_{2^{(R-4)}} &= \frac{1-2t'^{(R-4)} G_{2^{(R-3)}}- \omega^{(R-4)} G_0}{2 t^{(R-4)}}, &\qquad 
    G_{2^{(r)}} &= \frac{1-2t'^{(r)} G_{2^{(r+1)}}- \omega^{(r)} G_0}{2 t^{(r)}},
    \end{align}
\end{subequations}
are structurally the same. 

\subsection{\label{app:nnn-base-4}Base 4}
The choice of a quantics base is not unique and other bases, where the base corresponds to the number of sites that get rescaled in every step of the renormalization scheme (where the total system has $\mathrm{base}^R$ sites). As an example, the case of base 4 will be shown.

With base 4 (considering a system with $4^R$ grid points), we make use of the following 13 shifted equations
\begin{subequations}
    \begin{align} 
    &t' G_{i-8}+t G_{i-7} + \omega  G_{i-6}+ t G_{i-5} +t' G_{i-4}  = \delta_{i-6} + a  (\delta_{i-7} + \delta_{i-5})  \\
     &\qquad \qquad \qquad \qquad \qquad \qquad ....... \nonumber \\
     &t' G_{i+4}+t G_{i+5} + \omega  G_{i+6}+ t G_{i+7} +t' G_{i+8}  = \delta_{i+6} + a  (\delta_{i+5} + \delta_{i+7})  ,
\end{align}
\end{subequations}
where the coupling parameter $a$ was introduced for the same reason as in the quantics base. Now, we eliminate $G_{i \pm 7}, G_{i \pm 6}, G_{i \pm 5}, G_{i \pm 3}, G_{i \pm 2}, G_{i \pm 1}$ reducing the system to only one equation with $G_{i \pm 8}, G_{i \pm 4}$ terms. Completing the first renormalization step, we keep only the modulo 4 terms (this means also $\delta_{i \pm 4}$ and corresponds to renormalizing 4 sites) and we arrive at
\begin{align}
    &t'^{(1)} G_{4(i-2)} + t^{(1)} G_{4(i-1)} + \omega^{(1)} G_{4i} +t^{(1)} G_{4(i+1)}  + t'^{(1)} G_{4(i+2)} = \delta_i + a^{(1)} (\delta_{i-1} + \delta_{i+1}),
\end{align}
with the four renormalized couplings $t'^{(1)}, t^{(1)}, \omega^{(1)}, a^{(1)}$. This is consistent with Eq.~\eqref{eq:nnn-general-GF} and, thus, the renormalization scheme can be continued, where in the $r$-th renormalization step we arrive at 
\begin{align}
    &t'^{(r)} G_{4^r(i-2)} + t^{(r)} G_{4^r(i-1)} + \omega^{(r)} G_{4^ri} +t^{(r)} G_{4^r(i+1)}  + t'^{(r)} G_{4^r(i+2)}  = \delta_i + a^{(r)} (\delta_{i-1} + \delta_{i+1}),
\end{align}
with 
    \begin{subequations}
        \begin{align}
            t'^{(r+1)} &= -\frac{\left(t'^{(r)}\right)^4}{-2 t'^{(r)} \left(t^{(r)}\right)^2 + 2 \left[ \left(t'^{(r)}\right)^2 + \left(t^{(r)}\right)^2 \right] \omega^{(r)} - \left(\omega^{(r)}\right)^3 + 2 a^{(r)} t^{(r)} \left[ 2 \left(t'^{(r)}\right)^2 - \left(t^{(r)}\right)^2 - 2 t'^{(r)} \omega^{(r)} + (\omega^{(r)})^2 \right]},\\
            t^{(r+1)} &= -\frac{4 \left(t'^{(r)}\right)^4 - \left(t^{(r)}\right)^4 + 4 t'^{(r)} \left(t^{(r)}\right)^2 \omega^{(r)} - 2 \left(t'^{(r)}\right)^2 \left[ 2 \left(t^{(r)}\right)^2 + \left(\omega^{(r)}\right)^2 \right]}{-2 t'^{(r)} \left(t^{(r)}\right)^2 + 2 \left[ \left(t'^{(r)}\right)^2 + \left(t^{(r)}\right)^2 \right] \omega^{(r)} - \left(\omega^{(r)}\right)^3 + 2 a^{(r)} t^{(r)} \left[ 2 \left(t'^{(r)}\right)^2 - \left(t^{(r)}\right)^2 - 2 t'^{(r)} \omega^{(r)} + (\omega^{(r)})^2 \right]}, \\ 
            \omega^{(r+1)} &= -\frac{2 \left[ 3 \left(t'^{(r)}\right)^4 - 4 \left(t'^{(r)}\right)^2 \left(t^{(r)}\right)^2 + \left(t^{(r)}\right)^4 \right] + 8 t'^{(r)} \left(t^{(r)}\right)^2 \omega^{(r)} - 4 \left[ \left(t'^{(r)}\right)^2 + \left(t^{(r)}\right)^2\right] \left(\omega^{(r)}\right)^2 + \left(\omega^{(r)}\right)^4}{-2 t'^{(r)} \left(t^{(r)}\right)^2 + 2 \left[ \left(t'^{(r)}\right)^2 + \left(t^{(r)}\right)^2 \right] \omega^{(r)} - \left(\omega^{(r)}\right)^3 + 2 a^{(r)} t^{(r)} \left[ 2 \left(t'^{(r)}\right)^2 - \left(t^{(r)}\right)^2 - 2 t'^{(r)} \omega^{(r)} + (\omega^{(r)})^2 \right]}, \\
            a^{(r+1)} &= \frac{a^{(r)} t^{(r)} \left[ \left(t^{(r)}\right)^2 + 2 t'^{(r)} \left(t'^{(r)} - \omega^{(r)}\right)\right] + t'^{(r)} \left[ -\left(t^{(r)}\right)^2 + t'^{(r)} \omega^{(r)} \right]}{-2 t'^{(r)} \left(t^{(r)}\right)^2 + 2 \left[ \left(t'^{(r)}\right)^2 + \left(t^{(r)}\right)^2 \right] \omega^{(r)} - \left(\omega^{(r)}\right)^3 + 2 a^{(r)} t^{(r)} \left[ 2 \left(t'^{(r)}\right)^2 - \left(t^{(r)}\right)^2 - 2 t'^{(r)} \omega^{(r)} + (\omega^{(r)})^2 \right]}, 
        \end{align} 
        \label{eq:nnn-renormalization-base-4-couplings}
    \end{subequations}
    defined recursively. The system can be renormalized until $r=R-1$ is reached, where from setting $i=0,1,2,3$ we obtain the following set of equations 
    \begin{subequations}
        \begin{align}
        1 &= 2 t'^{(R-1)} G_{2 \cdot 4^{(R-1)}} +  t^{(R-1)} G_{3 \cdot 4^{(R-1)}} + \omega^{(R-1)} G_{0} + t^{(R-1)} G_{1 \cdot 4^{(R-1)}}  \\
        a^{(R-1)} &= 2 t'^{(R-1)} G_{3 \cdot 4^{(R-1)}} + t^{(R-1)} G_{0} + \omega^{(R-1)} G_{1 \cdot 4^{(R-1)}} + t^{(R-1)} G_{2 \cdot 4^{(R-1)}}  \\
        0 &= 2 t'^{(R-1)} G_{0} + t^{(R-1)} G_{1 \cdot 4^{(R-1)}} + \omega^{(R-1)} G_{2 \cdot 4^{(R-1)}} + t^{(R-1)}  G_{3 \cdot 4^{(R-1)}}  \\
        a^{(R-1)} &= 2 t'^{(R-1)} G_{1 \cdot 4^{(R-1)}} + t^{(R-1)} G_{2 \cdot 4^{(R-1)}} + \omega^{(R-1)} G_{3 \cdot 4^{(R-1)}} + t^{(R-1)}  G_0
    \end{align}
    \end{subequations}
    where the second and fourth equation enforce the periodic boundary conditions ($G_{1\cdot 4^{(R-1)}}=G_{-1 \cdot 4^{(R-1)}}$). This set of four equations determining four unknowns ($G_{0},G_{1\cdot 4^{(R-1)}},G_{2\cdot 4^{(R-1)}},G_{3\cdot 4^{(R-1)}}$) can be solved leading to the results  
    \begin{subequations}
        \begin{align}
        G_0 &= -\frac{-4 a^{(R-1)} t'^{(R-1)} t^{(R-1)} + 2\left(t^{(R-1)}\right)^2 - 2 t'^{(R-1)} \omega^{(R-1)} + 2 a^{(R-1)} t^{(R-1)} \omega^{(R-1)} - \left(\omega^{(R-1)}\right)^2}{\left[ - 2 t'^{(R-1)} + \omega^{(R-1)} \right] \left[ 4 \left( t'^{(R-1)}\right)^2 - 4 \left(t^{(R-1)}\right)^2 + 4t'^{(R-1)} \omega^{(R-1)} + \left(\omega^{(R-1)}\right)^2 \right]}, \label{eq:nnn-R-1-solution-1} \\
        G_{1\cdot 4^{(R-1)}} &= -\frac{-2 a^{(R-1)} t'^{(R-1)} + t^{(R-1)} - a^{(R-1)} \omega^{(R-1)}}{\left[ 4 \left( t'^{(R-1)}\right)^2 - 4 \left(t^{(R-1)}\right)^2 + 4t'^{(R-1)} \omega^{(R-1)} + \left(\omega^{(R-1)}\right)^2 \right]}, \\
        G_{2\cdot 4^{(R-1)}} &= -\frac{2 \left[ 2 \left(t'^{(R-1)}\right)^2 - 2 a^{(R-1)} t^{(R-1)} t'^{(R-1)} - \left(t^{(R-1)}\right)^2 + t'^{(R-1)} \omega^{(R-1)} + a^{(R-1)} t^{(R-1)} \omega^{(R-1)}\right]}{\left[ - 2 t'^{(R-1)} + \omega^{(R-1)} \right] \left[ 4 \left( t'^{(R-1)}\right)^2 - 4 \left(t^{(R-1)}\right)^2 + 4t'^{(R-1)} \omega^{(R-1)} + \left(\omega^{(R-1)}\right)^2 \right]}, \\
        G_{3\cdot 4^{(R-1)}} &= G_{1\cdot 4^{(R-1)}}. \label{eq:nnn-R-1-solution-4}
    \end{align}
    \end{subequations}
We can then continue in the same way as in the quantics base. Let us note that a vanishing of one of the denominators in Eqs.~\eqref{eq:nnn-renormalization-base-4-couplings} corresponds to the matrix $(\omega \delta_{ij} + t_{ij})$ becoming singular. 

\section{\label{app:n-general-renormalization}\texorpdfstring{$n$}{n}-th-nearest-neighbor renormalization procedure}
\subsection{\label{app:n-general-additional-couplings}Origin of coupling parameters \texorpdfstring{$a_1,...,a_{n-1}$}{a1 ... an} in consistent renormalization}
Inserting $n$-th-nearest-neighbor hopping ($t_{ij} := t_1 (\delta_{j,i-1}+\delta_{j,i+1}) + ... + t_n (\delta_{j,i-n}+\delta_{j,i+n})$) into Eq.~\eqref{eq:general-GF} leads to 
\begin{align} \label{eq:n-general-general-GF-naive}
     &t_n G_{i-n}+...+ t_2 G_{i-2} + t_1 G_{i-1} + \omega G_{i}+ t_1 G_{i+1}  +t_2 G_{i+2} + ...+ t_n G_{i+n} = \delta_i.
\end{align}
For renormalization procedure with cyclic reduction, we, first, aim at eliminating all Green's functions of the nearest-neighbors ($G_{i \pm 2m +1}, m \in Z$) and then derive a renormalized equation consistent with Eq.~\eqref{eq:n-general-general-GF-naive}. Hence, we shift the equation n times up and n times down and arrive at the $2n+1$ equations
\begin{subequations}
    \begin{align} 
     &t_n G_{i-2n}+t_{n-1} G_{i-2n+1}+...+ t_2 G_{i-n-2} + t_1 G_{i-n-1} + \omega G_{i-n}+ t_1 G_{i-n+1} +t_2 G_{i-n+2} + ...+t_{n-1} G_{i-1} + t_n G_{i}  &=& \delta_{i-n}  \\
    &t_n G_{i-2n+1}+t_{n-1} G_{i-2n+2}+...+ t_2 G_{i-n-1} + t_1 G_{i-n} + \omega G_{i-n+1}+ t_1 G_{i-n+2} +t_2 G_{i-n+3} + ...+t_{n-1} G_{i} + t_n G_{i+1} & =& \delta_{i-n+1} , \\
    &\qquad   \qquad  \qquad ...... \nonumber \\
    &t_n G_{i}+t_{n-1} G_{i+1}+...+ t_2 G_{i+n-2} + t_1 G_{i+n-1} + \omega G_{i+n}+ t_1 G_{i+n+1} +t_2 G_{i+n+2} + ...+t_{n-1} G_{i+2n-1} + t_n G_{i+2n} & =& \delta_{i+n} ,
    \end{align}
\end{subequations}
\label{eq:n-general-shifted-equations-naive}
of which we want to eliminate the $2n$ Green's functions $G_{i \pm 2m +1}, m \in Z$. It can be immediately seen that this elimination process with subsequent cyclic reduction will result in an equation not consistent with Eq.~\eqref{eq:n-general-general-GF-naive} due to the terms $\delta_{i \pm 2}, \delta_{i \pm 4}, ..., \delta_{i \pm n}$ for even $n$ and $\delta_{i \pm 2}, \delta_{i \pm 4}, ..., \delta_{i \pm n \mp 1}$ for uneven $n$. With this in mind, our first approach is to modify Eq.~\eqref{eq:n-general-general-GF-naive} by introducing $\lfloor \frac{n}{2} \rfloor$ additional parameters $a_i=0$ to
\begin{align} \label{eq:n-general-general-GF-naive-2}
     &t_n G_{i-n}+...+ t_2 G_{i-2} + t_1 G_{i-1} + \omega G_{i}+ t_1 G_{i+1}  +t_2 G_{i+2} + ...+ t_n G_{i+n}  = \delta_i + a_1 (\delta_{i-1} + \delta_{i+1})  + ... +  a_{\lfloor \frac{n}{2} \rfloor} (\delta_{i-\lfloor \frac{n}{2} \rfloor} + \delta_{i+\lfloor \frac{n}{2} \rfloor}),
\end{align}
where $\lfloor \frac{n}{2} \rfloor =  \frac{n}{2}$ for even $n$ and $\lfloor \frac{n}{2} \rfloor =  \frac{n}{2} -\frac{1}{2}$ for odd $n$. Repeating the shifts of Eqs.~\eqref{eq:n-general-shifted-equations-naive} the following modified equations are derived
\begin{subequations}
    \begin{align} 
     &t_n G_{i-2n}+t_{n-1} G_{i-2n+1}+...+ t_2 G_{i-n-2} + t_1 G_{i-n-1} + \omega G_{i-n}+ t_1 G_{i-n+1} +t_2 G_{i-n+2} + ...+t_{n-1} G_{i-1} + t_n G_{i} \nonumber \\
    &\quad = \delta_{i-n} + a_1 (\delta_{i-n-1} + \delta_{i-n+1}) + ... + a_{\lfloor \frac{n}{2}} \rfloor (\delta_{i-n-\lfloor \frac{n}{2} \rfloor} + \delta_{i-n+\lfloor \frac{n}{2} \rfloor}) , \\
    &t_n G_{i-2n+1}+t_{n-1} G_{i-2n+2}+...+ t_2 G_{i-n-1} + t_1 G_{i-n} + \omega G_{i-n+1}+ t_1 G_{i-n+2} +t_2 G_{i-n+3} + ...+t_{n-1} G_{i} + t_n G_{i+1} \nonumber \\
    &\quad = \delta_{i-n+1} + a_1 (\delta_{i-n} + \delta_{i-n+2}) + ... +  a_{\lfloor \frac{n}{2} \rfloor} (\delta_{i-n +1-\lfloor \frac{n}{2} \rfloor} + \delta_{i-n + 1+\lfloor \frac{n}{2} \rfloor}), \\
    &\qquad \qquad ...... \nonumber \\
    &t_n G_{i}+t_{n-1} G_{i+1}+...+ t_2 G_{i+n-2} + t_1 G_{i+n-1} + \omega G_{i+n}+ t_1 G_{i+n+1} +t_2 G_{i+n+2} + ...+t_{n-1} G_{i+2n-1} + t_n G_{i+2n} \nonumber \\
    &\quad = \delta_{i+n} + a_1 (\delta_{i+n-1} + \delta_{i+n+1}) + ... +  a_{\lfloor \frac{n}{2} \rfloor} (\delta_{i+n-\lfloor \frac{n}{2} \rfloor} + \delta_{i+n+\lfloor \frac{n}{2} \rfloor}).
\end{align}
\end{subequations}
Keeping only the $G_{i \pm 2m } \ ( m \in \mathbb{Z})$ Green's functions and eliminating the others, will lead to an equation containing $\delta_{i\pm n \pm \lfloor \frac{n}{2} \rfloor}, \delta_{i \pm n \pm \lfloor \frac{n}{2} \rfloor \mp 1} ,...$ terms of which the even contributions are kept in the cyclic reduction step. Hence, the resulting renormalized equation will not be consistent with Eq.~\eqref{eq:n-general-general-GF-naive-2} due to additional Kronecker delta terms such as $\delta_{i\pm \lfloor \frac{1}{2}(n \pm \lfloor \frac{n}{2} \rfloor) \rfloor}$. Therefore, the approach needs to be adjusted. In order to arrive at a fully consistent renormalization procedure, Eq.~\eqref{eq:n-general-general-GF-naive} needs to be modified by introducing $n-1$ additional coupling parameters $a_1,...,a_{n-1}=0$, which leads to Eq.~\eqref{eq:n-general-general-GF}
\begin{align}
     &t_n G_{i-n}+...+ t_2 G_{i-2} + t_1 G_{i-1} + \omega G_{i}+ t_1 G_{i+1}  +t_2 G_{i+2} + ...+ t_n G_{i+n} \nonumber \\
    &\quad = \delta_i + a_1 (\delta_{i-1} + \delta_{i+1})  + ... +  a_{n-1} (\delta_{i-n+1} + \delta_{i+n-1}).
\end{align}
Proceeding as before by shifting the equation $2n$ times, the following set of equations can be obtained:
\begin{subequations}
    \begin{align} 
     &t_n G_{i-2n}+t_{n-1} G_{i-2n+1}+...+ t_2 G_{i-n-2} + t_1 G_{i-n-1} + \omega G_{i-n}+ t_1 G_{i-n+1} +t_2 G_{i-n+2} + ...+t_{n-1} G_{i-1} + t_n G_{i} \nonumber \\
    &\quad = \delta_{i-n} + a_1 (\delta_{i-n-1} + \delta_{i-n+1}) + ... +  a_{n-1} (\delta_{i-2n+1} + \delta_{i-1}), \label{eq:nth-cyclic-reduction} \\
    &t_n G_{i-2n+1}+t_{n-1} G_{i-2n+2}+...+ t_2 G_{i-n-1} + t_1 G_{i-n} + \omega G_{i-n+1}+ t_1 G_{i-n+2} +t_2 G_{i-n+3} + ...+t_{n-1} G_{i} + t_n G_{i+1} \nonumber \\
    &\quad = \delta_{i-n+1} + a_1 (\delta_{i-n} + \delta_{i-n+2}) + ... +  a_{n-1} (\delta_{i-2n+2} + \delta_{i}), \\
    &\qquad \qquad ...... \nonumber \\
    &t_n G_{i}+t_{n-1} G_{i+1}+...+ t_2 G_{i+n-2} + t_1 G_{i+n-1} + \omega G_{i+n}+ t_1 G_{i+n+1} +t_2 G_{i+n+2} + ...+t_{n-1} G_{i+2n-1} + t_n G_{i+2n} \nonumber \\
    &\quad = \delta_{i+n} + a_1 (\delta_{i+n-1} + \delta_{i+n+1}) + ... +  a_{n-1} (\delta_{i+2n-1} + \delta_{i+1}).
\end{align}
\end{subequations}
Now, it can be seen that the furthest shifted Kronecker delta contributions are $a_{n-1} \delta_{i \pm 2n \mp 1}$. Eliminating the $G_{i \pm 2m +1} \ ( m \in \mathbb{Z})$ and then keeping only the even contributions in the cyclic reduction step leads to furthest shifted Kronecker delta contributions of $\delta_{2i \pm 2n \mp 2} = \delta_{2(i \pm n \mp 1)} = \delta_{i \pm n \mp 1}$ which are exactly the Kronecker deltas of the $a_{n-1}$ term in Eq.~\eqref{eq:n-general-general-GF}. This makes the renormalization step consistent and this leads to the following renormalized equation
\begin{align}
    &t_n^{(1)} G_{2(i-n)}+...+ t_2^{(1)} G_{2(i-2)} + t_1^{(1)} G_{2(i-1)} + \omega G_{2i}  + t_1^{(1)} G_{2(i+1)} +t_2^{(1)} G_{2(i+2)} + ...+ t_n^{(1)} G_{2(i+n)} \nonumber \\
    &\quad = \delta_i + a_1^{(1)} (\delta_{i-1} + \delta_{i+1})  + ... +  a_{n-1}^{(1)} (\delta_{i-n+1} + \delta_{i+n-1}),
\end{align}
containing $n-1$ additional rescaled couplings $a^{(1)}_i$. We have now derived a consistent renormalization scheme and $r$-th renormalization step, we get
\begin{align}
    &t_n^{(r)} G_{2^r(i-n)}+...+ t_2^{(r)} G_{2^r(i-2)} + t_1^{(r)} G_{2^r(i-1)} + \omega G_{(2n)^r i} + t_1^{(r)} G_{2^r(i+1)}  +t_2^{(r)} G_{2^r(i+2)} + ...+ t_n^{(r)} G_{2^r(i+n)} \nonumber \\
    &\quad = \delta_i + a_1^{(r)} (\delta_{i-1} + \delta_{i+1})  + ... +  a_{n-1}^{(r)} (\delta_{i-n+1} + \delta_{i+n-1}),
\end{align}
with $2n$ renormalized couplings ($\omega^{(r)},t_1^{(r)},...,t_n^{(r)}, a_1^{(r)}, ..., a_{n-1}^{(r)}$) for coarse-graining two sites into one in every step of the renormalization procedure. We want to note that as for the next-nearest-neighbor case the choice of a quantics basis is not unique and also a different base (with $\mathrm{base}^R$ lattice sites) can be chosen without affecting the number of renormalized couplings.

\subsection{Solutions}
In this section, we briefly want to show that although the complexity of the problem increases for the general $n$-th-nearest-neighbor hopping case, the renormalization procedure can simplify the analytical calculation of the Green's functions significantly. For example, continuing the coarse-graining until only two sites are remaining ($r=R-1$), the following two equations ($i=0,1$) are derived 
\begin{subequations}
    \begin{align}
    &(\omega^{(R-1)}+2t_2^{(R-1)}+2t_4^{(R-1)}+...+2t_{n-2}^{(R-1)}) G_0  + (2t_1^{(R-1)}+2t_3^{(R-1)}+...+2t_{n-1}^{(R-1)}) G_{2^{(R-1)}} \nonumber \\
    & = 1 + 2 a_2^{(R-1)} + 2 a_4^{(R-1)}+...+ 2 a_{n-2}^{(R-1)},  \\
    &(2t_1^{(R-1)}+2t_3^{(R-1)}+...+2t_{n-1}^{(R-1)}) G_0  + (\omega^{(R-1)}+2t_2^{(R-1)}+2t_4^{(R-1)}+...+2t_{n-2}^{(R-1)}) G_{2^{(R-1)}} \nonumber \\
    & = 2 a_1^{(R-1)}+2 a_3^{(R-1)}+...+ 2 a_{n-1}^{(R-1)},
    \end{align}
    \label{eq:n-general-renormalization-equation-R-1}
\end{subequations}
assuming $n$ is even otherwise the $n-1$ and $n-2$ subscripts in the two equations are exchanged. These equations can be easily solved leading to 
\begin{subequations}
    \begin{align}
    G_0 &= -\frac{\tilde a_2^{(R-1)} \tilde \omega^{(R-1)}-\tilde a_1^{(R-1)} \tilde t^{(R-1)}}{(\tilde t^{(R-1)})^2-(\tilde \omega^{(R-1)})^2}, \\
    G_{2^{(R-1)}} &= -\frac{\tilde a_1^{(R-1)} \tilde \omega^{(R-1)}-\tilde a_2^{(R-1)} \tilde t^{(R-1)}}{(\tilde t^{(R-1)})^2-(\tilde \omega^{(R-1)})^2}, 
\end{align}
\end{subequations}
where we introduced the following abbreviations
\begin{subequations}
    \begin{align}
    \tilde \omega^{(R-1)} &:= \omega^{(R-1)}+2t_2^{(R-1)}+2t_4^{(R-1)}+...+2t_{n-2}^{(R-1)}, \\
    \tilde t^{(R-1)} &:=2t_1^{(R-1)}+2t_3^{(R-1)}+...+2t_{n-1}^{(R-1)}, \\
    \tilde a_1^{(R-1)} &:= 2 a_1^{(R-1)}+2 a_3^{(R-1)}+...+ 2 a_{n-1}^{(R-1)}, \\
    \tilde a_2^{(R-1)} &:=1 + 2 a_2^{(R-1)} + 2 a_4^{(R-1)}+...+ 2 a_{n-2}^{(R-1)}.
\end{align}
\end{subequations}
Therefore, we can, for example, already obtain the local Green's function ($G_0$) just from identifying the recursive relations of the renormalized couplings and solving the two Eqs.~\eqref{eq:n-general-renormalization-equation-R-1}. Moreover, it is possible to determine all Green's functions by fine-graining the system step by step, while at the same time reducing the complexity of the problem, due to redundancies in the solvable equations caused by the insertion of the coarse-grained solutions. More importantly, the complexity of the problem is reduced because of the special structure of the Green's functions identified by the renormalization scheme, leading to the possibility to classify different Green's functions in groups of structurally equal ones, of which all can be derived easily if the most coarse-grained one is known.

\section{Nearest-neighbor QTT and renormalization}
\subsection{\label{app:nn-explicit-formula-M}Proof of Eq.~(\ref{eq:nn-qtt-mpo-M-explicit-formula})}
Here, we will prove that the circulant matrix $\mathbf{M}$
\begin{align}
        \mathbf{M} =  \omega \mathbf{\mathbb{I}}^{(R)} + t (\mathbf{P}_1^{(R)} + \mathbf{P}_{2^{R}-1}^{(R)}) = 
        \left(\begin{matrix}
        \omega & t& 0 & ... & 0 & 0 & t \\
        t & \omega& t & ... & 0 & 0 & 0 \\
        0 & t& \omega & ... & 0 & 0 & 0 \\
        ...& ...& ...& ...& ...& ...& ...\\
        0 & 0& 0 & ... & t & \omega & t \\
        t & 0& 0 & ... & 0 & t & \omega 
    \end{matrix} \right),
\end{align}
can be exactly decomposed into a QTT with bond dimension 3 in the following way (Eq.~\eqref{eq:nn-qtt-mpo-M-explicit-formula})
\begin{align}
    \mathbf{M}^{(R)} = \left[ \begin{matrix}
        I & K & K
    \end{matrix} \right] \lrtimes
    \left[ \begin{matrix}
        I & K & K \\
          & J &   \\
          &   & J'
    \end{matrix}\right]^{\lrtimes (R-2)} \lrtimes 
    \left[ \begin{matrix}
        \omega I + 2t H \\
        t J \\
        t J'
    \end{matrix}\right]. 
\end{align}

\textit{Proof.} We make use of the recursive block structure
\begin{align}
    M^{(l)} &= \left( \begin{matrix}
        M^{(l-1)}-tJ^{\otimes (l-1)}-tJ'^{\otimes (l-1)} & tJ^{\otimes (l-1)}+tJ'^{\otimes (l-1)} \\
        tJ^{\otimes (l-1)}+tJ'^{\otimes (l-1)} & M^{(l-1)}-tJ^{\otimes (l-1)}-tJ'^{\otimes (l-1)}
    \end{matrix}\right) \nonumber \\
    &= I \otimes M^{(l-1)} - t I \otimes J^{\otimes (l-1)} - t I \otimes J'^{\otimes (l-1)} + t J \otimes J'^{\otimes (l-1)}  + t J' \otimes J^{\otimes (l-1)} + t J \otimes J^{\otimes (l-1)} + t J' \otimes J'^{\otimes (l-1)} \nonumber \\
    &= I \otimes M^{(l-1)} + t K \otimes J^{\otimes (l-1)} + t K \otimes J'^{\otimes (l-1)},
\end{align}
which holds for $2 \leq l \leq R$. Hence, we have
\begin{align}
    \mathbf{M}^{(l)} = \left[ \begin{matrix}
        I & K & K
    \end{matrix} \right] \lrtimes
    \left[ \begin{matrix}
        \mathbf{M}^{(l-1)} \\
        t J^{\otimes (l-1)} \\
        t J'^{\otimes (l-1)}
    \end{matrix}\right], 
\end{align}
which we can expand to
\begin{align}
    \left[ \begin{matrix}
        \mathbf{M}^{(l)} \\
        t J^{\otimes (l)} \\
        t J'^{\otimes (l)}
    \end{matrix} \right] 
    = \left[ \begin{matrix}
        I & K & K \\
          & J &   \\
          &   & J'
    \end{matrix} \right] \lrtimes
    \left[ \begin{matrix}
        \mathbf{M}^{(l-1)} \\
        t J^{\otimes (l-1)} \\
        t J'^{\otimes (l-1)}
    \end{matrix}\right].
\end{align}
We can now recursively construct the QTT
\begin{align}
    \left[ \begin{matrix}
        \mathbf{M}^{(l)} \\
        t J^{\otimes (l)} \\
        t J'^{\otimes (l)}
    \end{matrix} \right] 
    & = \left[ \begin{matrix}
        I & K & K \\
          & J &   \\
          &   & J'
    \end{matrix} \right] \lrtimes
    \Bigg( \left[ \begin{matrix}
        I & K & K \\
          & J &   \\
          &   & J'
    \end{matrix} \right] \lrtimes
    \left[ \begin{matrix}
        \mathbf{M}^{(l-2)} \\
        t J^{\otimes (l-2)} \\
        t J'^{\otimes (l-2)}
    \end{matrix}\right] \Bigg) 
    = ... = \left[ \begin{matrix}
        I & K & K \\
          & J &   \\
          &   & J'
    \end{matrix} \right] \lrtimes
    \left[ \begin{matrix}
        I & K & K \\
          & J &   \\
          &   & J'
    \end{matrix} \right]^{\lrtimes (l-2)} \lrtimes
    \left[ \begin{matrix}
        \mathbf{M}^{(1)} \\
        t J \\
        t J'
    \end{matrix}\right] ,
\end{align}
where $\mathbf{M}^{(1)} = \omega I + 2t H = \left( \begin{matrix}
    \omega & 2 t \\
    2 t & \omega
\end{matrix}\right)$. By considering $l=R$ and selecting the first row in the first core, we conclude the proof for the formula in Eq.~\eqref{eq:nn-qtt-mpo-M-explicit-formula}. $\Box$

\subsection{\label{app:nn-renormalization-solutions}Solutions}
We will now show this explicitly. We go until $r=R-2$, where we get the QTT $\mathbf{G}_r=\mathcal{G}_1 \lrtimes \mathcal{\tilde G}_2$ from solving the equations
\begin{align}
    (\mathcal{A}_1 \underset{j_1}{\bullet} \mathcal{G}_1) \lrtimes (\mathcal{A}_3^{(R-2)} \underset{j_2}{\bullet} \mathcal{\tilde G}_2) = \Delta \lrtimes \Delta,
\end{align}
with 
\begin{align}
    \mathcal{G}_1 &= \left[\begin{matrix}
        G_{11}^{(1)} & G_{12}^{(1)} & ... & G_{1 D_1}^{(1)}
    \end{matrix} \right], \qquad 
    \mathcal{\tilde G}_2 = \left[\begin{matrix}
        \tilde G_{11}^{(2)} \\ \tilde G_{21}^{(2)} \\ ... \\ \tilde G_{D_1 1}^{(2)}
    \end{matrix} \right], \qquad
    G_{1 \alpha_1}^{(1)} &= \left(\begin{matrix}
        g_{0,1 \alpha_1}^{(1)} & g_{1,1 \alpha_1}^{(1)}
    \end{matrix} \right), \qquad 
    \tilde G_{\alpha_1 1}^{(2)} = \left(\begin{matrix}
        \tilde g_{0,\alpha_1 1}^{(2)} & \tilde g_{1,\alpha_1 1}^{(2)}
    \end{matrix} \right),
\end{align}
and the bond dimension $D_1$ of the first bond. This set of equations can be solved with $D_1=2$ leading to the result (one possibility, there are many solutions)
\begin{subequations}
    \begin{align}
    & g^{(1)}_{0,11} &=\frac{2 \left(t^{(R-2)}\right)^2 - \left(\omega^{(R-2)}\right)^2}{4 \tilde g^{(2)}_{0,11} \left(t^{(R-2)}\right)^2 \omega^{(R-2)}-\tilde g^{(2)}_{0,11} \left(\omega^{(R-2)}\right)^3}, \qquad 
    & g^{(1)}_{1,11} &=\frac{2 \left(t^{(R-2)}\right)^2}{-4 \tilde g^{(2)}_{0,11} \left(t^{(R-2)}\right)^2 \omega^{(R-2)}+\tilde g^{(2)}_{0,11} \left(\omega^{(R-2)}\right)^3}, \qquad 
    & g^{(1)}_{0,12} &= 0 \\
    & \tilde g^{(2)}_{1,11} &= \frac{\tilde g^{(2)}_{0,11} t^{(R-2)} \omega^{(R-2)}}{2 \left(t^{(R-2)}\right)^2 - \left(\omega^{(R-2)}\right)^2},\qquad 
    & \tilde g^{(2)}_{1,21} &= \frac{t^{(R-2)} }{g^{(1)}_{1,12} \left[ 2 \left(t^{(R-2)}\right)^2 - \left(\omega^{(R-2)}\right)^2 \right]},\qquad
    & \tilde g^{(2)}_{0,21} &= 0,
    \end{align}
\end{subequations}
with $g^{(1)}_{1,12}, \tilde g^{(2)}_{0,11}$ as not determined degrees of freedom(arbitrary except for 0). Following Eq.~\eqref{eq:nn-qtt-renormalization-r}, we arrive at 
\begin{align}
    &(\mathcal{A}_1 \underset{j_1}{\bullet} \mathcal{G}_1) \lrtimes \left(  (\mathcal{A}_2 \underset{j_2}{\bullet} \mathcal{G}_2) \lrtimes (\mathcal{A}_3^{(R-3)} \underset{j_{3}}{\bullet} \mathcal{\tilde G}_{3}) - (\mathcal{A}_3^{(R-2)} \underset{j_{2}}{\bullet} \mathcal{\tilde G}_{2}) \lrtimes \Delta \right) =0,  
\end{align}
with 
\begin{align}
    \mathcal{G}_2 &= \left[\begin{matrix}
        G_{11}^{(2)} & G_{12}^{(2)} & ... & G_{1 D_2}^{(2)} \\
        G_{21}^{(2)} & G_{22}^{(2)} & ... & G_{2 D_2}^{(2)}
    \end{matrix} \right], \qquad 
    \mathcal{\tilde G}_3 = \left[\begin{matrix}
        \tilde G_{11}^{(3)} \\ \tilde G_{21}^{(3)} \\ ... \\ \tilde G_{D_2 1}^{(3)}
    \end{matrix} \right], \qquad
    G_{\alpha_1 \alpha_2}^{(2)} &= \left(\begin{matrix}
        g_{0,\alpha_1 \alpha_2}^{(2)} & g_{1,\alpha_1 \alpha_2}^{(2)}
    \end{matrix} \right), \qquad 
    \tilde G_{\alpha_2 1}^{(3)} = \left(\begin{matrix}
        \tilde g_{0,\alpha_2 1}^{(3)} & \tilde g_{1,\alpha_2 1}^{(3)}
    \end{matrix} \right),
\end{align}
Inserting the previous results and solving these eight equations we find the following solution (again, one of many possible ones)
\begin{subequations}
    \begin{align}
    g^{(2)}_{0,11} &= \frac{\tilde g^{(2)}_{0,11}}{\tilde g^{(3)}_{0,11}} \\
    g^{(2)}_{1,11} &=\frac{-2 \tilde g^{(2)}_{0,11} \left(t^{(R-3)}\right)^4 + \tilde g^{(2)}_{0,11} \left(t^{(R-3)}\right)^2 \left(\omega^{(R-3)}\right)^2}{\tilde g^{(3)}_{0,11} \left[ 2 \left(t^{(R-3)}\right)^4 - 4 \left(t^{(R-3)}\right)^2 \left(\omega^{(R-3)}\right)^2 +\left(\omega^{(R-3)}\right)^4\right]} \\
    g^{(2)}_{0,12} &= \frac{\tilde g^{(2)}_{0,11}}{\tilde g^{(3)}_{1,21}} \left( -\frac{\tilde g^{(3)}_{1,11}}{\tilde g^{(3)}_{0,11}} + \frac{3 \left(t^{(R-3)}\right)^3 \omega^{(R-3)}- t^{(R-3)} \left(\omega^{(R-3)}\right)^3}{2 \left(t^{(R-3)}\right)^4 - 4 \left(t^{(R-3)}\right)^2 \left(\omega^{(R-3)}\right)^2 +\left(\omega^{(R-3)}\right)^4} \right) \\
    g^{(2)}_{1 ,12} &= -\frac{\tilde g^{(2)}_{0,11} \left(t^{(R-3)}\right)^2 \left( \tilde g^{(3)}_{0,11} t^{(R-3)} \omega^{(R-3)} + \tilde g^{(3)}_{1,11} \left[-2 \left(t^{(R-3)}\right)^2 + \left(\omega^{(R-3)}\right)^2 \right] \right)}{\tilde g^{(3)}_{0,11} \tilde g^{(3)}_{1,21}\left[ 2 \left(t^{(R-3)}\right)^4 - 4 \left(t^{(R-3)}\right)^2 \left(\omega^{(R-3)}\right)^2 +\left(\omega^{(R-3)}\right)^4\right]} \\
    g^{(2)}_{0,21} &= 0 \\
    g^{(2)}_{1,21} &= \frac{\left(t^{(R-3)}\right)^2 \omega^{(R-3)}}{\tilde g^{(3)}_{0,11}  g^{(1)}_{1,12}\left[ 2 \left(t^{(R-3)}\right)^4 - 4 \left(t^{(R-3)}\right)^2 \left(\omega^{(R-3)}\right)^2 +\left(\omega^{(R-3)}\right)^4\right]} \\
    g^{(2)}_{0,22} &= -\frac{\left(t^{(R-3)}\right)^3}{\tilde g^{(3)}_{1,12}  g^{(1)}_{1,12}\left[ 2 \left(t^{(R-3)}\right)^4 - 4 \left(t^{(R-3)}\right)^2 \left(\omega^{(R-3)}\right)^2 +\left(\omega^{(R-3)}\right)^4\right]} \\
    g^{(2)}_{1,22} &= -\frac{t^{(R-3)} \left[ -\tilde g^{(3)}_{1,11} t^{(R-3)} \omega^{(R-3)} + \tilde g^{(3)}_{0,11} \left(t^{(R-3)} - \omega^{(R-3)}\right) \left(t^{(R-3)} + \omega^{(R-3)}\right) \right]}{\tilde g^{(3)}_{0,11} \tilde g^{(3)}_{1,12}  g^{(1)}_{1,12}\left[ 2 \left(t^{(R-3)}\right)^4 - 4 \left(t^{(R-3)}\right)^2 \left(\omega^{(R-3)}\right)^2 +\left(\omega^{(R-3)}\right)^4\right]} \\
    \tilde g^{(3)}_{0,21} &= 0,
\end{align}
\end{subequations}
    
where $\tilde g^{(3)}_{1,11}, \tilde g^{(3)}_{0,11}, \tilde g^{(3)}_{1,12}$ can be chosen arbitrarily (except for 0). In the next step ($r=R-4$), we now structurally get the same equation as in Eq.~\eqref{eq:nn-qtt-renormalization-r}
\begin{align}
    &(\mathcal{A}_1 \underset{j_1}{\bullet} \mathcal{G}_1) \lrtimes (\mathcal{A}_2 \underset{j_2}{\bullet} \mathcal{G}_2) \lrtimes \left(  (\mathcal{A}_2 \underset{j_{3}}{\bullet} \mathcal{G}_{3}) \lrtimes (\mathcal{A}_3^{(R-4)} \underset{j_{4}}{\bullet} \mathcal{\tilde G}_{4}) - (\mathcal{A}_3^{(R-3)} \underset{j_{3}}{\bullet} \mathcal{\tilde G}_{3}) \lrtimes \Delta \right) =0. 
\end{align}
Inserting the previous results and the renormalized couplings into these 16 equations gives the following solution (again one of many possible ones)
\begin{subequations}
    \begin{align}
    g^{(3)}_{0,11} &= \frac{\tilde g^{(3)}_{0,11}}{\tilde g^{(4)}_{0,11}} \\
    g^{(3)}_{1,11} &= \frac{\tilde g^{(3)}_{1,11}}{\tilde g^{(4)}_{0,11}} \\
    g^{(3)}_{0,12} &= -\frac{\left(\tilde g^{(3)}_{0,11} + \tilde g^{(3)}_{1,11}\right) \tilde g^{(4)}_{0,11} t^{(R-4)} +\tilde g^{(3)}_{0,11} \tilde g^{(4)}_{1,11} \omega^{(R-4)}}{\tilde g^{(4)}_{0,11} \tilde g^{(4)}_{1,21} \omega^{(R-4)}} \\
    g^{(3)}_{1,12} &=  \frac{\left(\tilde g^{(3)}_{0,11} + \tilde g^{(3)}_{1,11}\right) \tilde g^{(4)}_{0,11} \left(t^{(R-4)}\right)^2 - \tilde g^{(3)}_{1,11} \tilde g^{(4)}_{1,11} \omega^{(R-4)} t^{(R-4)}-\tilde g^{(3)}_{1,11} \tilde g^{(4)}_{0,11} \left(\omega^{(R-4)}\right)^2}{\tilde g^{(4)}_{0,11} \tilde g^{(4)}_{1,21} \omega^{(R-4)} t^{(R-4)}} \\
    g^{(3)}_{0,21} &= 0
    \end{align}
    \begin{align}
    g^{(3)}_{1,21} &= \frac{\tilde g^{(3)}_{1,21}}{\tilde g^{(4)}_{0,11}} \\
    g^{(3)}_{0,22} &= -\frac{\tilde g^{(3)}_{1,21}t^{(R-4)}}{\tilde g^{(4)}_{1,21}\omega^{(R-4)}} \\
    g^{(3)}_{1,22} &=  \frac{\tilde g^{(3)}_{1,21} \left[ \tilde g^{(4)}_{0,11} \left(t^{(R-4)}\right)^2 - \tilde g^{(4)}_{1,11} t^{(R-4)} \omega^{(R-4)} - \tilde g^{(4)}_{0,11} \left(\omega^{(R-4)}\right)^2 \right] }{\tilde g^{(4)}_{0,11} \tilde g^{(4)}_{1,21} \omega^{(R-4)} t^{(R-4)}} \\
    \tilde g^{(4)}_{0,21} &= 0
\end{align}
\end{subequations}    
with three degrees of freedom ($\tilde g^{(4)}_{0,11}, \tilde g^{(4)}_{1,11}, \tilde g^{(4)}_{1,21}$) leading to the bond dimension $D_3 = 2$. Since the bond dimension saturates and the equations of the next step $r=R-5$ are now structurally the same 
\begin{align}
    &(\mathcal{A}_1 \underset{j_1}{\bullet} \mathcal{G}_1) \lrtimes (\mathcal{A}_2 \underset{j_2}{\bullet} \mathcal{G}_2) \lrtimes (\mathcal{A}_2 \underset{j_3}{\bullet} \mathcal{G}_3) \lrtimes  \left(  (\mathcal{A}_2 \underset{j_{4}}{\bullet} \mathcal{G}_{4}) \lrtimes (\mathcal{A}_3^{(R-5)} \underset{j_{5}}{\bullet} \mathcal{\tilde G}_{5}) - (\mathcal{A}_3^{(R-4)} \underset{j_{4}}{\bullet} \mathcal{\tilde G}_{4}) \lrtimes \Delta \right) =0, 
\end{align}
we get the same results as in the previous step, just with every superscript ${(4)}$ exchanged with ${(5)}$ and ${(3)}$ substituted by ${(4)}$ and ${(R-4)}$ changed to ${(R-5)}$. This now holds for every arbitrary step, giving the solution of the $r=R-s$-th renormalization step ($s \geq 4$) as
\begin{subequations}
    \begin{align}
    g^{(s-1)}_{0,11} &= \frac{\tilde g^{(s-1)}_{0,11}}{\tilde g^{(s)}_{0,11}} \\
    g^{(s-1)}_{1,11} &= \frac{\tilde g^{(s-1)}_{1,11}}{\tilde g^{(s)}_{0,11}} \\
    g^{(s-1)}_{0,12} &= -\frac{\left(\tilde g^{(s-1)}_{0,11} + \tilde g^{(s-1)}_{1,11}\right) \tilde g^{(s)}_{0,11} t^{(R-s)} +\tilde g^{(s-1)}_{0,11} \tilde g^{(s)}_{1,11} \omega^{(R-s)}}{\tilde g^{(s)}_{0,11} \tilde g^{(s)}_{1,21} \omega^{(R-s)}} \\
    g^{(s-1)}_{1,12} &=  \frac{\left(\tilde g^{(s-1)}_{0,11} + \tilde g^{(s-1)}_{1,11}\right) \tilde g^{(s)}_{0,11} \left(t^{(R-s)}\right)^2 - \tilde g^{(s-1)}_{1,11} \tilde g^{(s)}_{1,11} \omega^{(R-s)} t^{(R-s)}-\tilde g^{(s-1)}_{1,11} \tilde g^{(s)}_{0,11} \left(\omega^{(R-s)}\right)^2}{\tilde g^{(s)}_{0,11} \tilde g^{(s)}_{1,21} \omega^{(R-s)} t^{(R-s)}} \\
    g^{(s-1)}_{0,21} &= 0 \\
    g^{(s-1)}_{1,21} &= \frac{\tilde g^{(s-1)}_{1,21}}{\tilde g^{(s)}_{0,11}} \\
    g^{(s-1)}_{0,22} &= -\frac{\tilde g^{(s-1)}_{1,21}t^{(R-s)}}{\tilde g^{(s)}_{1,21}\omega^{(R-s)}} \\
    g^{(s-1)}_{1,22} &=  \frac{\tilde g^{(s-1)}_{1,21} \left[ \tilde g^{(s)}_{0,11} \left(t^{(R-s)}\right)^2 - \tilde g^{(s)}_{1,11} t^{(R-s)} \omega^{(R-s)} - \tilde g^{(s)}_{0,11} \left(\omega^{(R-s)}\right)^2 \right] }{\tilde g^{(s)}_{0,11} \tilde g^{(s)}_{1,21} \omega^{(R-s)} t^{(R-s)}} \\
    \tilde g^{(s)}_{0,21} &= 0,
\end{align}
\end{subequations}    
with three degrees of freedom ($\tilde g^{(s)}_{0,11}, \tilde g^{(s)}_{1,11}, \tilde g^{(s)}_{1,21}$) and bond dimension 2 along all bonds, for $4 \leq s \leq R$. This means that we can now determine the Green's function for arbitrarily large grids, because we now have the general solution for every tensor in our QTT and we see that the bond dimension saturates at exactly 2 equal to the number of couplings.

\section{Next-nearest-neighbor QTT and renormalization}
\subsection{\label{app:nnn-proof-M}Proof of Eqs.~(\ref{eq:nnn-qtt-mpo-M-explicit-formula-1}, \ref{eq:nnn-qtt-mpo-M-explicit-formula-2})}
Here, we will prove that the circulant matrix $\mathbf{M}$
\begin{align}
        \mathbf{M} =  \omega \mathbf{\mathbb{I}}^{(R)} + t (\mathbf{P}_1^{(R)} + \mathbf{P}_{2^{R-1}}^{(R)}) + t' (\mathbf{P}_2^{(R)} + \mathbf{P}_{2^{R-2}}^{(R)}) = 
        \left(\begin{matrix}
        \omega & t& t' & 0 &  ... &0 & 0 & t' & t \\
        t & \omega& t & t'&... &0 & 0 & 0 & t' \\
        t' & t& \omega & t&... &0& 0 & 0 & 0 \\
        ...& ...& ...& ...& ...& ...& ...\\
        t' & 0& 0 & 0&... &t'& t & \omega & t \\
        t & t'& 0 &0& ... &0& t' & t & \omega 
    \end{matrix} \right),
\end{align}
can be exactly decomposed into a QTT with bond dimension 5 in the following way (Eqs. (\ref{eq:nnn-qtt-mpo-M-explicit-formula-1}, \ref{eq:nnn-qtt-mpo-M-explicit-formula-2}))
\begin{align}
    \mathbf{M}^{(R)} &= \left[ \begin{matrix}
        I & K & K & K & K 
    \end{matrix} \right] \lrtimes
    \left[ \begin{matrix}
        I & K & K & K & K \\
          & J &   &   &   \\
          &   & J'&   &   \\
          &   &   & J &   \\
          &   &   &   & J'
    \end{matrix}\right]^{\lrtimes (R-2)}  \lrtimes 
    \left[ \begin{matrix}
        (\omega + 2t') I + 2t H \\
        t J \\
        t J' \\
        t' I \\
        t' I
    \end{matrix}\right]. 
\end{align}

\textit{Proof.} We make use of the recursive block structure
\begin{align}
    M^{(l)} &= \left( \begin{matrix}
        M^{(l-1)}-tJ^{\otimes (l-1)}-tJ'^{\otimes (l-1)} -t'J^{\otimes (l-2)}\otimes I-t'J'^{\otimes (l-2)}\otimes I & tJ^{\otimes (l-1)}+tJ'^{\otimes (l-1)} +t'J^{\otimes (l-2)}\otimes I+t'J'^{\otimes (l-2)}\otimes I\\
        tJ^{\otimes (l-1)}+tJ'^{\otimes (l-1)} +t'J^{\otimes (l-2)}\otimes I+t'J'^{\otimes (l-2)}\otimes I & M^{(l-1)}-tJ^{\otimes (l-1)}-tJ'^{\otimes (l-1)} -t'J^{\otimes (l-2)}\otimes I-t'J'^{\otimes (l-2)}\otimes I 
    \end{matrix}\right) \nonumber \\
    &= I \otimes M^{(l-1)} + t K \otimes J^{\otimes (l-1)} + t K \otimes J'^{\otimes (l-1)} + t' K \otimes J^{\otimes (l-2)}\otimes I + t' K \otimes J'^{\otimes (l-2)} \otimes I,
\end{align}
which holds for $2 \leq l \leq R$. Hence, we have
\begin{align}
    \mathbf{M}^{(l)} = \left[ \begin{matrix}
        I & K & K & K & K
    \end{matrix} \right] \lrtimes
    \left[ \begin{matrix}
        \mathbf{M}^{(l-1)} \\
        t J^{\otimes (l-1)} \\
        t J'^{\otimes (l-1)} \\
        t'  J^{\otimes (l-2)}\otimes I\\
        t'  J'^{\otimes (l-2)}\otimes I
    \end{matrix}\right], 
\end{align}
which we can expand to
\begin{align}
    \left[ \begin{matrix}
        \mathbf{M}^{(l)} \\
        t J^{\otimes (l)} \\
        t J'^{\otimes (l)} \\
        t'  J^{\otimes (l-1)}\otimes I\\
        t'  J'^{\otimes (l-1)}\otimes I
    \end{matrix} \right] 
    = \left[ \begin{matrix}
        I & K & K & K & K \\
          & J &   &   &   \\
          &   & J'&   &   \\
          &   &   & J &   \\
          &   &   &   & J'
    \end{matrix} \right] \lrtimes
    \left[ \begin{matrix}
        \mathbf{M}^{(l-1)} \\
        t J^{\otimes (l-1)} \\
        t J'^{\otimes (l-1)} \\
        t'  J^{\otimes (l-2)}\otimes I\\
        t'  J'^{\otimes (l-2)}\otimes I
    \end{matrix}\right].
\end{align}
We can now recursively construct the QTT
\begin{align}
    \left[ \begin{matrix}
        \mathbf{M}^{(l)} \\
        t J^{\otimes (l)} \\
        t J'^{\otimes (l)} \\
        t'  J^{\otimes (l-1)}\otimes I\\
        t'  J'^{\otimes (l-1)}\otimes I
    \end{matrix} \right] 
    & = \left[ \begin{matrix}
        I & K & K & K & K \\
          & J &   &   &   \\
          &   & J'&   &   \\
          &   &   & J &   \\
          &   &   &   & J'
    \end{matrix} \right] \lrtimes
    \Bigg( \left[ \begin{matrix}
        I & K & K & K & K \\
          & J &   &   &   \\
          &   & J'&   &   \\
          &   &   & J &   \\
          &   &   &   & J'
    \end{matrix} \right] \lrtimes
    \left[ \begin{matrix}
        \mathbf{M}^{(l-2)} \\
        t J^{\otimes (l-2)} \\
        t J'^{\otimes (l-2)} \\
        t'  J^{\otimes (l-3)}\otimes I\\
        t'  J'^{\otimes (l-3)}\otimes I
    \end{matrix}\right] \Bigg) \nonumber \\
    &= ... = \left[ \begin{matrix}
         I & K & K & K & K \\
          & J &   &   &   \\
          &   & J'&   &   \\
          &   &   & J &   \\
          &   &   &   & J'
    \end{matrix} \right] \lrtimes
    \left[ \begin{matrix}
         I & K & K & K & K \\
          & J &   &   &   \\
          &   & J'&   &   \\
          &   &   & J &   \\
          &   &   &   & J'
    \end{matrix} \right]^{\lrtimes (l-2)} \lrtimes
    \left[ \begin{matrix}
        \mathbf{M}^{(1)} \\
        t J \\
        t J'\\
        t' I\\
        t' I
    \end{matrix}\right] ,
\end{align}
where $\mathbf{M}^{(1)} = (\omega+2t') I + 2t H = \left( \begin{matrix}
    (\omega+2t') & 2 t \\
    2 t & (\omega+2t')
\end{matrix}\right)$. By considering $l=R$ and selecting the first row in the first core (and insertion of the renormalized couplings $\omega^{(r)},t^{(r)},t'^{(r)}$ for the derivation of $\mathbf{M}_r$), we conclude the proof for the formulas in Eqs. (\ref{eq:nnn-qtt-mpo-M-explicit-formula-1}, \ref{eq:nnn-qtt-mpo-M-explicit-formula-2}). $\Box$

\subsection{\label{app:nnn-proof-Delta}Proof of Eqs.~(\ref{eq:nnn-qtt-mps-D-explicit-formula-1}, \ref{eq:nnn-qtt-mps-D-explicit-formula-2})}
Here, we will prove that the vector $\mathbf{D} = \left(1,a,0,...,0,a \right)^T$ can be exactly decomposed into a QTT with bond dimension 2 in the following way (Eqs. (\ref{eq:nnn-qtt-mps-D-explicit-formula-1}, \ref{eq:nnn-qtt-mps-D-explicit-formula-2}))
\begin{align}
    \mathbf{D}^{(R)} = \left[ \begin{matrix}
        C & F  
    \end{matrix} \right] \lrtimes \left[ \begin{matrix}
        C & F  \\
          & B 
    \end{matrix}\right]^{\lrtimes (R-2)} \lrtimes \left[ \begin{matrix}
        C + 2aB \\
        a B 
    \end{matrix}\right],
\end{align}

\textit{Proof.} We make use of the recursive block structure
\begin{align}
    \mathbf{D}^{(l)} &= \left( \begin{matrix}
        \mathbf{D}^{(l-1)} - a B^{\otimes (l-1)} \\
        a B^{\otimes (l-1)}
    \end{matrix}\right) = C \otimes \mathbf{D}^{(l-1)} + a F \otimes B^{\otimes (l-1)}. 
\end{align}
which holds for $2 \leq l \leq R$. Hence, we have
\begin{align}
    \mathbf{D}^{(l)} = \left[ \begin{matrix}
        C & F
    \end{matrix} \right] \lrtimes
    \left[ \begin{matrix}
        \mathbf{D}^{(l-1)} \\
        a B^{\otimes (l-1)} 
    \end{matrix}\right], 
\end{align}
which we can go further to
\begin{align}
    \left[ \begin{matrix}
        \mathbf{D}^{(l)} \\
        a B^{\otimes (l)}
    \end{matrix} \right] = 
    \left[ \begin{matrix}
        C & F \\
          & B
    \end{matrix} \right] \lrtimes
    \left[ \begin{matrix}
        \mathbf{D}^{(l-1)} \\
        a B^{\otimes (l-1)} 
    \end{matrix}\right], 
\end{align}
We can now recursively construct the QTT
\begin{align}
    \left[ \begin{matrix}
        \mathbf{D}^{(l)} \\
        a B^{\otimes (l)}
    \end{matrix} \right] &= 
    \left[ \begin{matrix}
        C & F \\
          & B
    \end{matrix} \right] \lrtimes \Bigg(
    \left[ \begin{matrix}
        C & F \\
          & B
    \end{matrix} \right] \lrtimes
    \left[ \begin{matrix}
        \mathbf{D}^{(l-2)} \\
        a B^{\otimes (l-2)} 
    \end{matrix}\right] \Bigg) 
    = ... =  
    \left[ \begin{matrix}
        C & F \\
          & B
    \end{matrix} \right] \lrtimes 
    \left[ \begin{matrix}
        C & F \\
          & B
    \end{matrix} \right]^{\lrtimes (l-2)} \lrtimes
    \left[ \begin{matrix}
        \mathbf{D}^{(1)} \\
        a B 
    \end{matrix}\right]
\end{align}
where $\mathbf{D}^{(1)} = C+2aB = \left( \begin{matrix}
    1 \\
    2a
\end{matrix}\right)$. By considering $l=R$ and selecting the first row in the first core (and insertion of the renormalized coupling $a^{(r)}$ for the derivation of $\mathbf{D}_r$), we conclude the proof for the formulas in Eqs. (\ref{eq:nnn-qtt-mps-D-explicit-formula-1}, \ref{eq:nnn-qtt-mps-D-explicit-formula-2}). $\Box$

\section{\label{app:error-bounds}Error bounds}
\subsection{\label{app:general-error-bounds}General error bounds}
In this section, the error bound in Eq.~\eqref{eq:general-error-bound} will be derived. Making use of
\begin{align}
    \frac{1}{x+\delta} = \frac{1}{x} - \frac{1}{x^2}\delta +....
\end{align}
for the inversion of the circulant matrix $A=B+R$, with decaying couplings $t_n$, where $B$ contains all couplings $t_n, n \leq m$ and $R$ all $t_n, n >m$ leads to the following derivation of the error bound of the approximation of $A^{-1}$ by $B^{-1}$
\begin{align}
    \epsilon &= ||A^{-1}-B^{-1}|| = ||(B+R)^{-1}-B^{-1}||  = ||B^{-1} - B^{-2}R + \frac{1}{4}B^{-3}R^2-\frac{1}{36}B^{-4}R^3 + ... - B^{-1}|| \nonumber \\
    &\quad = \Big| \Big| \sum_{j=2}^{\infty} (-1)^{j-1} B^{-j} R^{j-1} \cdot \left( \frac{1}{(j-1)!} \right)^2 \Big| \Big|  \leq \sum_{j=2}^{\infty} || B^{-j}||\cdot ||R^{j-1}|| \cdot \left( \frac{1}{(j-1)!} \right)^2 \nonumber \\
    &\quad \leq ||B^{-2}|| \cdot ||R|| \cdot \sum_{j=0}^{\infty} || B^{-j}||\cdot ||R^{j}|| \cdot \left( \frac{1}{(j+1)!} \right)^2  \leq ||B^{-1}||^2 \cdot ||R|| \cdot \sum_{j=0}^{\infty} || B^{-1}||^j \cdot ||R||^j \cdot \left( \frac{1}{j!} \right) \nonumber \\
    &\quad \leq ||R|| \cdot ||B^{-1}||^2 \cdot \exp(||R|| \cdot ||  B^{-1}||). 
\end{align}
Let us briefly elaborate on the tightness of the bound. In order to bring the result to exponential form, the strongest approximation is applied in the third line of the derivation, where, among others, we use $\left( \frac{1}{(j+1)!} \right)^2 \leq \left( \frac{1}{j!} \right)$. Since these factors are multiplied by $|| B^{-1}||^j \cdot ||R||^j$ in the sum, the smaller $|| B^{-1}|| \cdot ||R||$ the tighter the bound will be. Therefore, we expect a tighter bound for exponentially decaying couplings, where the factor $|| B^{-1}|| \cdot ||R||$ is expected to be more suppressed than in the case of a power-law decay as can be seen in Fig.~\ref{fig:error-bound}.

\subsection{\label{app:error-bounds-exponential}Exponentially decaying couplings}
In the case of exponentially decaying couplings $t_n = t_{-n} = C e^{-\alpha n}, C,\alpha \in \mathbb{R}^{+}, n \in \mathbb{N}, t_0=\omega$ we aim at obtaining a priori error bounds for truncating the hoppings $|n|>m (t_n =0)$in the matrix $B$, from which a QTT with maximum bond dimension $2m$ can be constructed. We start by calculating the Fourier series for the matrix $B$
\begin{align}
    &f_B(\phi) = \sum_{n=-\infty}^{\infty} t_n e^{i \phi n} = \omega + C \left( \sum_{n=1}^{m} (e^{-\alpha+ i \phi})^n + \sum_{n=1}^{m} (e^{-\alpha- i \phi})^n \right) 
    = \omega + C \left( -2 + \sum_{n=0}^{m} (e^{-\alpha+ i \phi})^n + \sum_{n=0}^{m} (e^{-\alpha- i \phi})^n \right) \nonumber \\
    &= \omega + C \left( -2 + \frac{1 - (e^{i \phi - \alpha})^{m+1}}{1 - e^{-\alpha + i \phi}} + \frac{1 - (e^{-i \phi - \alpha})^{m+1}}{1 - e^{-\alpha - i \phi}} \right) 
    = \omega + \frac{2 C \left( -1 + e^{\alpha} \cos(\phi) + e^{-\alpha m} \left[ \cos(m \phi) - e^{\alpha} \cos( (m+1) \phi)\right] \right)}{1 - 2 e^{\alpha} \cos(\phi) + e^{2 \alpha}} 
\end{align}
By looking at the error bound derived in Eq.~\eqref{eq:general-error-bound}, we need to find the lower bound on the eigenvalues of $B$ and then invert this value. Hence, we need to find the minimum of $f_B(\phi)$. We find the minimum of $f_B(\phi)$ at $\phi=\pi$, for which we get
\begin{align}
    m_{f_B} = f_B(\phi=\pi) = \omega + \frac{2C \left[-1 + e^{-\alpha m} \cos(m \pi)\right]}{1+e^{\alpha}}
\end{align}
as the lower bound $m_{f_B}$ on the spectrum of $B$. Similarly, we can derive the upper bound $M_{f_R}$ on the spectrum of $R$.
\begin{align}
    f_R(\phi) &= \sum_{n=-\infty}^{\infty} t_n e^{i \phi n} =  C \left( \sum_{n=m+1}^{N} (e^{-\alpha+ i \phi})^n + \sum_{n=m+1}^{N} (e^{-\alpha- i \phi})^n \right)  \\
    &=  \frac{2C}{1-2 e^{-\alpha} \cos(\phi) + e^{-2\alpha}} \left[ e^{-\alpha(m+1)} \cos((m+1)\phi)  - e^{-\alpha(N+1)} \cos((N+1)\phi) - e^{-\alpha(m+2)} \cos(m\phi)   + e^{-\alpha(N+2)} \cos(N\phi) \right] \nonumber
\end{align}
Since $||R||_2 = max_{\lambda_i \in sp(R)} |\lambda_i|$, we are interested in an upper bound for the maximum eigenvalue of $R$. We find $M_{f_R}$ as the maximum value of $f(\phi)$ at $\phi=0$.
\begin{align}
    M_{f_R}= f(\phi=0) = \frac{2C \left(e^{-\alpha m} - e^{-\alpha N}\right)}{e^{\alpha}-1}
\end{align}

\subsection{\label{app:error-bounds-power-law}Power-law decay}
Now we consider the model with power-law hoppings $t_n = t_{-n} = C n^{-\alpha}, C \in \mathbb{R}, n \in \mathbb{N}, \alpha > 1, t_0=\omega$. In the same way as before, we can calculate the Fourier series for the matrices B and R and look for the respective minimum and maximum. 
\begin{align}
    &f_B(\phi) = \sum_{n=-\infty}^{\infty} t_n e^{i \phi n} 
    = \omega + C \left( \sum_{n=1}^{m} n^{-\alpha} e^{i n \phi} + \sum_{n=1}^{m} n^{-\alpha} e^{-i n \phi} \right) 
    = \omega + 2C \sum_{n=1}^{m} n^{-\alpha} \cos (n \phi)
\end{align}
We get the minimum at $\phi=\pi$.
\begin{align}
    m_{f_B} = f_B(\phi=\pi) = \omega + 2C \sum_{n=1}^{m} (-1)^n n^{-\alpha}
\end{align}
Similar for the matrix $R$, we get
\begin{align}
    &f_R(\phi) = \sum_{n=-\infty}^{\infty} t_n e^{i \phi n} 
    = C \left( \sum_{n=m+1}^{N} n^{-\alpha} e^{i n \phi} + \sum_{n=m+1}^{N} n^{-\alpha} e^{-i n \phi} \right) 
    =  2C \sum_{n=m+1}^{N} n^{-\alpha} \cos (n \phi).
\end{align}
The Fourier series takes the maximum at $\phi=0$
\begin{align}
    M_{f_R}= f(\phi=0) = 2C \sum_{n=m+1}^{N} n^{-\alpha} .
\end{align}

\subsection{\label{app:error-bounds-extended-figure}Extension of Fig.~\ref{fig:error-bound}}
In this section, in Fig.~\ref{fig:error-bound-reply},  we repeat Fig.~\ref{fig:error-bound}  
with extended $\alpha$-range in panels (a) and (c), which makes it more evident that the derived error bound (red curve) indeed serves as an upper bound on the error.

\begin{figure}[H]
    \centering
    \includegraphics[width=0.5\linewidth]{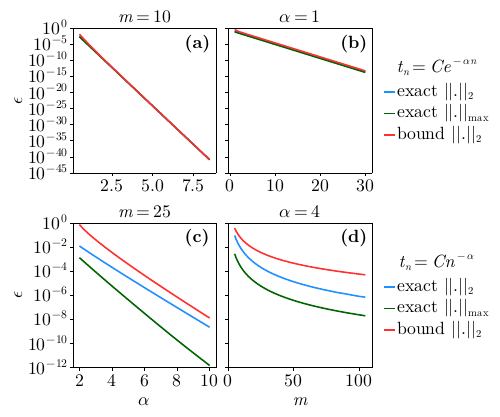}
    \caption{ Approximation error of $G=A^{-1}:=(\omega \mathbb{I} -H)^{-1}$ by a QTT with maximum bond dimension \Dmax$=2m$ with increased $\alpha$-range. (a) and (b) show the case of exponentially decaying couplings $t_n=Ce^{-\alpha n}$, while (c) and (d) present results for power-law decaying hoppings $t_n=Cn^{-\alpha}$ for the tight-binding model with $C=1, \omega=2$ and 1001 sites, where $m$ denotes the number of nonzero hoppings in the approximating matrix. (a) and (c) show the errors for fixed $m$ as a function of $\alpha$ and (b) and (d) for fixed $\alpha$ as a function of $m$. The calculated a priori error bound (red) and the exact approximation error (blue), both with respect to the spectral norm, as well as the exact maximum normalized error (green) are shown.}
    \label{fig:error-bound-reply}
\end{figure}

\section{\label{app:spectral-function}Spectral function of 2d interacting model}
In Fig.~\ref{fig:spectral-function}, the k-integrated spectral function $A(\omega) = -\frac{1}{\pi} \underset{\eta \rightarrow 0 }{\lim} \int dk_x \int dk_y G(\mathbf{k},\omega_n \rightarrow \omega + i \eta)$ of the 2d interacting case of Sec.~\ref{sec:interacting-model} [cf. Fig.~\ref{fig:Dmax_tol1e-5_interdim}(b)] is shown, where $\omega$ denote real and $\omega_n$ Matsubara frequencies. It can be seen that at $\mu=\frac{U}{2}+4=6$  spectral weight is shifted to $\omega=0$ resulting in a metal. Such metal to insulator transition as a function of the chemical potential $\mu$ is similar to the one observed in the Falicov-Kimball model~\cite{Falicov1969,Freericks2003}, where the self-energy also does not depend on the chemical potential and the spectrum just shifts with the change of $\mu$.

\begin{figure}[H]
    \centering
    \includegraphics[width=\linewidth]{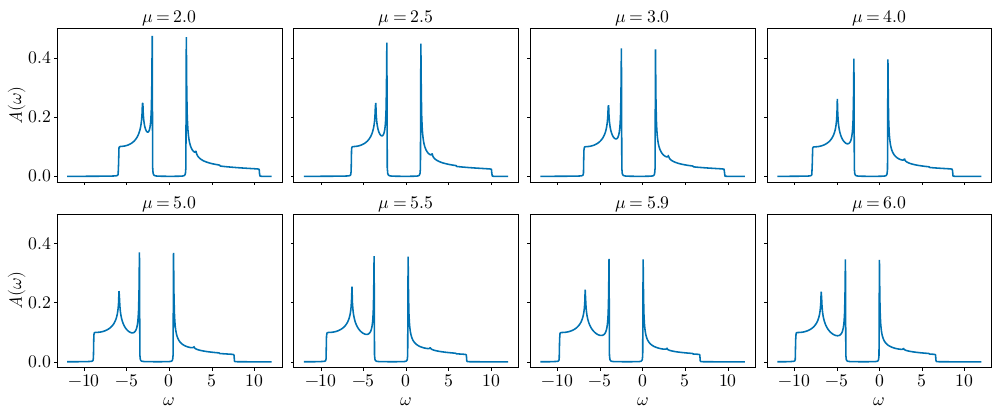}
    \caption{Local (momentum integrated) spectral function $A(\omega)$ of the 2d interacting model [the same as in in Fig.~\ref{fig:Dmax_tol1e-5_interdim}(b)] for different values of the chemical potential $\mu$ at $U=4$ with $\eta=10^{-2}$.}
    \label{fig:spectral-function}
\end{figure}

\end{widetext}

\bibliography{Bibliography_QTTrenormalization}
\end{document}